\documentclass[aps,pre,groupedaddress,showkeys,showpacs,amsmath]{revtex4-1}
\usepackage{amssymb}
\usepackage{amsmath}
\usepackage{graphicx}
\usepackage{amsbsy}
\usepackage{bm}
\usepackage{xcolor}
\usepackage{gensymb}
\usepackage{caption}
\usepackage{subcaption}
\usepackage[space]{grffile}
\definecolor{cream}{RGB}{222,217,201}
\usepackage{siunitx}
\DeclareSIUnit[number-unit-product = {\,}]
\cal{cal}
\DeclareSIUnit\kcal{\kilo\cal}
\DeclareSIUnit\kcal{\kilo\joule\per\mole}
\DeclareSIUnit\molar{\mole\per\cubic\deci\metre}
\DeclareSIUnit\Molar{\textsc{m}}
\usepackage{multirow}

\usepackage{url} 
\usepackage{multirow}
\usepackage{hyperref}
\usepackage{lineno}
\usepackage{soul}
\usepackage{xr}
\usepackage[english]{babel}
\usepackage{comment}

\begin{document}

\title{Computational Pipeline to probe NaV1.7 gain-of-functions variants in neuropathic painful syndromes}

\author{Alberto Toffano}
\affiliation{Universit\'a Ca' Foscari Venezia, Dipartimento di Scienze Molecolari e Nanosistemi, Venezia-Mestre, Italy}
\author{Giacomo Chiarot}
\affiliation{Universit\'a Ca' Foscari Venezia, Dipartimento di Scienze Ambientali, Informatica e Statistica,  Venezia-Mestre, Italy}
\author{Stefano Zamuner}
\affiliation{\`Ecole Polytechnique F\`ed\`erale de Lausanne (EPFL), Laboratory of Statistical Biophysics, Institute of Physics, School of Basic Sciences, Lausanne, Switzerland}
\author{Margherita Marchi}
\affiliation{Fondazione IRCCS Istituto Neurologico “Carlo Besta”, Neuroalgology Unit, Milan, Italy}
\author{Erika Salvi}
\affiliation{Fondazione IRCCS Istituto Neurologico “Carlo Besta”, Neuroalgology Unit, Milan, Italy}
\author{Stephen G. Waxman}
\affiliation{Center for Neuroscience and Regeneration Research, VA Connecticut Healthcare System and Yale Medical School, West Haven, USA}
\author{Catharina G. Faber}
\affiliation{Maastricht University, MHeNs school for Mental Health and Neuroscience, Maastricht, The Netherlands}
\affiliation{Maastricht University Medical Center, Department of Neurology, Maastricht, The Netherlands}
\author{Giuseppe Lauria}
\affiliation{Fondazione IRCCS Istituto Neurologico “Carlo Besta”, Neuroalgology Unit, Milan, Italy}
\affiliation{University of Milan, Department of Biomedical and Clinical Sciences “Luigi Sacco”, Milan, Italy} 
\author{Achille Giacometti}
\affiliation{Universit\'a Ca' Foscari Venezia, Dipartimento di Scienze Molecolari e Nanosistemi, Venezia-Mestre, Italy}
\affiliation{European Centre for Living Technology (ECLT), Venice, Italy}
\author{Marta Simeoni}
\email{simeoni@unive.it}
\affiliation{Universit\'a Ca' Foscari Venezia, Dipartimento di Scienze Ambientali, Informatica e Statistica,  Venezia-Mestre, Italy}
\affiliation{European Centre for Living Technology (ECLT), Venice, Italy}

\date{\today}


\begin{abstract}
Applications of machine learning and graph theory techniques to neuroscience have witnessed an increased interest in the last decade due to the large data availability and unprecedented technology developments. Their employment to investigate the effect of mutational changes in genes encoding for proteins modulating the membrane of excitable cells, whose biological correlates are assessed at electrophysiological level, could provide useful predictive clues. We apply this concept to the analysis of variants in sodium channel NaV1.7 subunit found in patients with chronic painful syndromes, by the implementation of a dedicated computational pipeline empowering different and complementary techniques including homology modeling, network theory, and machine learning. By testing three templates of different origin and sequence identities, we provide an optimal condition for its use. Our findings reveal the usefulness of our computational pipeline in supporting the selection of candidates for cell electrophysiology assay and with potential clinical applications.
\end{abstract}

\maketitle
%
%
%


\section{Introduction}
\label{sec:introduction}
\begin{figure}[htbp]
\begin{subfigure}{0.48\textwidth}
    \includegraphics[width=\textwidth]{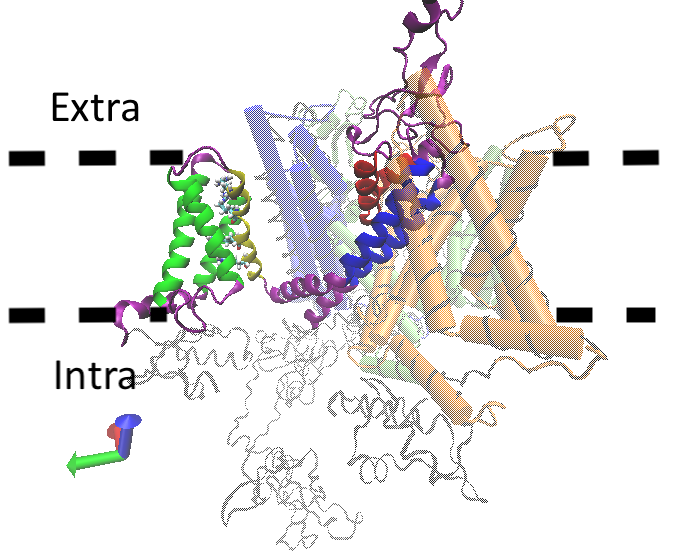} 
    \caption{}
\end{subfigure}
\begin{subfigure}{0.48\textwidth}
    \includegraphics[width=\textwidth]{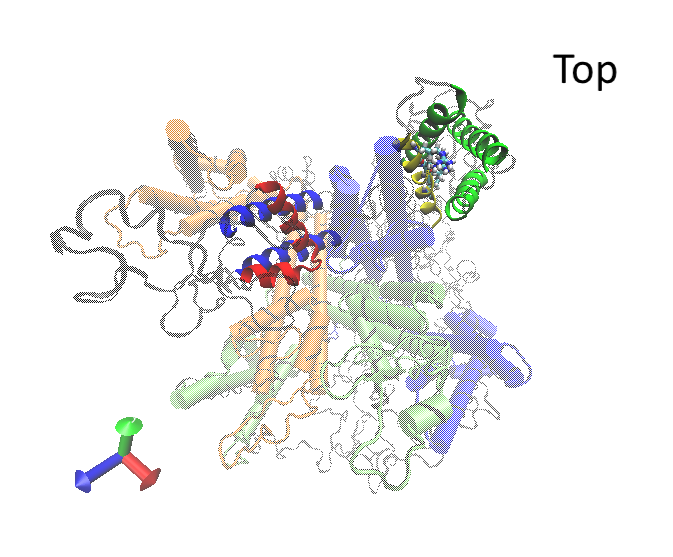}
    \caption{}
\end{subfigure}
\caption{Snapshot of the Nav1.7 wild type. (a) Side view reporting high details of the Domain I: transmembrane helical segments S1, S2, and S3 are colored in green, segment S4 in yellow, segments S5 and S6 constituting the Pore Domain are identified by blue color, and the P-loop between S5 and S6 that helps to form the Selectivity Filter is colored in red. Thick dotted lines labelled as Intra and Extra identify the transmembrane region; (b) Top view, Domain I is displayed in full colors as in (a), and the coordination of the four Domains is clearly visible. Here shaded blue, green and orange identify Domains II, III, and IV, respectively. }
    \label{fig:snapshot}
\end{figure}

NaV1.7 is responsible of the propagation of the pain stimuli through the peripheral nervous system. It belongs to the family of Voltage Gated Sodium Channels (VGSCs) proteins expressed both in the prokaryotic and the eukaryotic realms. The most known biophysical activity is carried out by the $\alpha$-subunits that, for eukarya, is formed by the three-dimensional rearrangement of a single polypeptide chain of almost 2,000 amino acids. This chain is wrapped in four domains (DI-DIV), each composed of six transmembrane helices, with the first four S1-S4 forming the voltage-sensing domain (VSD), and the last two S5-S6, as well as the extracellular linkers (P-loop) between S5 and S6, forming the selectivity filter (SF) and the central aqueous pore domain (PD), as shown in Figure \ref{fig:snapshot}. The reaction to the change of membrane potential is mediated by helix S4, thanks to the positively charged amino acids arginine (R) and lysine (K) located along the segment. 
Note that the positive charges are wrapped and stabilized by amino acids present in the helices from S1 to S3 \cite{Catterall201435}. Depolarization of the membrane is believed to produce a motion toward the extracellular side of S4 segments of each domain, and this motion is transferred to the pore domain via intracellular linkers between the segments S4 and S5. The conformational change results in the opening of the channel pore. The segments S5 and S6 are scaffold of the pore channel, while the extracellular linkers are responsible for the selectivity filter to Na ions that is highly conserved and composed of aspartate (D) in DI, glutamate (E) in DII, lysine (K) in DIII, and alanine (A) in DIV  forming a ring, which is the geometrically narrowest region of the ion pore \cite{DeLeraRuiz20157093}, see Figure \ref{fig:domains}. In mammals, there are nine different isoforms of $\alpha$-subunits, Nav1.1‚-Nav1.9, and their ratio of expression changes among different type of tissues. The isoform NaV1.7 we consider, which is encoded by \textit{SCN9A}, is highly expressed in nociceptors, where it has a pivotal role in subthreshold membrane depolatization. Here, single aminoacid substitutions in patients diagnosed with inherited erythromelalgia (IEM), paroxysmal extreme pain disorder (PEPD), small fiber neuropathy (SFN) and painful diabetic neurophaty (PDN) \cite{Emery} were found to induce a gain-of-function of the channel.

A significant challenge that is posed to the above scenario is given by the large number of possible case studies, as well as by the difficulties involved in performing electrophysiology studies at large scales. On the other hand, technologies such as artificial intelligence can be applied to gain insights from multiple data sources and rationalize them to improve clinical decision-making ability and improve patient outcomes. This is particularly true in neuroscience where recent findings  \cite{Ullman2019} have show the power of these tools both as classificators and predictors, thus envisaging the use of computational pipelines in workflows fully integrated into clinical applications. Indeed, it was recently suggested \cite{Waxman14} that the available computerized predictive algorithms “are not infallible” and a more holistic approach was in order.

 Building upon past work by our group \cite{Dimos}, in this study we provide one such computational pipeline combining homology modeling \cite{Lee1992}, Residue Interaction Networks (RINs) \cite{web:RING2.0, RINGpaper} and machine learning techniques \cite{GraphKernels,RotaBulo2017}, to determine whether there is a common structural pattern linking gain-of-function pathogenic mutations to each other.  

Starting from a given template, we identify a set of sequences representative of some known gain-of-function pathogenic mutations (\textbf{PAT}), as well as other neutral (\textbf{NEUTRAL}) variants, and use homology modeling to generate the corresponding three dimensional structures. In order to make the relative comparison of these structures feasible, the information of the topology of each structure is then used to map them into their corresponding graphs using RINs, thus reducing the dimensionality of the problem and hence its complexity. This step then allows for a pairwise comparison of all the obtained networks via graph kernel techniques, and for classifying them through machine learning techniques that are able to identify common structural patterns. 
Note that here \textbf{NEUTRAL} refers to genetic variants selected either from annotated sequence substitutions not known to affect a protein’s function and possibly contribute to genetic disease or from pseudo mutations between orthologous NaV1.7 proteins in closely related mammals.

Three different templates are presented as case studies. The first one is the same template used in past work Ref.\cite{Dimos} and it will be used as a benchmark for our analysis. The other two templates derive from two other homologous proteins with greater identities with the original sequence. All three templates depict the sodium channel protein in its closed state and have different sequence identities. The comparative results from these three different case studies will provide useful information on the general potentialities of the proposed workflow, the optimal condition for its use, as well as the limiting factors that need to be addressed to make it an useful tool in clinical applications.

In short, the main aim of our computational workflow is to predict whether a mutation is expected to be deleterious and be able to be distinguished from all others. We then benchmark our method by applying it to the 85  \textit{SCN9A} mutations reported in Figure \ref{fig:mutations} (a) and see whether it is able to classify \textbf{PAT} mutations and discriminate them from \textbf{NEUTRAL} ones. 

It is important to emphasize that the present work considers only gain-of-function mutations and relies upon the critical assumption that the functional impact of a point mutation is caused by changes in residue-residue interaction network only, while several additional effects can be induced by point mutation and affect the functionality of the channel. Indeed ion-channels function is determined by many factors besides the shift in gating kinetics. At least 30 mutations have been described in scientific literature for impairing the Nav1.7 function \cite{HGMD}, either by haploinsufficiency effect or by hindering the selective permeability of the pore. However, these kinds of null-mutations result in a lower quantity of functional Nav1.7 on the membrane, and are associated with Congenital Insensitivity to Pain (CIP) a disease characterized by pain signaling defects \cite{Cox2006,Drenth2007}.

\section{Results and Discussion}
\label{sec:results}
\begin{figure}[htbp]
    \centering
    \includegraphics[width=0.8\textwidth]{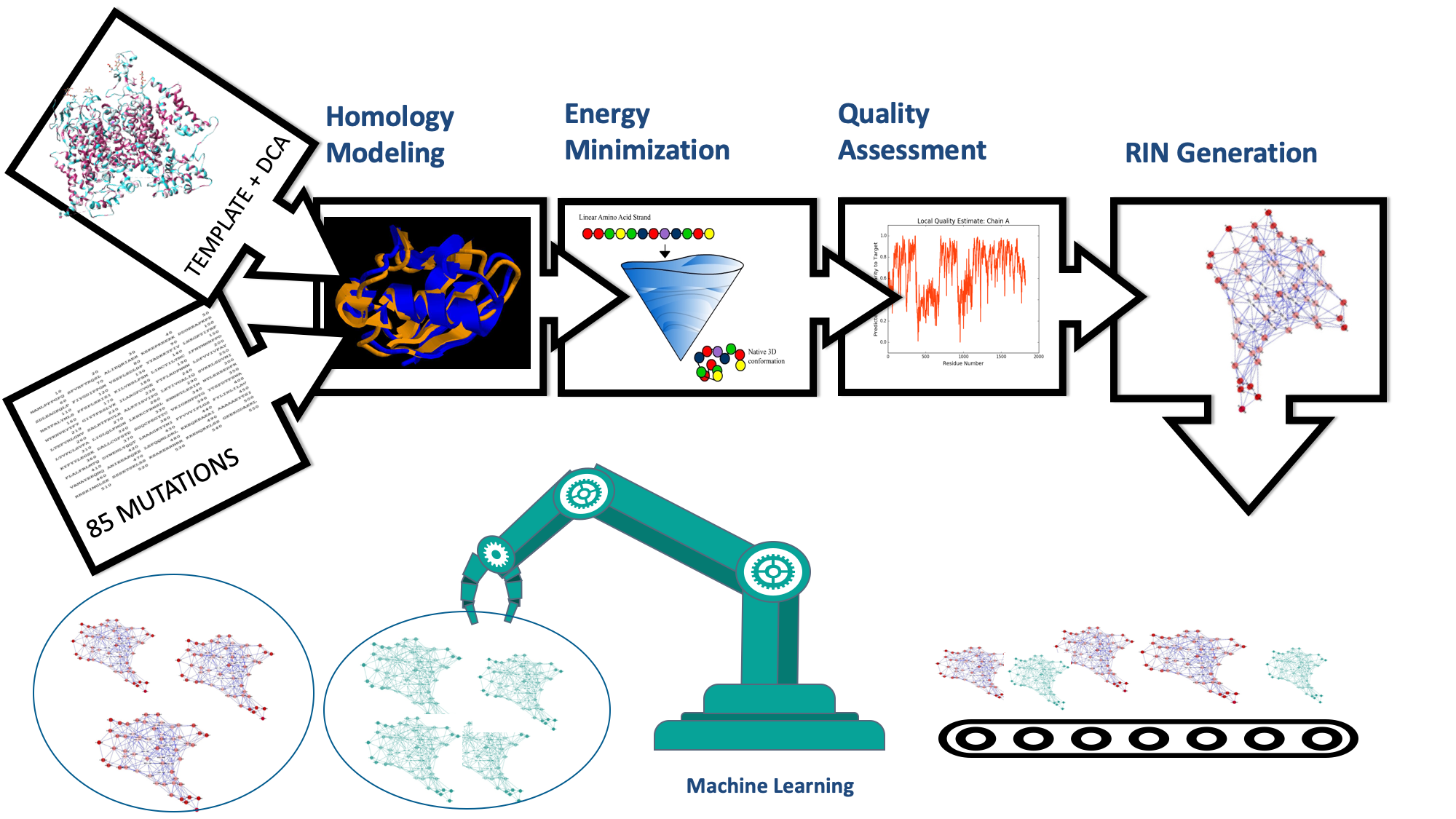}
    \caption{The computational pipeline. Starting from a template and 85 genetic variants, homology models are used to identify the corresponding three-dimensional structures, followed by energy minimization and quality assessment to refine them. RINs are then implemented to map them into their representing graphs and machine learning techniques are used to analyze them and identify patterns.}
    \label{fig:workflow}
\end{figure}

\begin{figure}[htbp]
  \centering
  \includegraphics[scale=0.32]{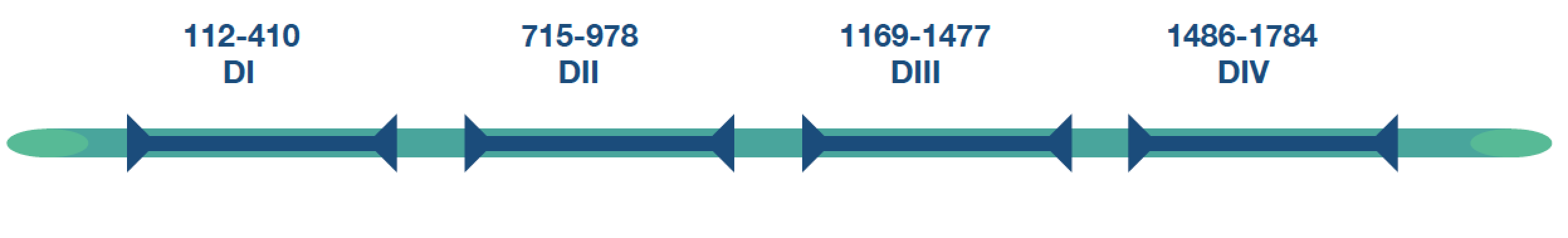}
  \includegraphics[scale=0.235]{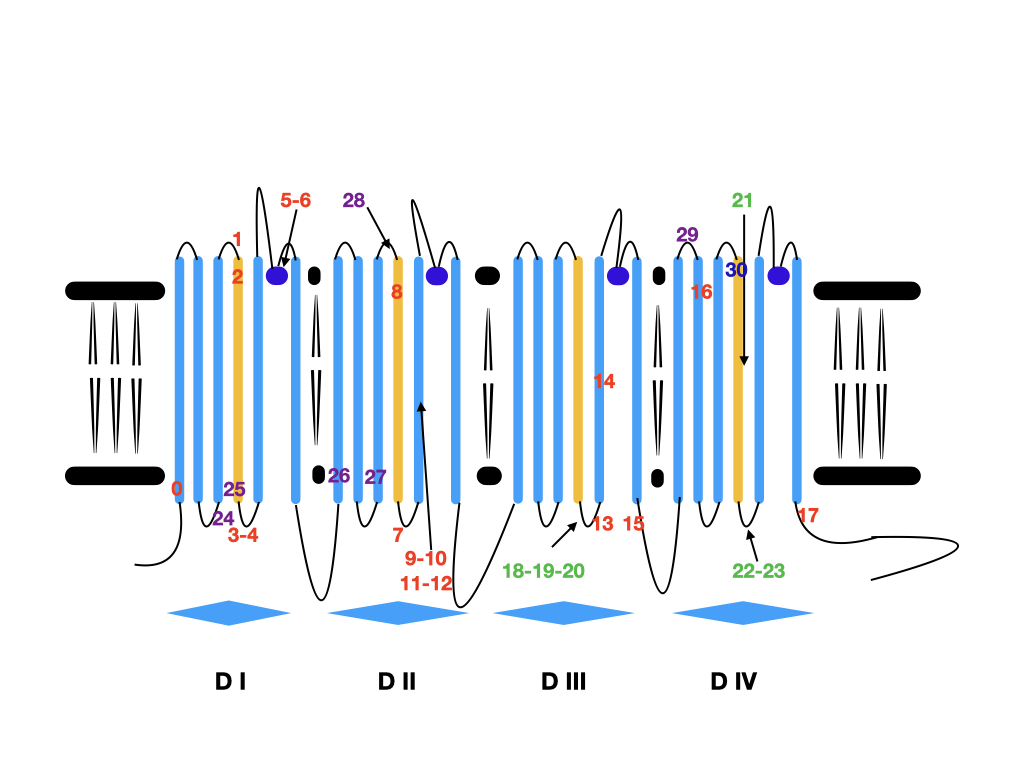}
    ~~~~
    \includegraphics[scale=0.235]{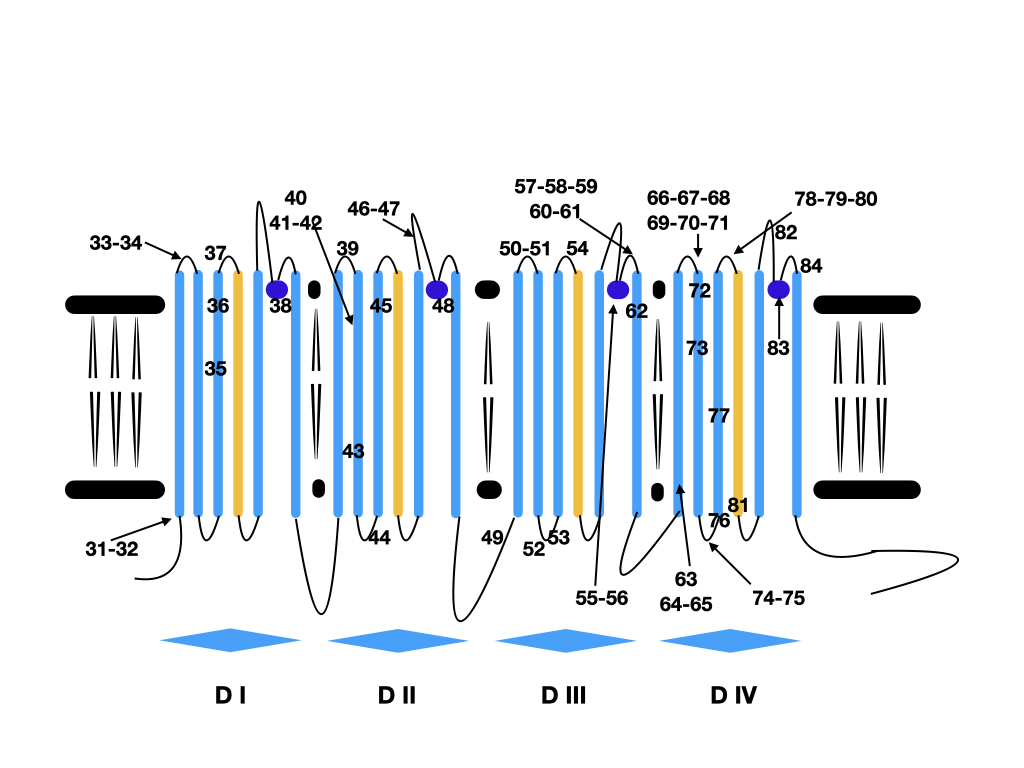}    
    \caption{(Top) Primary structure and domains positions; (Bottom) Schematic illustration of the  poly-peptide chain structure and localization of pathogenic mutations (\textbf{PAT}) associated with pain conditions (left) and not pathogenic variants (\textbf{NEUTRAL}) (right). Colored numbers (left) highlight the four different pathologies, color-coded according to the list shown in Figure~\ref{fig:mutations}}
    \label{fig:domains}
\end{figure}

Our computational pipeline is sketched in Figure \ref{fig:workflow} in the specific case of human protein Nav1.7 
As in Ref. \cite{Dimos}, the NCBI sequence \textbf{NP\_002968.1} has been chosen as the Wild Type (WT) sequence for our study. It is classified as the first isoform of the transcript variant 1, identification code NM\_002977.3 for the chromosomal sequence \cite{NCBI}. It is composed by four domains DI-DIV as pictured in Figure \ref{fig:domains} (top), located at positions DI (112-410), DII (715-978), DIII (1169-1477), DIV (1488-1784). Figure \ref{fig:domains} (bottom) provides its pictorial representation along the sequence, and Figure \ref{fig:snapshot} its snapshot both as side (a) and top (b) views.

\begin{figure}[htbp]
\begin{subfigure}{0.55\textwidth}
    \centering
    \scriptsize
    \begin{tabular}{|lll|llll|}
    \hline
    \multicolumn{3}{|c|}{\textbf{PAT}} &  
    \multicolumn{4}{|c|}{\textbf{NEUTRAL}} \\ 
    \hline
    disease &id & mutation & id & mutation & id & mutation\\
    \hline
    \multirow{18}{*}{\textcolor{red}{IEM}} 
    &0 & \textcolor{red}{I136V}\cite{Lee2007,I136V.2,I136V.3}  & 31 & S126A & 62 & \textbf{V1428I}$^{*}$\\ 
    &1 & \textcolor{red}{S211P}\cite{S211P}  & 32 & L127A & 63 & \textbf{A1505V} \\
    &2 & \textcolor{red}{F216S}\cite{F216S.1,F216S.2}  & 33 & \textbf{M145L}& 64 & S1509A\\
    &3 & \textcolor{red}{I234T}\cite{I234T}  & 34 & N146S & 65 & S1509T\\ 
    &4 & \textcolor{red}{S241T}\cite{S241T.1,S241T.2,S241T.3}  & 35 & V194I & 66 & Q1530D\\
    &5 & \textcolor{red}{N395K}\cite{F216S.1,N395K}  & 36 & \textbf{L201V}& 67 & \textbf{Q1530K}\\ 
    &6 & \textcolor{red}{V400M}\cite{S241T.1,V400M}  & 37 & \textbf{N206D} & 68 & Q1530P\\ 
    &7 & \textcolor{red}{L823R}\cite{L823R,L823R.2} & 38 & \textbf{T370M} & 69 & \textbf{H1531Y}\\ 
    &8 & \textcolor{red}{I848T}\cite{I848T.1,I848T.2,I848T.3,Yue06,I848T.5} & 39 & E759D & 70 & \textbf{M1532V}\\ 
    &9 & \textcolor{red}{L858H}\cite{I848T.1, I848T.2, I848T.3}  & 40 & A766T & 71 & E1534D\\
    &10 & \textcolor{red}{L858F}\cite{L858F.1, F216S.1, L858F.3}  & 41 & \textbf{A766V} & 72 & Y1537N\\
    &11 & \textcolor{red}{A863P}\cite{A863P} & 42 & \textbf{I767V} & 73 & T1548S\\ 
    &12 & \textcolor{red}{V872G}\cite{V872G}& 43 & \textbf{T773S} & 74 & H1560C\\ 
    &13 & \textcolor{red}{P1308L}\cite{P1308L} & 44 & V795I & 75 & H1560Y\\
    &14 & \textcolor{red}{V1316A}\cite{V1316A.1, V1316A.2} & 45 & A815S & 76 & V1565I\\
    &15 & \textcolor{red}{F1449V}\cite{F1449V.1, S241T.1} & 46 & D890E  & 77 & I1577L\\
    &16 & \textcolor{red}{W1538R}\cite{W1538R} & 47 & D890V &  78 & \textbf{D1586E}\\ 
    &17 & \textcolor{red}{A1746G}\cite{W1538R} & 48 & \textbf{T920N}$^{*}$ & 79 & \textbf{T1590K}\\  
    \multirow{6}{*}{\textcolor{teal}{PEPD}}
    &18 & \textcolor{teal}{V1298D}\cite{V1298D} & 49 & K1176R &  80 & T1590R\\  
    &19 & \textcolor{teal}{V1298F}\cite{V1298D, V1298F.2, P1308L} & 50 & \textbf{R1207K} & 81 & \textbf{V1613I}\\ 
    &20 & \textcolor{teal}{V1299F}\cite{V1298D,V1298F.2,I848T.3} & 51 & T1210N & 82 & V1662A\\ 
    &21 & \textcolor{teal}{G1607R}\cite{G1607R} & 52 & I1235V & 83 & G1674A\\ 
    &22 & \textcolor{teal}{M1627K}\cite{V1298D,M1627K.2,P1308L,M1627K.4} & 53 & \textbf{N1245S}$^{*}$ & 84 & K1700A\\     
    &23 & \textcolor{teal}{A1632E}\cite{A1632E} & 54 & \textbf{L1267V}$^{*}$ &  &\\     
    \multirow{6}{*}{\textcolor{violet}{SFN}}
    &24 & \textcolor{violet}{R185H}\cite{R185H} & 55 & \textbf{T1398M} &  &  \\     
    &25 & \textcolor{violet}{I228M}\cite{R185H,I228M.2} & 56 & I1399D && \\    
    &26 & \textcolor{violet}{I739V}\cite{R185H,I739V.2} & 57 & D1411N && \\    
    &27 & \textcolor{violet}{G856D}\cite{G856D} & 58 & K1412E &&\\      
    &28 & \textcolor{violet}{M932L}\cite{R185H} & 59 & K1412I &&\\    
    &29 & \textcolor{violet}{M1532I}\cite{R185H} & 60 & K1415I &&\\  
\textcolor{blue}{PDN} &30 & \textcolor{blue}{T1596I}\cite{t1596i.1,t1596i-l1267v} & 61 & \textbf{S1419N} &&\\  
    \hline  
    \end{tabular}
    \caption{The considered genetic mutations and corresponding ids. Color coding is according to Figure \ref{fig:domains}}
    \end{subfigure}  
\begin{subfigure}{0.44\textwidth}
  \centering
  \includegraphics[scale=0.1]{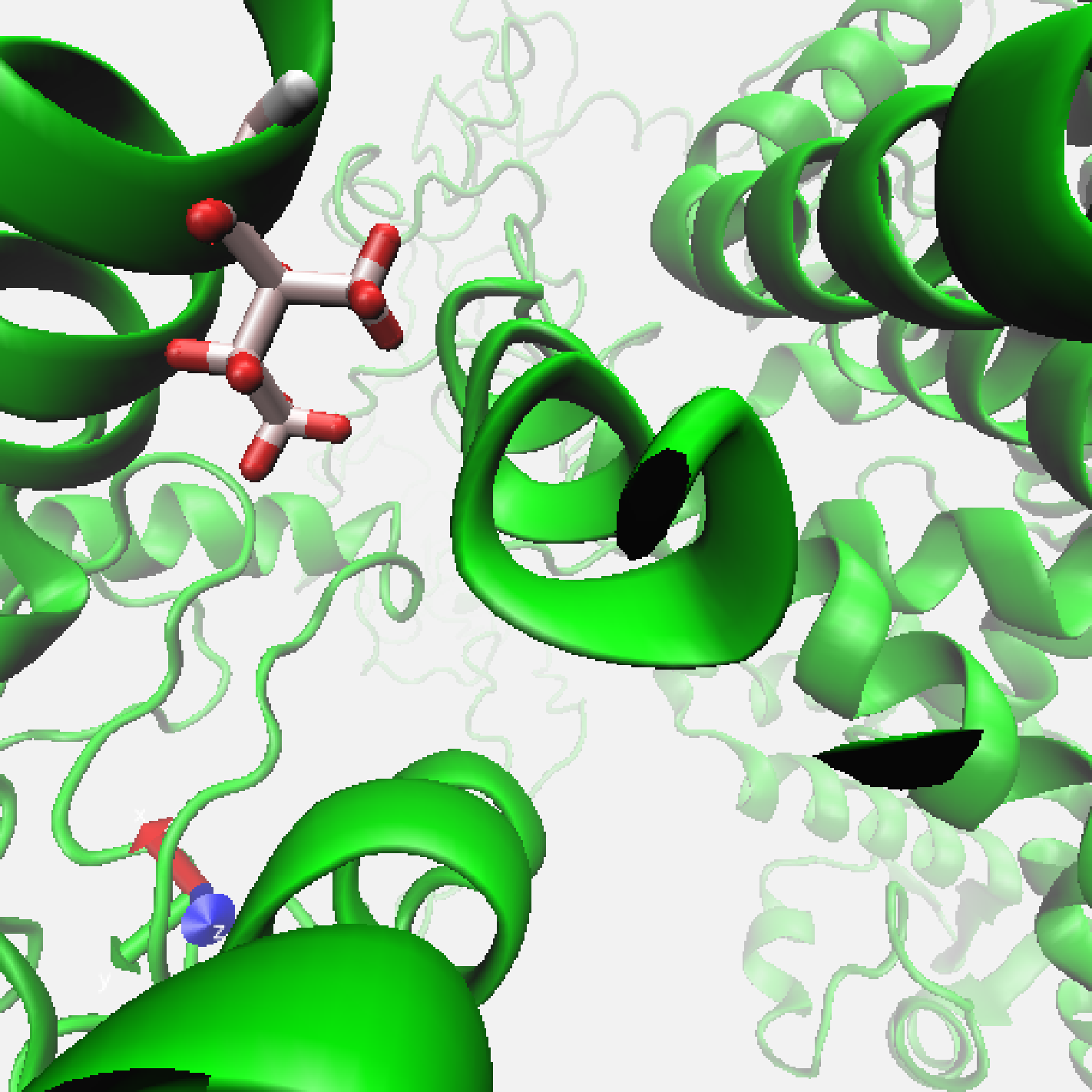}
  \caption{\textcolor{red}{\textbf{I136V}}: Isoleucine}
  ~\\
  \includegraphics[scale=0.1]{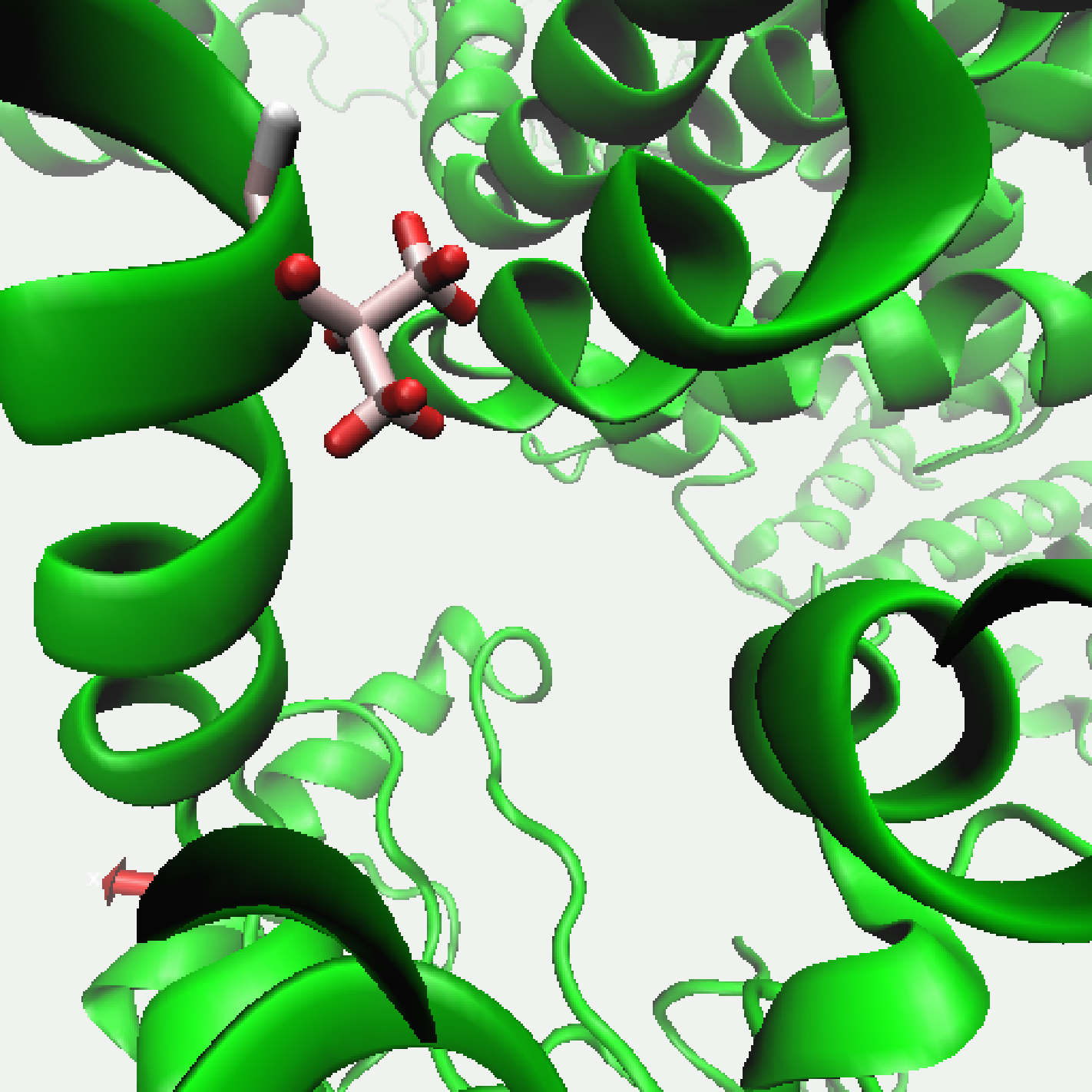}
  \caption{\textcolor{red}{\textbf{I136V}}: Valine}
  \end{subfigure}
  \caption{[Left] (a) \textbf{PAT}  and  \textbf{NEUTRAL} genetic variants. \textbf{PAT}  mutations are further divided by disease and
  highlighted with different colors according to Figure \ref{fig:domains}. Among the \textbf{NEUTRAL} variants, the 18 known human variants are highlighted in bold, whereas the 4 not causing biophysical abnormalities are also starred. [Right] Mutation I136V: the
  initial Isoleucine (b) in position 136 is turned into a Valine (c).}
  \label{fig:mutations}
\end{figure}

We then create 85 copies of the WT sequence each one with a single amino-acid mutation, divided into pathogenic (referred to as \textbf{PAT} in the following)  and non-pathogenic (henceforth \textbf{NEUTRAL}) variants. The first 31 \textbf{PAT} mutations 
have been proved to be associated with gain-of-function of the protein by cell electrophysiology assay,
see Figure \ref{fig:mutations}(a).
The additional control group of 54 \textbf{NEUTRAL} mutations is composed by 4 variants not causing biophysical abnormalities (nABN), 17 genetic variants from dbSNPs with uncertain significance or benign that do not alter the biophysical properties of the channel,  and 33 pseudo mutations between orthologous proteins in closely related species missense variants already considered in Ref.\cite{Dimos} and identified between SCN9a homologous  genes from mammalian species sharing >90\% nucleotide sequence identity, as commonly accepted in the community \cite{Sunyaev01,Yue06,Care07,Adzhubei10}.

Among the gain-of-function \textbf{PAT} mutations, there are 18 causing Inerithed Erythromelalgia (IEM), 6  associated with Small Fibre Neurophathy (SFN), 6  causing Paroxysmal Extreme Pain Disorder (PEPD) and 1 related with painful neuropathy in diabetic patients (PDN).   The list of all the considered genetic variants is shown in  Figure \ref{fig:mutations} (a), where the first 31 pathogenic are those involved in the represented disease, color-coded accordingly. As an example, Figures \ref{fig:mutations} (b) and \ref{fig:mutations} (c)  display a blow up of the three-dimensional structure on the I136V mutation, where the isoleucine in position 136 is mutated into a valine residue.

Three different template structures, describing  the closed state of the sodium channel, have been selected. We have used Clustal Omega \cite{ClustalOmega} to assess the sequence identity between each template and the original WT. The NavAb  is derived from the cryo-EM structure of the ortholog protein  of  \textit{Acrobacter butzleri} (PDB code 3RVY) , reporting the channel captured in a closed-pore conformation with four activated voltage sensors at 2.7 \AA resolution, and modified as described in Ref. \cite{Dimos}. We shall refer to it as MOESM3 \cite{PSZC2011} hereafter.  The sequence identity of the \textit{Acrobacter butzleri} sequence with the original WT is $27\%$ while the derived MOESM3 template increases the identity to $50.8\%$. The second template NavPaS has been obtained from the eukaryotic \textit{American Periplaneta} with the spider neurotoxin Dc1a, via cryo-electron microscopy (cryo-EM)(PDB code 6A90)  \cite{NavPas} (hereafter  6A90). Comparisons between ligand-free protein and protein-neurotoxin do not show any conformational differences, except for the VSD2. The template, therefore, refers to the closed conformation with the VDS2 in the 'up' conformation. This template shows a global sequence identity of approximately 32$\%$ compared to the WT sequence. Finally, the last selected template hNav1.7 is  derived from the eukaryotic protein NaV1.7 of \textit{Homo Sapiens} via cryo-EM (PDB code 6J8J)  \cite{Shen1303} and will be referred to as 6J8J. It represents an inactivated state with all four Voltage Sensitive domains in the “up” conformation and the intracellular gate closed, meaning that the channel is closed and not available for activation. The alignments of the WT sequence and this template is $97\%$. The detailed alignments of all used templates are reported in the Appendix.


When the primary sequence and the function of the protein are known, we use Direct Coupling Analysis (DCA) \cite{MorcosE1293}, homology modeling \cite{Lee1992}, and energy minimization to obtain the three dimensional structure of the mutated sequence. DCA provides an assessment on the quality of a given template and hinges upon the idea that the mutation of any amino acid is constrained by the need of conserving protein function, so that only mutations conserving the original physical properties are allowed. Then, homology modeling allows the transfer of a protein's quaternary structure from one sequence to another one by homology. The final step is energy minimization to remove possible steric clashes and inconsistencies that might arise during this process . We have used a in-house implementation of DCA, Swiss-Model \cite{Swiss-Model, web:Swiss-Model} for the homology modelling and FG-MD \cite{FG-MD} for energy minimization. 
Once the structures have been generated we use QMEANBrane \cite{QMEANBrane}  and RAMPAGE \cite{RAMPAGE} to evaluate the quality of the obtained structure in terms of their Ramachandran angles $\Phi-\Psi$ along the peptide sequence. Note that in general, this is known to be an important step that cannot be bypassed \cite{Huang2017}. See Section Methods for details on the above techniques.

After the quality assessment, a set of 85 three-dimensional structures, representing the 85 genetic variants reported in Figure \ref{fig:mutations}, are available and need to be compared with each other to test for differences. This is in general a very challenging task when carried out in a full fledged three-dimensional representation, where the comparison is carried out residue by residue. A possible way around is to map each structure into its graph representation. In this way we clearly loose the three-dimensional information on the spatial position of each amino acid, but we can still compare the relative topologies that is crucial to identify common patterns.
This step can be done via the Residue Interaction Network (RIN) technique, where nodes represent amino acids and edges represent their non-covalent interactions (see last step at the top of Figure \ref{fig:workflow}). Here, we use RING 2.0  \cite{web:RING2.0, RINGpaper}  that is able to distinguish six different types of interactions: Hydrogen bond, Van der Waals, Ionic bridges, $\pi-\pi$ Stacking, $\pi$-cation and Disulfide bridges. See Section Methods for more details.  

The final step of the computational pipeline is to compare and classify the obtained networks, as shown in the lower part of Figure \ref{fig:workflow}. To accomplish these tasks we employ Graph Kernels \cite{GraphKernels}  to perform pairwise RINs comparison, 
and then both unsupervised \cite{RotaBulo2017} and supervised \cite{goos_support_2001} machine learning to identify common patterns.
We used two different kernels to compare RINs, the
Vertex Histogram (VH) and  Weisfeiler-Lehman (WL) subtree kernels, implemented by the GraKel Python library \cite{GraKel}.
Both of them are based on node labels comparison, where node labels are set as the position of each node -- i.e. amino acid, in the protein sequence. The essential difference between the two methods is that the VH method makes a pairwise comparison of the nodes labels only, whereas the WL method compares also the corresponding subtrees, hence providing a more complete and general contrast.

As for the unsupervised learning, we consider the Dominant Set clustering method \cite{RotaBulo2017} where a set of elements are organized in clusters, in such a way that each cluster satisfies an internal homogeneity and expresses an external inhomogeneity. We also apply the Average-Linkage hierarchical clustering \cite{ELL01}. Supervised learning has also been considered using the Support Vector Machine (SVM)  method \cite{goos_support_2001} implemented in the \textit{Scikit learn} library \cite{sklearn_api}.

Details on all the above calculations can be found in Section Methods and in the Appendix.

\begin{figure}
  \begin{subfigure}{0.40\textwidth}
     \centering{\includegraphics[width=\linewidth]{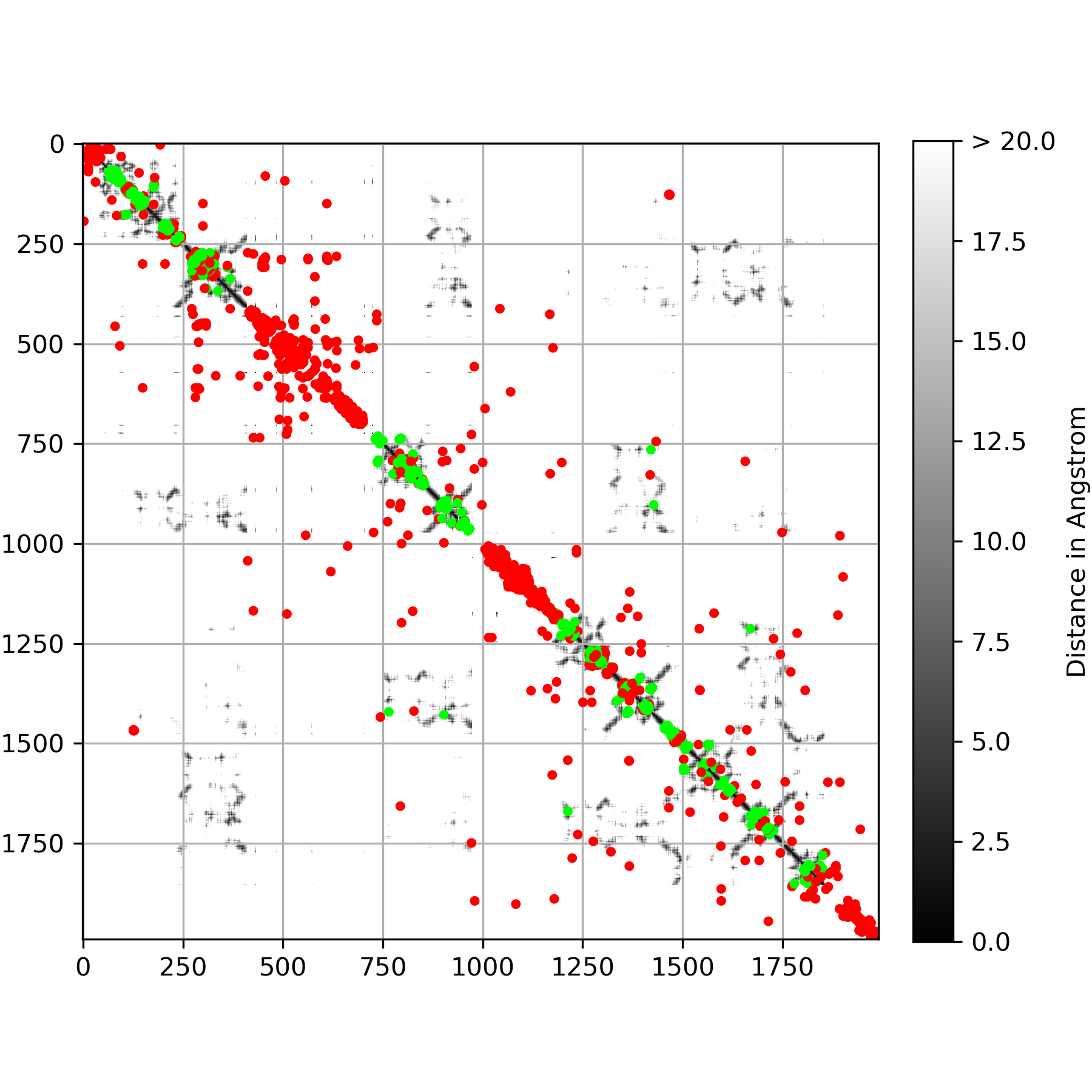}}
    \caption{6A90 - full sequence}
   \end{subfigure}
   \begin{subfigure}{0.40\textwidth}     \centering{\includegraphics[width=\linewidth]{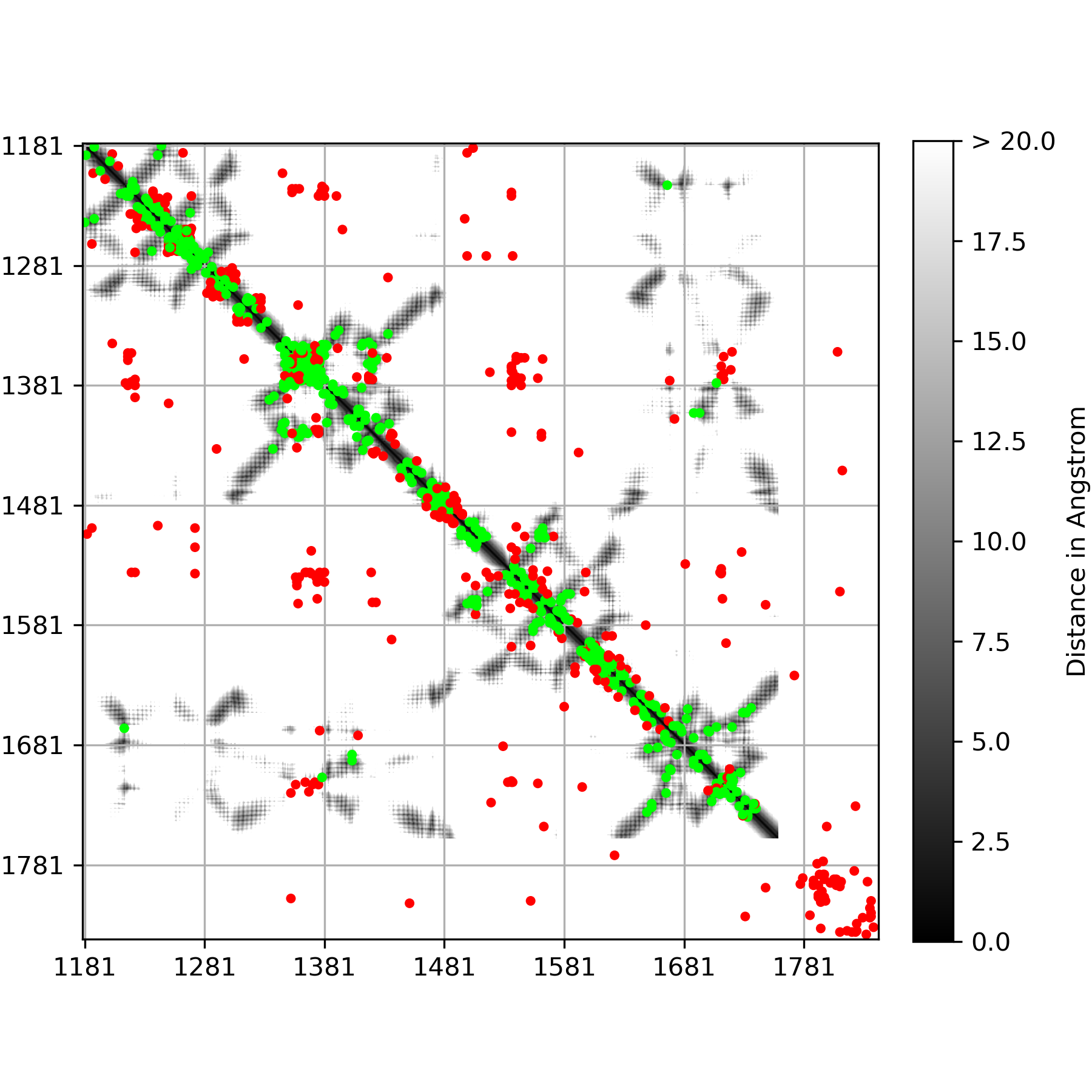}}
    \caption{6J8J - full sequence}
   \end{subfigure}
   \begin{subfigure}{0.40\textwidth}
     \centering{\includegraphics[width=\linewidth]{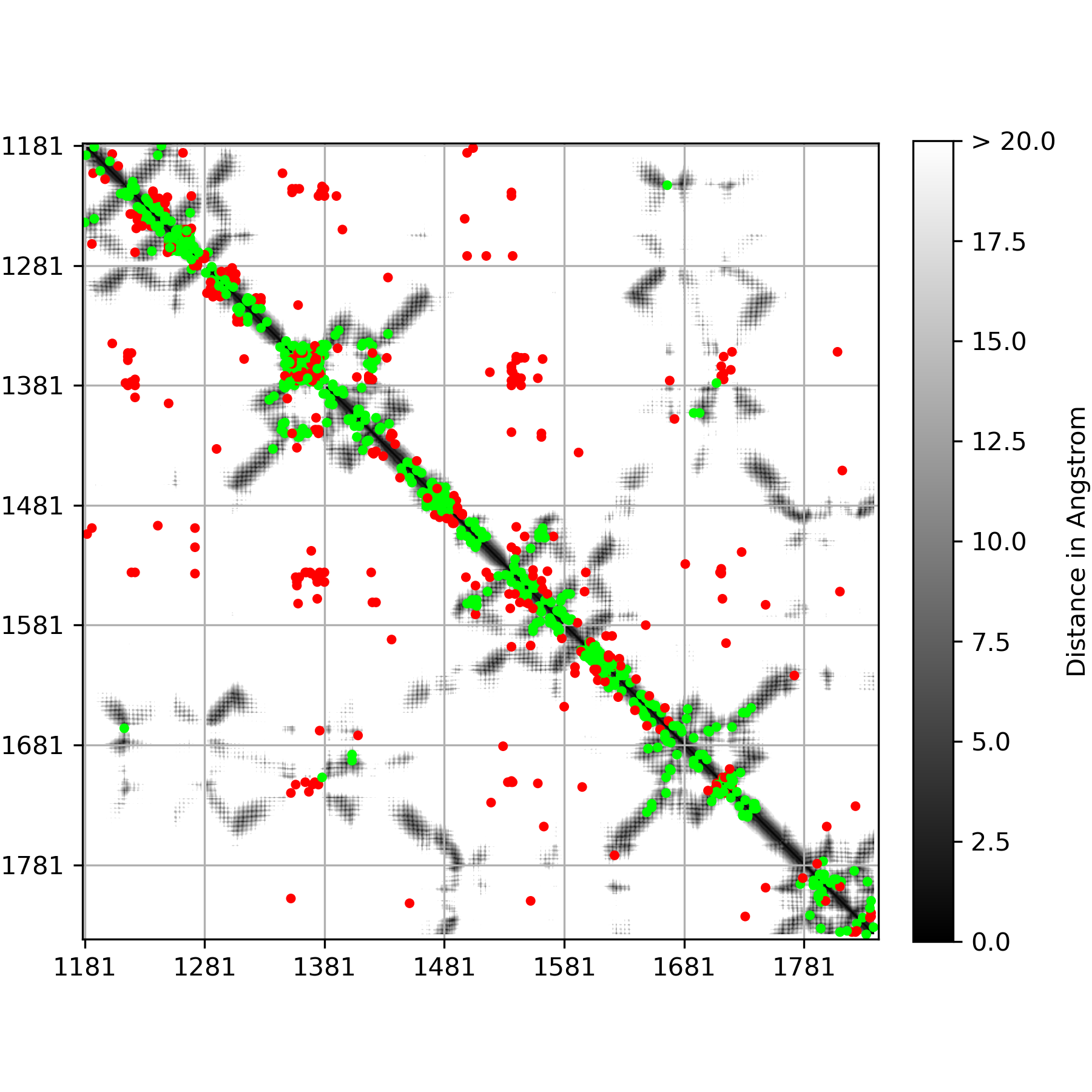}}
    \caption{6A90 - interval 1181-1851}
   \end{subfigure}
   \begin{subfigure}{0.40\textwidth}     \centering{\includegraphics[width=\linewidth]{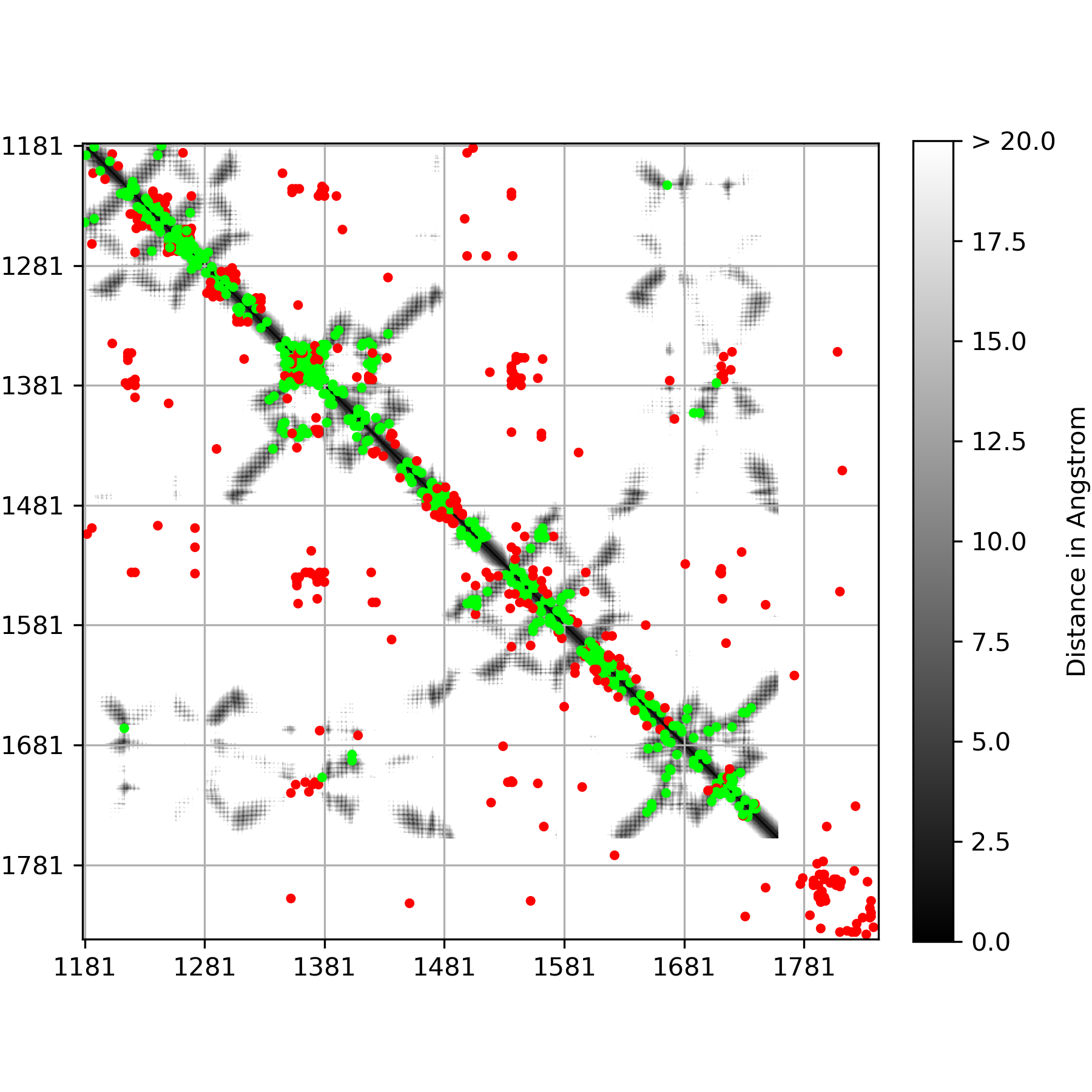}}
    \caption{6J8J - interval 1181-1851}
   \end{subfigure}
   \caption{DCA analysis for templates 6A90 and 6J8J where green points are representing the contact map of the original template and the red points the corresponding DCA representations. (a) and (b) full sequence; (c) and (d): interval 1181-1851}
    \label{fig:DCA}
\end{figure}

\begin{figure}[htbp]
   \begin{subfigure}[l]{0.55\textwidth}
     \includegraphics[width=\linewidth]{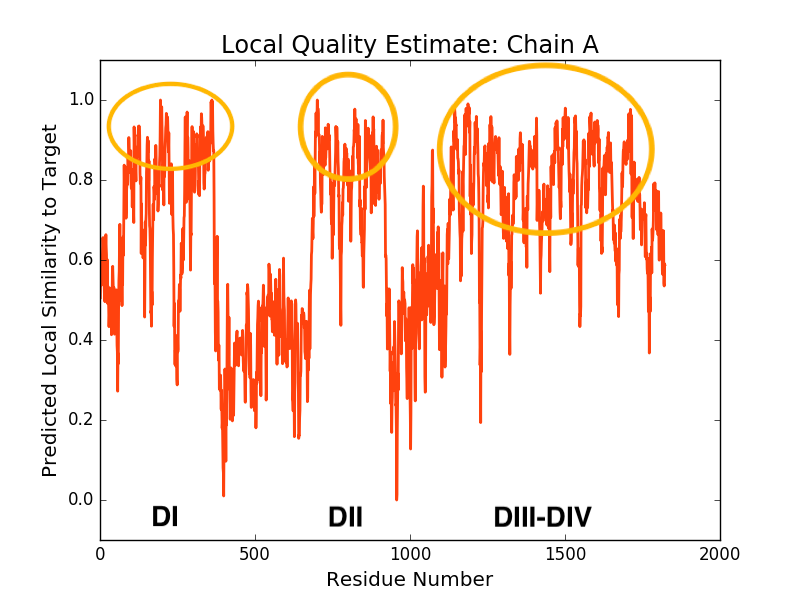} 
     \caption{}
   \end{subfigure}
  \begin{subfigure}[l]{0.44\textwidth}
     \includegraphics[width=\linewidth]{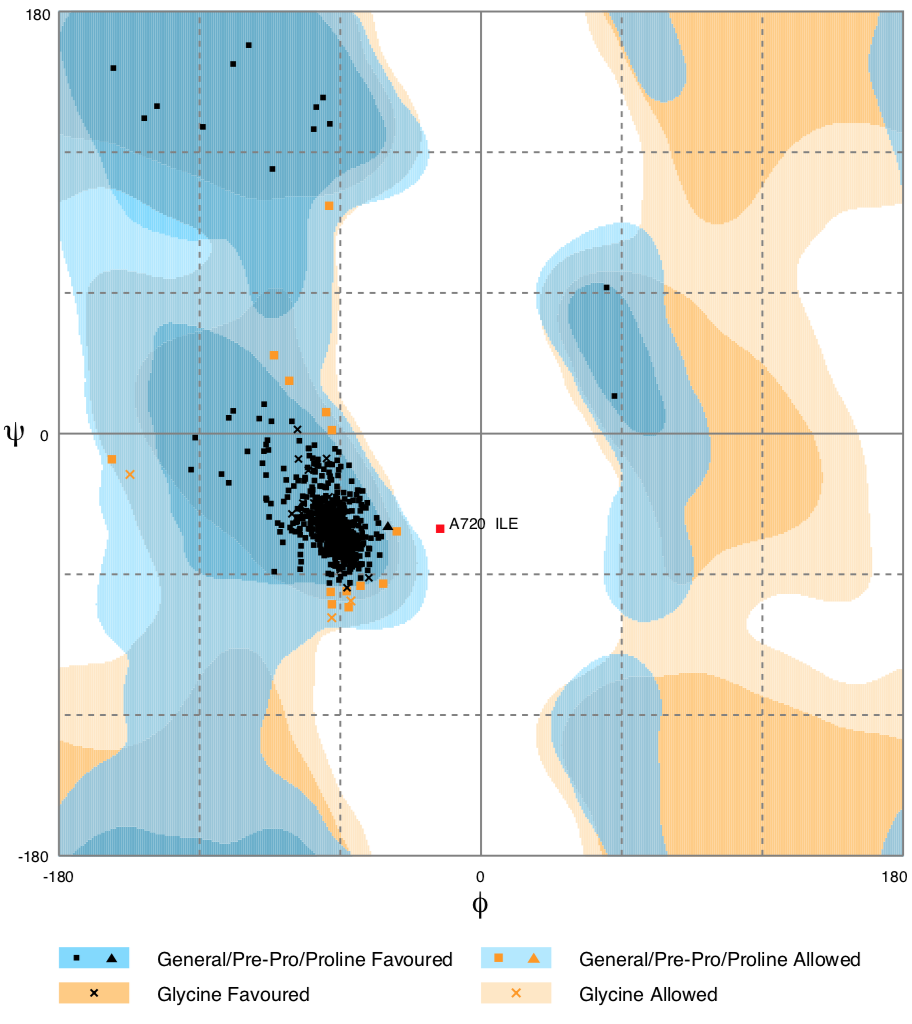}
    \caption{}
   \end{subfigure}
   \caption{Quality assessments for the 6A90 Wild Type. (a) Predicted local similarity along the sequence: regions with high similarity are circled and correspond to the domains are DI-DIV; (b) Ramachandran plots where each point represents an amino acid and where characteristic regions are highlighted in different colours. Regions colored in white are considered forbidden.}
   \label{fig:quality6A90}
\end{figure}

\begin{figure}[htbp]
   \begin{minipage}[l]{0.70\textwidth}
   \begin{subfigure}[b]{0.49\textwidth}
     \includegraphics[width=\linewidth]{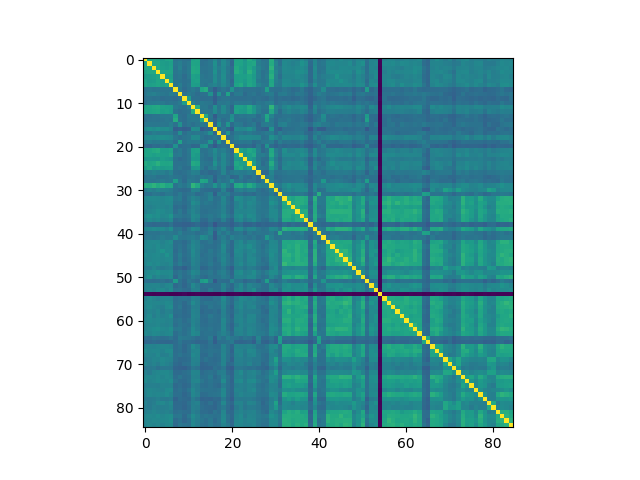}
          \caption{MOESM3: WL  kernel}
  \end{subfigure}  
   \begin{subfigure}[b]{0.49\textwidth}
     \includegraphics[width=\linewidth]{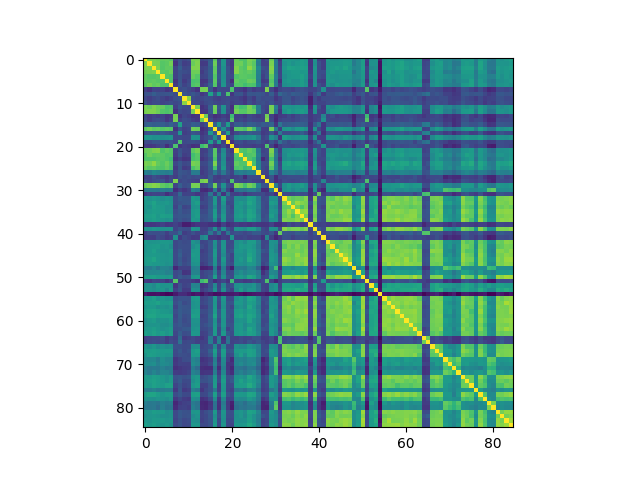}
     \caption{MOESM3: VH kernel}
  \end{subfigure}  
   \begin{subfigure}[b]{0.49\textwidth}
     \includegraphics[width=\linewidth]{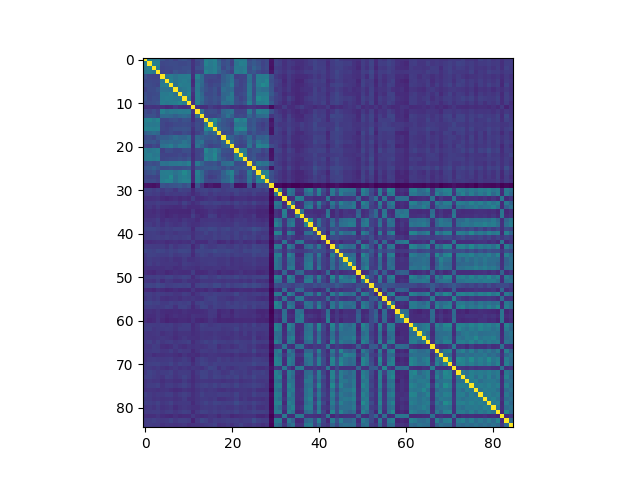}
     \caption{6A90: WL kernel}
   \end{subfigure}
   \begin {subfigure}[b]{0.49\textwidth}
     \includegraphics[width=\linewidth]{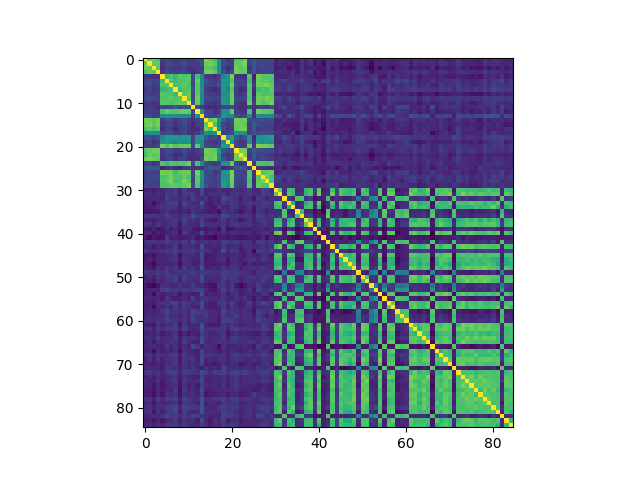}
     \caption{6A90: VH kernel}   
   \end{subfigure} 
   \begin{subfigure}[b]{0.49\textwidth}
     \includegraphics[width=\linewidth]{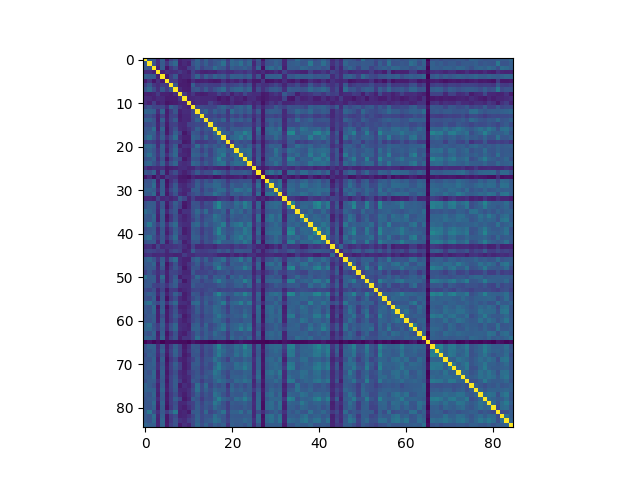}
     \caption{6J8J: WL kernel}
   \end{subfigure}
   \begin{subfigure}[b]{0.49\textwidth}
     \includegraphics[width=\linewidth]{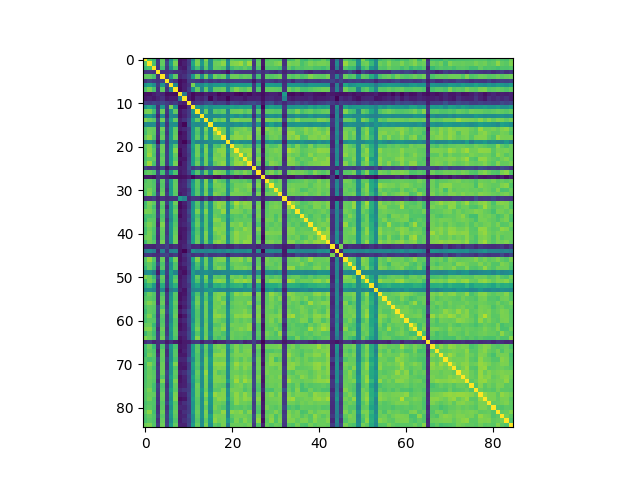}
     \caption{6J8J: VH kernel}
   \end{subfigure}    
   \end{minipage}
   \begin{minipage}[r]{0.29\textwidth}
   \begin{subfigure}[c]{0.48\textwidth}
    \centering{
    \includegraphics[angle=-90,origin=c, scale= 0.17]{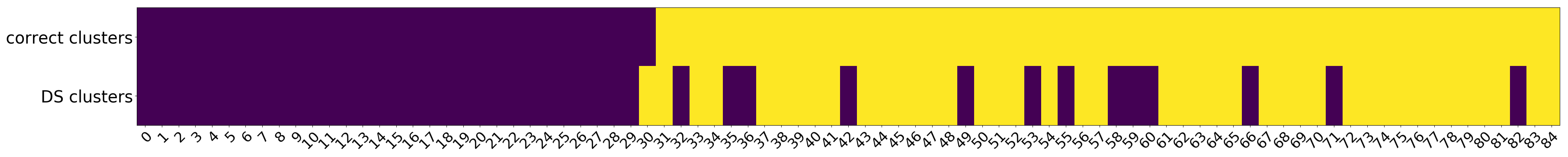}
    }
    \caption{}
    \end{subfigure}  
    \begin{subfigure}[c]{0.48\textwidth} 
       \centering
       \includegraphics[angle=-90,origin=c, scale= 0.17]{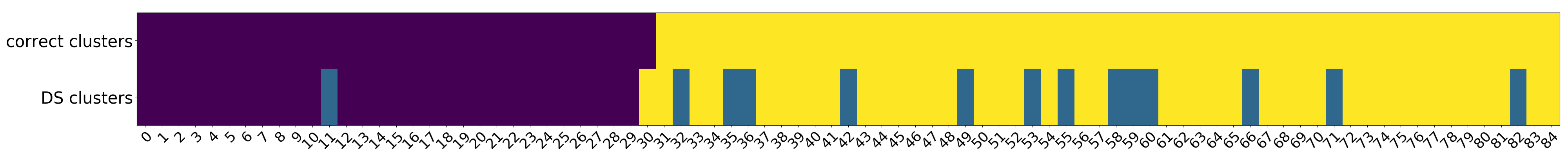}
       \caption{}
    \end{subfigure}
    \end{minipage}
   \caption{[Left] Similarity matrices of the  Weisfeiler-Lehman (5 iterations) (a)-(c)-(e) and Vertex Histogram (b)-(d)-(f) kernels applied to RINs resulting from MOESM3, 6A90 and 6J8J templates. 
   [Right] Dominant set classification for the WL similarity matrix (c) of 6A90 template: first (g) and second (h) iterations.} 
   \label{fig:kernelResults}
\end{figure}

In short, our computational pipeline (Figure \ref{fig:workflow}) includes the following steps.
\begin{description}
\item[Step 1] Reconstruction of the three dimensional structures of the mutated sequences via DCA, homology modelling, and energy minimization. See Figures \ref{fig:domains} and \ref{fig:mutations};
\item[Step 2] Quality assessment of the obtained structures. See Figures \ref{fig:DCA} and \ref{fig:quality6A90}; 
\item[Step 3] Mapping of each three-dimensional structure into a network, comparison between networks via Graph Kernels techniques, and identification of pathogenic mutations via machine learning techniques. See Figure \ref{fig:kernelResults}. 
\end{description}


We stress the particular importance of obtaining high quality structures from the homology modeling, because any drawbacks at this stage could affect all downstream analyses, affecting network representations and making comparison and classification questionable. As we shall see below, however, this step is quite delicate. Indeed other possible templates, in addition to those reported here, could have been selected in principle. For instance, Huang and collaborators \cite{Huang2017} used the recently determined \cite{Wu2015,Wu2016} atomic structure of the  rabbit voltage-gated calcium channel Cav1.1 as template for Nav1.7 studies. Moreover, other human templates have also been proposed recently \cite{Xu2019}. However, as detailed in the Appendix (see Figures \ref{fig:6J8J-6N4I} and \ref{fig:6N4I-WT}), it turns out that they could not be used in our case for either low resolution in the original crystal or for a limited extension compared with our WT.

While the reliability of the MOESM3 has already been assessed \cite{Dimos}, those of 6A90 and 6J8J have not. As a preliminary step, we then first performed a DCA analysis of the 6A90 and 6J8J templates whose predictions are shown in Figure~\ref{fig:DCA}.
In particular, Figure \ref{fig:DCA} displays the comparison of the contact maps for the 6A90 and 6J8J templates (green points) with the corresponding DCA predictions (red points). Figures \ref{fig:DCA}(a)-(b)  show this comparison for the full sequence of the 6A90 and the 6J8J, respectively. Here, the alignment contains a relatively small number of residues and quite large fraction of gaps, and the predicted contact map is relatively noisy.  On the other hand, if we restrict the analysis to the region of the alignment for which the fraction of gaps is less than 30\%, which mainly correspond to the region from residue 1181 to residue 1851, then the predictions display a good agreement with the deposited structures 6A90 and 6J8J as illustrated in  Figures \ref{fig:DCA}(c)-(d). 
Several reasons could explain the above discrepancy between the experimental deposited structure and DCA predictions. Firstly, the predicted rearrangement could be due to an allosteric conformation of the protein different from the one seen in the experiments. Secondly, these interactions could be important for a large fraction of the sequences in our alignment, but not necessarily for the crystallized protein. Finally, the discrepancy could be due to crystal artifacts, not uncommon for membrane proteins.

Bearing this information in mind, we next proceed to the quality analysis of the WT template 6A90 with QMEANBrane and RAMPAGE, as shown in Figure \ref{fig:quality6A90}. The quality analysis for the other templates has also been performed and is reported in the Appendix, see Figures \ref{fig:qualityMOESM3} and \ref{fig:quality6J8J}. Moreover, for each template and each point mutation we reported the quality value resulting from the QMEANBrane analysis, see Tables \ref{tab:MutQMEANBrane}, \ref{tab:hSNPM3Brane}, \ref{tab:MutQMEANBrane6A}, \ref{tab:hSNPBrane6A}, \ref{tab:MutQMEANBrane6J8J}, \ref{tab:hSNPBrane6J8J} \ref{tab:MutRamachandranPlot} and \ref{tab:hSNPRamachandranPlot}.
It is worth noticing that the analysis in Figure \ref{fig:quality6A90} is mainly relevant for the protein segments corresponding to the transmembrane regions, i.e. all the $\alpha$-helices (see Figure \ref{fig:snapshot}), because this is the critical region as previously discussed.

Figure \ref{fig:quality6A90} (a) shows the predicted local similarity to the target structure, and highlights how the produced models have high quality within each domain area, and low quality in the inter-domains area, where the reliability of the models is significantly lower. However, this is completely acceptable as inter-domains region are formed by unstructured loops outside the membrane area (see Figures \ref{fig:snapshot} and \ref{fig:mutations}), whereas all the considered mutations fall within the high-quality domain regions. As it will be further elaborated below, the low-quality loop regions enter in the network analysis without however jeopardizing the soundness of the results for the important high-quality domains region.

Additional insights can be obtained  by performing a Ramachandran plot analysis, to assess whether the obtained structures obey all characteristic constraints provided by both steric hindrance and quantum chemistry. This is done in Figure \ref{fig:quality6A90} (b), and shows that indeed 97.9\%  of the produced amino acids fall within a favorable region, 2\% fall within a permitted region and only one amino acid falls in the forbidden region (the white region).

Armed by all these information, we now consider the issue of comparing the obtained different structures. As anticipated, our strategy hinged upon a mapping of each three-dimensional structure into a corresponding network (univocally identified by its topology) and then compare different networks with each other. We use RIN analysis to perform this mapping
and both Vertex Histogram  (VH) kernel and  Weisfeiler-Lehman (WL) subtree kernel with 5 iterations, to compare them, as illustrated in Section Methods. 
The similarity between all considered graphs can be visually assessed via the similarity matrix rendering reported in Figure \ref{fig:kernelResults} for all three MOESM3, 6A90 and 6J8J templates. 
Each depicted similarity matrix of Figure \ref{fig:kernelResults} (a)-(f) (left panel) has rows and columns labeled according the mutation ids $0-84$ of  Figure~\ref{fig:mutations}. Accordingly, $0-30$ correspond to \textbf{PAT} mutations while $30-84$ 
are \textbf{NEUTRAL} variants.
Each cell $(i,j)$ in a matrix shows the similarity value between the $i$-th and $j$-th RINs color-coded so that lighter colors correspond to RINs with high degree of similarity (from yellow to blue). Clearly the main diagonal shows always the lightest color, being the result of the comparison of a graph with itself.
Three main points emerge from the analysis of the results of Figures \ref{fig:kernelResults} (a)-(f):
\begin{itemize}
\item In both the MOEMS3 and 6A90 templates two well-defined clusters are clearly visible. The first one grouping together 
\textbf{PAT} mutations (ids $0-29$) that have a high degree of similarity within each other and a low degree of similarity with  
\textbf{NEUTRAL} mutations  (ids $30-84$). Likewise \textbf{NEUTRAL} mutations are similar to each others but different from \textbf{PAT} ones.

\item No such division in clusters is visible for the 6J8J template, where there is no clear distinction between \textbf{PAT} and \textbf{NEUTRAL} mutations. Although a few genetic variants appear to display a more marked difference, the majority show a high degree of similarity with each other. The reasons for this marked contrast with what has been observed for templates MOESM3 and 6A90 is unclear.  One possible explanation stems from the observation that 6J8J is a human protein template representing the native state of the NaV1.7 sodium channel in closed state. The considered WT sequence of NaV1.7 and the 6J8J primary sequence are then nearly identical likely for all the 85 considered genetic variants (each variant differ from the WT for just one amino acid). Hence reconstructed models via homology modeling and energy minimization will also provide nearly identical RINs whose differences can hardly be captured by any pattern recognition algorithm such as that employed here. This is not the case of the two other templates MOESM3 and 6A90 where generated three dimensional structures via homology models might show a significant structural difference with one another so that energy minimization might drive them into different local minima. See the Appendix, Figure \ref{fig:enrscape}.

\item For all the three considered templates, the WL and VH kernels show a similar pattern. This means that the pairwise comparison of the nodes labels alone, as done by the VH kernel, is already sufficient to capture the main features. The WL kernel performs a more general comparison since also the subtree structure of each node is considered. Hence, while a pairwise comparison of the nodes labels is sufficient to discriminate the \textbf{PAT} and \textbf{NEUTRAL} classes, the subtree structure and labels is also found to be coherent with this classification; 
\end{itemize}

We also examined the role of each interaction (H-bond, Van der Waals, and Ionic) separately and found the same pattern observed in the whole RINs.  All these results can be found in the Appendix, see Figures \ref{fig:kernelResultsMOESM3separated}, \ref{fig:kernelResults6A90separated}.

Further support to the above results can be obtained using unsupervised machine learning techniques. To this aim we apply the Dominant Set (DS) algorithm \cite{RotaBulo2017} to the above similarity matrices. The results for 
matrix (c) of the 6A90 template are reported the right panel of Figure \ref{fig:kernelResults}. 
The diagrams (g) and (h)  are composed of 85 rows labeled with the ids of considered genetic variants and two columns. In the first column (labeled correct clusters) the first 31 \textbf{PAT} mutations are color-coded in black, and the remaining \textbf{NEUTRAL} variants are color-coded in yellow. The second column (labeled DS clusters) reports the classification according to the DS algorithm.
Diagram (g) is the result of one iteration of the DS algorithm. Remarkably, the algorithm captures the correct classification of 96.8\% of the pain related \textbf{PAT} mutations, color-coded in black.
Only one pain mutation (T1596I, id 30) is not correctly classified. This result is not surprising: looking at matrix (c) it is evident that mutation 30 is very different from the ones in the \textbf{PAT} group and instead very similar to the ones in the \textbf{NEUTRAL} group. 
Among the \textbf{NEUTRAL} variants, color-coded in yellow, 13 variants are not classified correctly, 4 of which are human variants. An additional iteration of the DS algorithm within the variants originally classified in the black group further splits it in two clusters as shown in diagram (g): the first cluster, shown in blue, contains the 13 neutral variants plus the deleterious mutation (A863P, id 11) and the second one, shown in black, contains all the other deleterious mutations. Note that the lower degree of similarity of mutation (A863P, id 11)  and the rest of the \textbf{PAT} group is also visually evident from the similarity matrix (c) and can be due to a lower quality of the structure model. 

\begin{figure}
    \centering
    \includegraphics[width=0.90\textwidth]{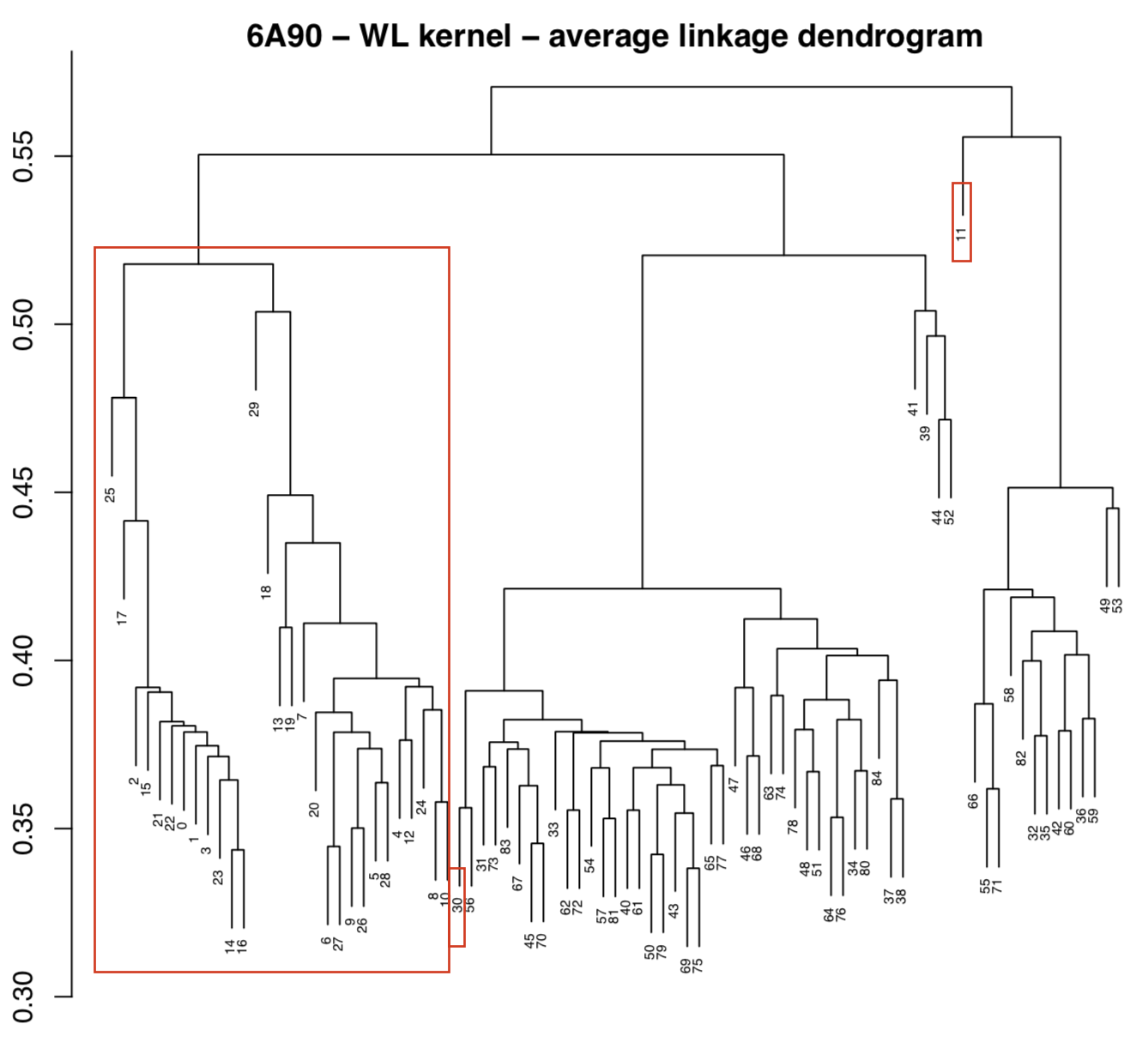}
    \caption{Average-linkage hierarchical clustering of the 6A90 template. The leaves of the dendrogram are labeled with the ids of the considered genetic variants. The scale on the left shows the distance among variants. The input distance matrix has been derived from the WL kernel similarity matrix of the 6A90 template. The big red box highlights the cluster of the pain \textbf{PAT} mutations. The small red boxes highlight mutations ids 11 and 30.}
    \label{fig:6A90averagedendrogram}
\end{figure}

Consistent findings are found by applying the average-linkage hierarchical clustering \cite{ELL01} to the distance matrix derived from matrix (c), as presented in Figure \ref{fig:6A90averagedendrogram}. Here, the leaves of the dendrogram are labeled with the ids of the considered genetic variants and the scale on the left shows the distance among clusters.  Note that the \textbf{PAT} mutations are clearly separated from the \textbf{NEUTRAL} variants, as for the DS algorithm, but mutations 11 and 30 are further singled out, as suggested by the second iteration of the DS algorithm, diagram (g).

 The clustering results for templates MOESM3 and 6J8J can be found in the Appendix, see Figure \ref{fig:dominantSetResults}.

We also applied supervised machine learning techniques in the form of the Support Vector Machine (SVM) method with cross-validation to determine its prediction ability. Cross-validation is a standard  technique used to evaluate the ability of prediction models, see Section Methods for a detailed description. It is particularly useful when a small dataset is available, as in our case.
The accuracy of the predictions made for the three considered templates are shown in Table \ref{table:svm}.
As visible, the WL kernel predictions outperforms VH for all templates and the accuracy results are in line with the kernels results: the ability of the VH and WL kernels to discriminate between pain related mutations and non pain ones is reflected also by the SVM learning method. In particular, the prediction accuracy is very good for templates MOESM3 and  6A90 and rather poor for 6J8J, in agreement with previous findings.

A word of caution is in order here. At first sight, it might appear striking the contradiction between the high sequence identity of the 6J8J hNav1.7 human template with the WT and the corresponding low accuracy of the prediction. However, this is in line with past work, as anticipated earlier, and can be ascribed to the difficulties  of homology modelling when comparing sequences with too high identity (see e.g. Ref. \cite{Haddad2020}), whereas performs better when comparing different species. A good example is provided by a recent study by Huang et al \cite{Huang2017} where the atomic structure of the rabbit voltage-gated calcium channel Cav1.1 provides a very good template for homology-based structural modeling of Nav channels notwithstanding a comparatively low value ($21\%$) of sequence identity. This is more true within our computational pipeline because we are performing a pairwise comparison of structures to identify emerging global patterns. Additional human templates with lower sequence identity have also been recently proposed \cite{Xu2019}, but the corresponding reduced identity can there be ascribed to a significantly lower quality of the original deposited structure and to the much lower number of represented amino acids, as detailed in the Appendix.
\begin{table}[htb]
    \centering
    \begin{tabular}{|c|c|c|}
         \hline 
         Template & kernel method & Prediction Accuracy \\
         \hline
         MOESM3 & VH & 63.5\%\\
         6A90   & VH & \textbf{95.14\%}\\
         6J8J   & VH & 63.5\%\\
         \hline         
         MOESM3 & WL & \textbf{94.30}\%\\
         6A90   & WL & \textbf{97.64}\%\\
         6J8J   & WL & 63.5\%\\
         \hline
    \end{tabular}
    \caption{Prediction accuracy of the SVM method for the three considered templates} 
    \label{table:svm}
\end{table}

The general scenario emerging from previous analysis strongly suggests the existence of some critical mutations that lead to deleterious effect. This is an important issue with far reaching consequences in the case of human genome \cite{Lohmueller08}.
Even at the level of globular proteins, it is well-known how single mutations of specific amino acids can change folds and functions in a very controlled way \cite{Alexander07,Alexander09}. 
However, this case appears to be more subtle. For instance, it is interesting to note that M1532I is a \textbf{PAT} genetic variant and M1532V is a \textbf{NEUTRAL} genetic variant notwithstanding the fact that Isoleucine (I) and Valine (V) are both hydrophobic amino acids with quite similar chemical structure. On this basis, both M1532I and M1532V were expected to behave similarly. Yet, there is significant evidence in the literature for the former to be a pathogenetic variant \cite{Faber2012}, whereas  M1532V is reported in dbSNP \cite{dbSNP}, lacking information about the association to a clinical condition and about minor allele frequency. The case of these two closely localized variants with a different predicted impact, points out the importance to implement this study with new variants in order to highlight even very subtle differences between residues, which can discriminate between high-impact and neutral variants.
While this particular case is particularly striking, it turns out to be not the only one (see e.g. Huang et al \cite{Huang2017}). In fact, we did not find any well-defined correlation between the chemical physical properties of the mutated amino acids and their putative pathogenic impact.

We can build on this idea in our case by presenting two different analyses that provide concurring results supporting the above predicted scenario.

The first analysis is prompted by the ability of the VH kernel to discern between \textbf{PAT} and \textbf{NEUTRAL} mutations, and hinges on the node frequency analysis reported in Section Methods. 
On recalling that each node represents an amino acid of the protein sequence, 
labeled by its position in the sequence itself, and that the VH kernel is based on nodes labels comparison,
we compute the frequency of each node label within the \textbf{PAT} and the \textbf{NEUTRAL} groups. Consider a node with label $l$ (i.e. the amino acid in position $l$ in the protein sequence), we then look to all RINs associated to \textbf{PAT} group -- that can originate from any mutation of a specific node $l^{\prime}=1,\ldots 2000$ including $l^{\prime}=l$, and count the number of times that that $l$ is involved in the RINs of that group. A frequency 1 then means that $l$ has non-covalent bonds in all mutations belonging to the \textbf{PAT} group, with a frequency 0 indicating that $l$ has no bonds except the covalent ones. The same analysis can then clearly be performed for the  \textbf{NEUTRAL} group, with similar interpretation.  
\begin{figure}[htbp]
    \centering
    \includegraphics[width=1.0\textwidth]{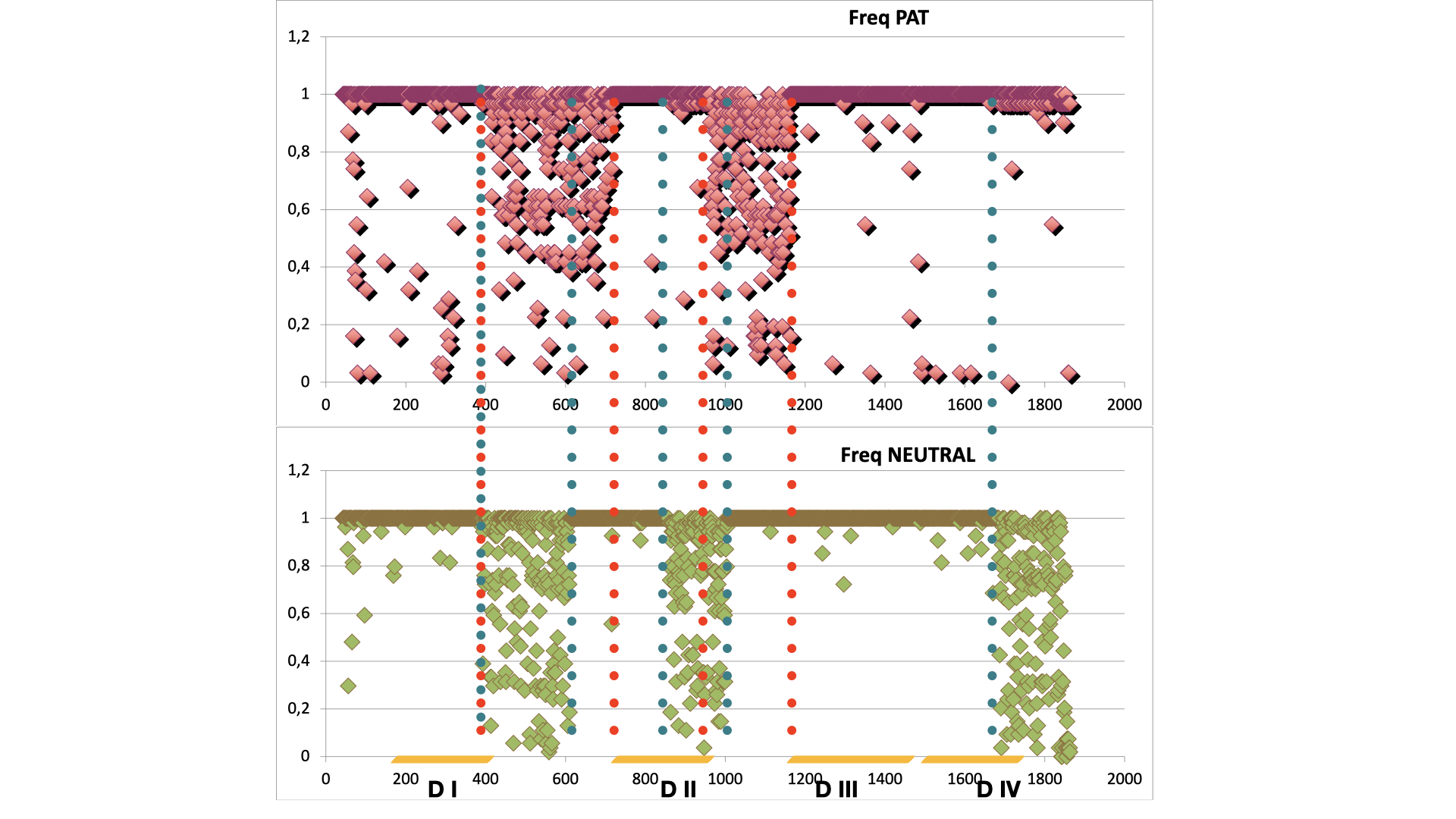}
    \caption{Relative frequency of nodes for pathogenic (\textbf{PAT}) and non pathogenic (\textbf{NEUTRAL}) mutations. Vertical dotted lines identify regions of different behaviour between the two groups, color-coded according to their specificity (see text)}
    \label{fig:frequencyAnalysis}
\end{figure}
Figure \ref{fig:frequencyAnalysis} shows the frequency of nodes along the protein sequence for the \textbf{PAT} and \textbf{NEUTRAL} groups. The range of the four domains are highlighted in orange along the sequence axis and the dashed lines identify the intervals exhibiting a frequency variation in one of the two groups and not in the other.
Note that the majority of nodes have frequency one in both groups, meaning that all the corresponding RINs are very similar for large parts of the protein sequence. 
For the \textbf{PAT} group, this includes almost entirely the regions involved in the formation of the four domains, except for the terminal traits of D II and D IV.
When not with frequency one, the \textbf{PAT} and \textbf{NEUTRAL} groups appear to be characterized by two substantially different patterns: the \textbf{PAT} group shows two broad bands located in the regions between 400-720 and 950-1150; while the \textbf{NEUTRAL} group is characterized by having three more picked bands: 400-610, 850-1000, and from 1700 onwards.
Both these bands are highlighted with thick dotted lines in Figure \ref{fig:frequencyAnalysis}, red for \textbf{PAT} group, green for the \textbf{NEUTRAL} group.


\begin{figure}[htbp]
    \centering
    \includegraphics[width=0.8\textwidth]{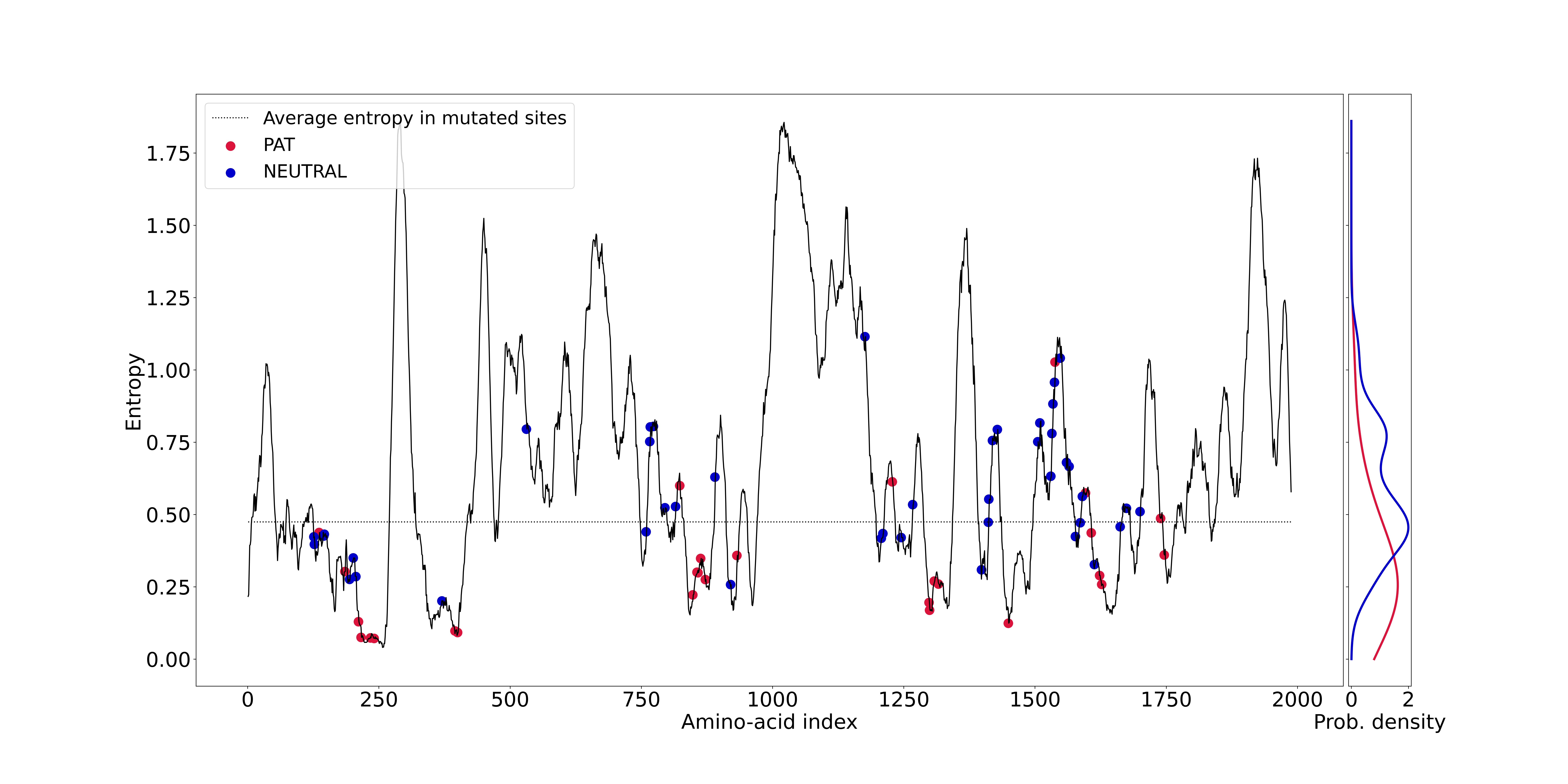}
    \caption{Entropy profile of the full alignment: red dots highlight \textbf{PAT} mutations and blue dots the \textbf{NEUTRAL} ones. The horizontal line is the mean entropy value of the alignment in the mutated sites.}
    \label{fig:entropy_profile}
\end{figure}

The second analysis is based on the calculation of the entropy profile that can be obtained by the MSA calculation performed within the DCA approach, as outlined in Section Methods.
The main idea is that a low-entropy is characteristic of conserved position, so we expect low entropy values for \textbf{PAT} mutations and high entropy values for \textbf{NEUTRAL} mutations.
Note that 'low' and 'high' are here defined with respect to the average value of the entropy in the mutated sites. Note that here, unlike the node frequency analysis, the entropy value refers to the $l$-th node (amino acid) that has been muted. 

As Figure \ref{fig:entropy_profile} shows, only four \textbf{PAT} mutations occurs in amino acids with higher-than-average entropy, so the overwhelming majority of these mutations occur in conserved regions with low entropy. By contrast,
most of the \textbf{NEUTRAL} mutations occurs in higher-than-average entropy, indicating that those mutations occur in regions with high mutation propensity. Not surprisingly, \textbf{PAT} mutations with low entropy bear mutations of residues with indices located in DI ($\approx 250$), DII ($\approx 800$), DIII ($\approx 1300$), and DIV ($\approx 1700$) domains, in agreement with the intuitive fact that evolution has optimized sequences that try to avoid \textbf{PAT} mutations. When contrasted with the analysis of Figure \ref{fig:frequencyAnalysis} we note that the second and the forth of the above regions are again located in the VSD domain, but the other two are not.

It would be interesting to  pursue a better characterization of those mutations that do not follow this simple rule.
However, a much larger dataset than that available at the present time would be needed to properly discuss the pathogenicity of some particular mutation. Ideally a detailed characterization of the effect of all possible mutations for that particular wild type residue would be desirable.

\section{Conclusions}
\label{sec:conclusions}
 In short, we have implemented a computational pipeline to analyze the effects of a single residue mutation in human NaV1.7 channel. Our aim was to identify emerging patterns associated with gain-of-function mutations that lead to diseases by impairment of protein activity.
The general workflow includes the following steps. 
Start with a specific template, perform a preliminary test via DCA analysis for its reliability, and then use homology modeling and energy minimization to find all different three-dimensional structures associated with each of the single amino acid variants; quality assessment of the obtained structures; a mapping of the three-dimensional structures into the corresponding topological network via RIN creation; machine learning of the similarity index of different RINs to identify emerging patterns.

We have applied this workflow to the specific case of different mutations of protein NaV1.7 that is involved in the propagation of nociception signals. We examined 31 gain-of-function mutations (\textbf{PAT}) that have been shown to cause a functional impairment of the channel  demonstrated by cell electrophysiology assay  and 21 
benign or with uncertain significance variants from dbSNPs and 33 pseudo mutations identified among SCN9A homologous genes \cite{Sunyaev01,Yue06,Care07,Adzhubei10} from mammalian species sharing >90\% nucleotide sequence identity.
In the context of inter-species differences, it important to remark that wild-type (normal) rodent (rat) and human Nav1.8 display substantial biophysical differences, and these make DRG neurons carrying the human Nav1.8 more excitable than DRG neurons carrying the rat channel \cite{Han15}. Compared to rat, the human wild-type Nav1.8 could be judged to be “pathogenic”.  
By contrast, the application of our computational pipeline to the set of the selected 85 mutations for a specific template, unambiguously identified the class of \textbf{PAT} mutations as deleterious mutations and clearly distinguished them from the \textbf{NEUTRAL} ones. Comparing the patterns of each structure relative to the considered mutations, we were able to recognize those mutations having the same physiological meaning and sharing a common pattern, supporting the reliability of our computational pipeline as a predictive tool for deleterious mutations.

While the starting point was the MOESM3 template in closed state as in past work by some of us \cite{Dimos}, we extended it in many and different aspects. Firstly, we took full advantage of the arsenal of the techniques recently implemented in the realm of artificial intelligence. This includes
a full rather than point-like analysis of RINs, as well as both unsupervised and supervised machine learning analysis.
Secondly, the specific analysis was expanded by including 2 additional templates, the 6A90 and the 6J8J also for the closed state of the channel.
Interestingly, we found a prediction accuracy of the order of $95\%$ for NavAb MOESM3 that has a $51\%$ sequence identity with the original WT, a prediction accuracy of the order of $97\%$ for NavPaS 6A90 that has a $31\%$ sequence identity with the original WT, and finally a prediction accuracy of $63\%$ for hNav1.7 6J8J template that has a sequence identity of $97\%$ with the original WT. This is in line with past studies \cite{Huang2017} and shows that a carefully selected template with a non-human template with intermediate sequence identity is the optimal choice for the present computational pipeline. More generally, our findings indicate that proposed computational pipeline might be accurate and precise also in other similar cases, thus paving the way toward the possibility of using this tool as a preliminary analysis to predict the pathogenicity of \textit{SCN9A} mutations and better address candidate variants to cell electrophysiology confirmatory assay. Within this framework, it is worth emphasizing that human pain processing is very complex, involving higher levels of processing than in these lower species such as bacteria. As a result of that, the human channel is so finely tuned that perhaps, even the most sophisticated computational predictions fail to capture subtle and  small but functionally important changes \cite{Han15,Waxman14}.

In perspective, there are a number of shortcomings that are clearly emerging from our analysis. Firstly, the lack of a robust and extended dataset of human variants that have been found to be not related to any pathology that forced the use of different types of variants. Secondly, the variable resolution of the experimental deposited structures, that requires a preliminary analysis of the template in order to pursue a reliable analysis. This was clearly highlighted by our DCA analysis, which showed how all three used templates presented a significant number of gaps (i.e. missing residues).
Finally, the use of so many and so different techniques, while instructive, may be cumbersome when applied to a large number of different inputs. A simpler and more direct workflow would be clearly desirable.

It would be extremely interesting to be able to extend the present study of the sodium channel protein by analyzing different conformational states besides the closed one, in order to understand if the ability to distinguish \textbf{PAT} mutations is preserved also after a structural change. We hope to be able to address this and other points in a future dedicated study.

In conclusions, this study represents a pilot application of prediction methods to separate deleterious from neutral genetic variants in Nav1.7 sodium channel, supporting the selection of candidates for cell electrophysiology assay.

\section{Methods}
\label{sec:methods}
\subsection{Direct Coupling Analysis (DCA)}
\label{subsec:DCA}

Roughly speaking, the idea underlying DCA is that the mutation of any amino acid is constrained by the need of conserving protein function. In particular, a mutation of a residue involved in an important interaction with a second one can only be fixed during evolution if the new mutated residue has similar physical properties to the first one or if the second residue is also mutated in a concerted way. 
This mechanism gives rise to patterns of correlated mutations that can be detected in a multiple sequence alignment (MSA) of proteins of the same family and used to infer pair of residues that are most likely interacting in the tertiary structure.
DCA has been proved to be able to correctly predict with good accuracy the contact map of proteins belonging to sufficiently large protein families.

For this study we used an in-house code \cite{web:DCA},  which compute DCA using the pseudolikelihood approximation \cite{Malinverni2019}. 
The inference was performed for only the regions of the alignment for which the fraction of gaps is less than 30\%, which mainly correspond to the region from residue 1280 to residue 1840.

In particular, DCA analysis was obtained by performing an initial alignment comprising sequences Q15858(\textit{Homo sapiens}), Q62205(\textit{Mus musculus}), O08562(\textit{Rattus norvegicus}) and Q28644(\textit{Oryctolagus cuniculus}) from the Uniprot\cite{uniprot} database. The alignment has been used as a seed for searching homologous sequences in Uniprot and Metaclust\cite{Steinegger2018} databases.
An hidden markov model profile of the seed has been built with hmmbuild with default parameters.
We searched the profile using hmmsearch with default parameters. We found 45447 matches on Uniprot and 104741 on Metaclust.
We concatenated the sequences obtained from the two databases into a single MSA and filtered out all those sequences with more than 50\% of gaps.
The filtered MSA only contained 6637 sequences.
In order to select a region of the protein for which we could obtain a lower fraction of gaps we selected from the full MSA only those columns with less than 30\% of gaps, and we filtered out again those sequences with more than 70\% gaps. This restricted MSA contains 19045 sequences.

Uniprot database has been downloaded on 2018/11/27, while we used Metaclust release $2018-06$ for metagenomic data.

\subsection{Homology Modelling}
\label{subsec:homology}
Homology modeling  relies on evolutionary related structures (templates) to generate a structural model of protein of interest (target). For this task we used Swiss-Model \cite{web:Swiss-Model}, which is widely known and used in the literature. It uses a sequence S and a template T (in PDB format) as input. The output generated is a tentative three-dimensional (3d) structure (in PDB format) associated with the sequence S. 
Swiss-model includes also an energy minimization step to resolve small structural distortions, unfavourable interactions or clashes introduced during the modelling process. Besides the 3d structure, also the quality of the obtained model is evaluated.

\subsection{Energy Minimization}
\label{subsec:minimization}
The structures produced by homology modeling are often closer to the template than to their native structures. It is therefore necessary to perform an energy minimization step using dedicate tools. To this purpose we employed FG-MD \cite{FG-MD}: it uses a multiple templates approach to reshape the energy landscape from golf-course-like to funnel-like ones and drive the energy minimization closer to native state. 
\subsection{Quality Assessment}
\label{subsec:quality}
QMEANBrane \cite{QMEANBrane} exploits the increasing availability of deposited high definition membrane protein structures to adapt knowledge-based methods to this class of proteins.
It is known that the properties of membrane proteins are strongly influenced by their interaction with phospholipid tails, but a clear division into a membrane region and a soluble region does not adequately reflect the variation in molecular properties along the membrane axis.
To capture these differences, QMEANBrane divides the study into three parts: an interface zone consisting of all those residues whose $C_\alpha$ are at a distance of 5\AA~from the defined membrane plane; a membrane region enclosed by all those residues that are more than 5\AA~between the two planes; finally, a region of soluble protein consisting of the remaining amino acids.  
\subsection{Residue Interaction Networks (RIN)}
\label{subsec:RIN}
Given a protein structure, its corresponding RIN is generated by inspecting its atoms and looking for non-covalent bonds between them. We used RING 2.0\cite{web:RING2.0} to perform this task. Given the PDB representation of a protein, the RING-2.0 algorithm generates the graph in two steps. The first identifies a list of residue-residue pairs eligible to undergo an interaction based merely on distance measurements. The second characterizes every contact by identifying the specific type of interaction, which can be Hydrogen bond, Van der Waals, Ionic bridges, $\pi$-$\pi$-Stacking,$\pi$-cation and Disulfide bridges.

RING 2.0 allows the RIN generation according to  various parameters that the user can select. All RINs produced in this study have been generated using the default options. 
In particular, for the \textit{Network policy} parameter we used the \textit{Closest} option, where all atoms of a residue-residue pair are considered to measure the distance; for the \textit{Interaction type} we used the \textit{Multiple} option, where RING reports multiple interactions per residue pair but only one interaction per interaction type; for the \textit{Distance threshold} we used the \textit{Strict} option, where each type of interaction has a specific threshold depending on its strength, as reported in Table \ref{tab:distanceThreshold}.
\begin{table}[th]
  \centering 
  \small
  \begin{tabular}{|c|c|}
  \hline
  Interaction type     &  \textit{Strict} threshold (\AA) \\  
  \hline
  Hydrogen bonds       &  3.5 \\
  Van der Waals        &  0.5 \\
  Ionic Bridge         &  4.0 \\
  $\pi$-$\pi$ stacking &  6.5 \\
  $\pi$-cation         &  5.0 \\
  \hline
    \end{tabular}
    \caption{Threshold distance values for the  \textit{Strict} option of RING 2.0}
    \label{tab:distanceThreshold}
\end{table}

\subsection{Graph Kernels}
\label{subsec:graph}
Graph kernels can be intuitively understood as functions measuring the similarity of pairs of graphs. We used two different kernels to compare RINs, the Vertex Histrogram (VH) and  Weisfeiler-Lehman (WL) subtree kernels, implemented by the GraKel Python library \cite{GraKel}.
Both of them are based on node labels comparison, where node labels are set as the position of each node, i.e. amino acid, in the protein sequence. The essential difference between the two methods is that the VH method makes a pairwise comparison of the nodes labels only, whereas the WL method compares also the corresponding subtrees, hence providing a more complete and general contrast. 
Formal introduction to graph kernels can be found e.g. in \cite{NSV2019,shervashidze2011weisfeiler,SB2015}. Here follows a brief description for the VH and WL kernels:    
 \begin{itemize}
     \item \textbf{Vertex Histogram kernel}:  
     Given a graph $G$, its vertex histogram is defined as a vector counting, for each 
     possible node label, the number of nodes in $G$ having exactly that label.
     Hence, all input graphs on the same set of node labels can be compared through
     their vertex histrograms and the kernel function summarizes the result of
     such comparison;
     \item \textbf{Weisfeiler-Lehman subtree kernel}: 
     the key idea of the Weisfeiler-Lehman algorithm is to replace the label of each vertex with a multiset label consisting of the original label of the vertex and the sorted set of labels of its neighbors. The resultant multiset is then compressed into a new, short label. Such new label reflects the knowledge of the node and its neighborhood. This relabeling process is then repeated for $h$ iterations.  
     By performing this procedure simultaneously on all input graphs, it follows that two vertices from different graphs will get identical new labels if and only if they have identical multiset labels. 
     The kernel function in this case compare the node labels of the graphs resulting after 
     each iteration and summarizes the comparison with a real number. It can be shown that this is equivalent to comparing the number of shared subtrees between the two input graphs (the kernel considers all subtrees up to height $h$).
\end{itemize}

\subsection{Clustering methods}
\label{subsec:clustering}
We apply two different clustering methods: Dominant Set (DS) \cite{RotaBulo2017} and Average-Linkage \cite{ELL01}. 
The DS method tries to answer the problem of organizing a set of elements in clusters, in such a way that each group satisfies an internal homogeneity and expresses an external inhomogeneity between the groups. The algorithm behind this approach does not require any assumption underlying the data representation and does not require prior knowledge on the number of clusters to be obtained, as it is able to determine them in sequence.

The Average Linkage is a well-known clustering method that creates clusters by proceeding in a bottom-up way. Given in input the distance matrix of a set of objects, it starts by assigning to each object its own cluster and then proceeds by joining at each step the two most similar clusters. 
In Average Linkage hierarchical clustering, the distance between two clusters is defined as the average distance between each object in one cluster to every object in the other cluster.
We used the R Package \cite{Rpackage} implementation of the Average Linkage clustering and displayed the result as a dendrogram where all input objects are shown on the leaves and the scale represents the distance between clusters. 

\subsection{Support Vector Machine}
\label{subsec:SVM}
 In the basic configuration, Support Vector Machine (SVM) works with two classes and considers linearly separable problems, that is data points which can be separated by an hyperplane. Considering data as n-dimensional labeled points, SVM searches the hyperplane that separates points of different classes, maximizing the distance between the boundary and data points. 
For unsupervised learning, model's prediction accuracy must be evaluated. To this aim we use the cross-validation technique, as implemented in the \textit{Scikit learn} library \cite{sklearn_api}. 

Cross-validation is one of the standard techniques to check the prediction ability of the SVM supervised learning method.
Roughly speaking, given a dataset of known data (the \emph{training set}), cross validation divides the  dataset into $k$ partitions and, for $k$ times, one partition at a time is used as validation set, while the rest of the dataset remains as training set. At each iteration a value of prediction accuracy is measured and, at the end, the mean value of such measures will be the final prediction accuracy result. 
In our case the training set consists of (the representation of) the 85 considered genetic variants and we set $k = 10$, that is, we repeat the partition into validation set and training set $10$ times.

Cross-validation is useful to avoid overfitting problems in accuracy evaluation. The boundary found by SVM to divide two classes is maintained by a small subset of points, called support vectors. These points are the closest to the boundary, so adding more data does not affect predictions unless the new data is considerably closer to the boundary than the support vectors. Since we don't have a large data set, the use of cross-validation helps to perturb the boundary. In fact, for each step, the support vectors are composed of different points. If cross validation produces a good result for accuracy it means that the problem is highly linearly separable and the model has a good level of generalization. 
Even having a large dataset, using cross-Validation gives more reliable results, that are not strictly related to a particular partition of training and testing datasets.

\subsection{Node Frequency Analysis}
\label{subsec:node}

This analysis takes in consideration RINs nodes.
We remind that each node represents an amino acid of the protein sequence, labeled by its position in the sequence itself. 
The idea is then to compute the frequency of each node with the aim of highlighting the nodes role in pathogenic and non pathogenic mutations.

In particular, we computed the frequency of each node label within the pathogenic mutations (\textbf{PAT} group) and, separately, within the normal variants (\textbf{NEUTRAL} group). 
For instance, a node label $l$ (i.e. the amino acid in position $l$ in the protein sequence) that is present in all RINs of the pathogenic mutations has frequency one for the \textbf{PAT} group. If the same node label $l$ is also present in all RINs of the normal variants, its frequency is one as well for the \textbf{NEUTRAL} group. By comparing the two frequencies we could argue that $l$ is not a sensible node position w.r.t. the \textbf{PAT} and \textbf{NEUTRAL} genetic variants.

\subsection{Entropy Analysis}
\label{subsec:entropy}
The MSA calculated for DCA analysis can be also used to calculate the entropy profile of the protein, with the idea of shedding light on which amino acids, along the chain, are most likely to change during evolution. The entropy profile $S_i$, of our MSA, has been calculated by first computing the frequency $f_i(\alpha)$ of each non-gap symbols $\alpha$ for a given position $i$, and finally summing $S_i=-\sum_{\alpha}\;f_i(\alpha)\ln(f_i(\alpha))$ according with the Shannon formula. A high entropy is indicative of a high mutation propensity, and vice versa low entropy is characteristic of conserved positions. 

\acknowledgments{We thank Dimos Kapetis and Paolo De Los Rios for useful discussions.  The use of the SCSCF multiprocessor cluster at  the Universit\`{a} Ca' Foscari Venezia  is gratefully acknowledged. The work was supported by MIUR PRIN-COFIN2017 \textit{Soft Adaptive Networks} grant 2017Z55KCW (A.G).}
\section*{Author contributions statement}
M.S., G.L. and A.G. conceived the research. A.T., G.C., S.Z., and M.S. performed the research. A.T, A.G., and M.S. analyzed the data and wrote the paper. All authors discussed the results and reviewed the manuscript.
\clearpage
\appendix
\section{Rin Generation}

RING 2.0 \cite{web:RING2.0} allows the RIN generation according to  various parameters that the user can select.
All RINs produced for the three considered templates and the 85 considered mutations have been generated according to the following parameters:
\begin{itemize}
\item \textbf{Network policy:} we used the \textbf{Closest}
option (the default option), where all atoms of a residue-residue pair are considered to measure the distance;
\item\textbf{Interaction type:} we used the \textbf{Multiple} option (the default one) where
    RING reports multiple interactions per residue pair but only one interaction per interaction type.
\item \textbf{Distance threshold:} we used the \textbf{Strict} option, where each type of interaction has a specific threshold depending on its strength, as are reported in Table \ref{tab:distanceThreshold_app}:
\end{itemize}
\begin{table}[th]
  \centering 
  \small
  \begin{tabular}{|c|c|}
  \hline
  Interaction type     &  \textbf{Strict} threshold (\AA) \\  
  \hline
  Hydrogen bonds       &  3.5 \\
  Van der Waals        &  0.5 \\
  Ionic Bridge         &  4.0 \\
  $\pi$-$\pi$ stacking &  6.5 \\
  $\pi$-cation         &  5.0 \\
  \hline
    \end{tabular}
    \caption{threshold distance values for the  \textbf{Strict} option of RING 2.0}
    \label{tab:distanceThreshold_app}
\end{table}
\section{DCA analysis}
\label{S2_Appendix}

\begin{figure}
   \begin{subfigure}{0.40\textwidth}
     \centering{\includegraphics[width=\linewidth]{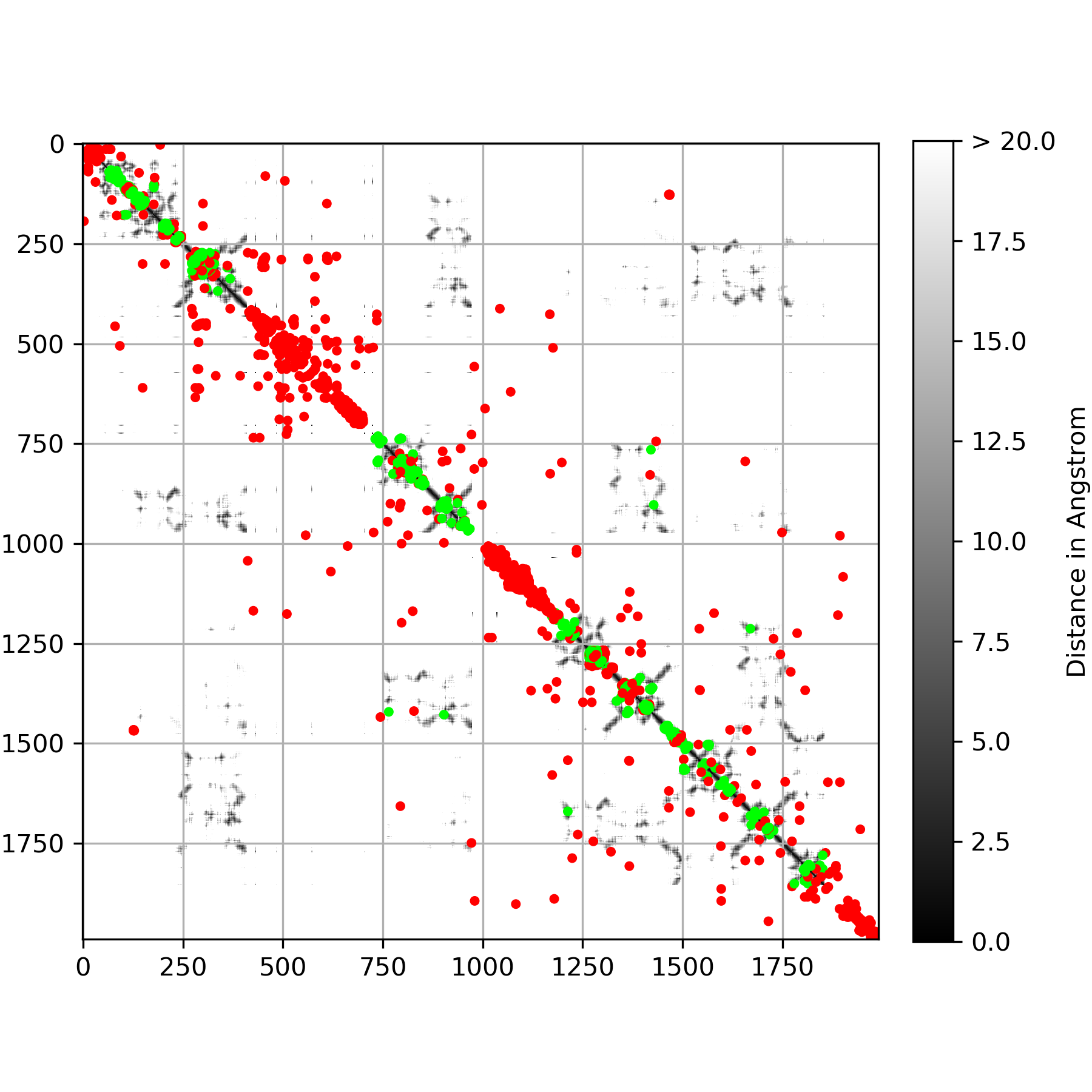}}
    \caption{6A90 - full sequence}
   \end{subfigure}
   \begin{subfigure}{0.40\textwidth}     \centering{\includegraphics[width=\linewidth]{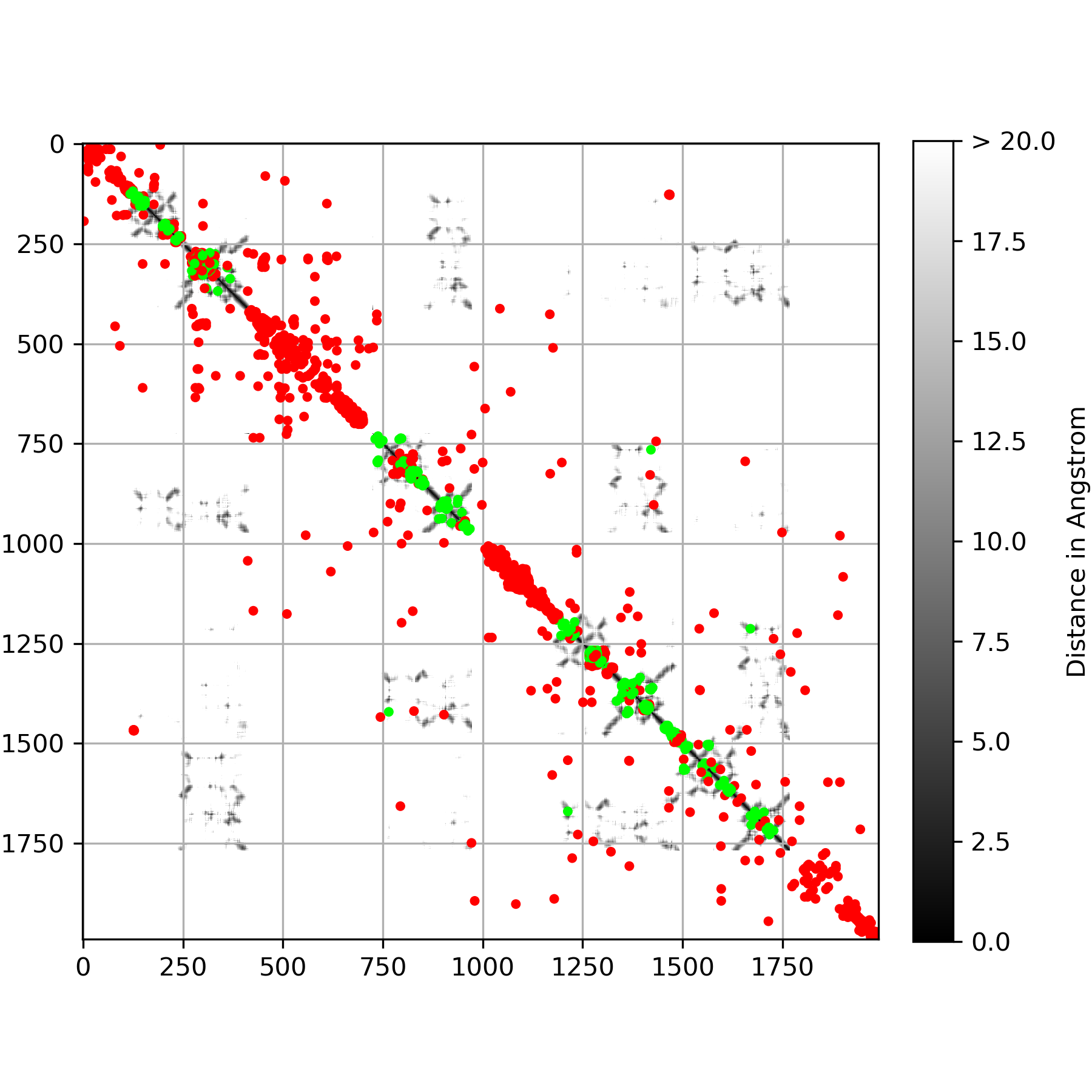}}
    \caption{6J8J - full sequence}
   \end{subfigure}
   \begin{subfigure}{0.40\textwidth}
     \centering{\includegraphics[width=\linewidth]{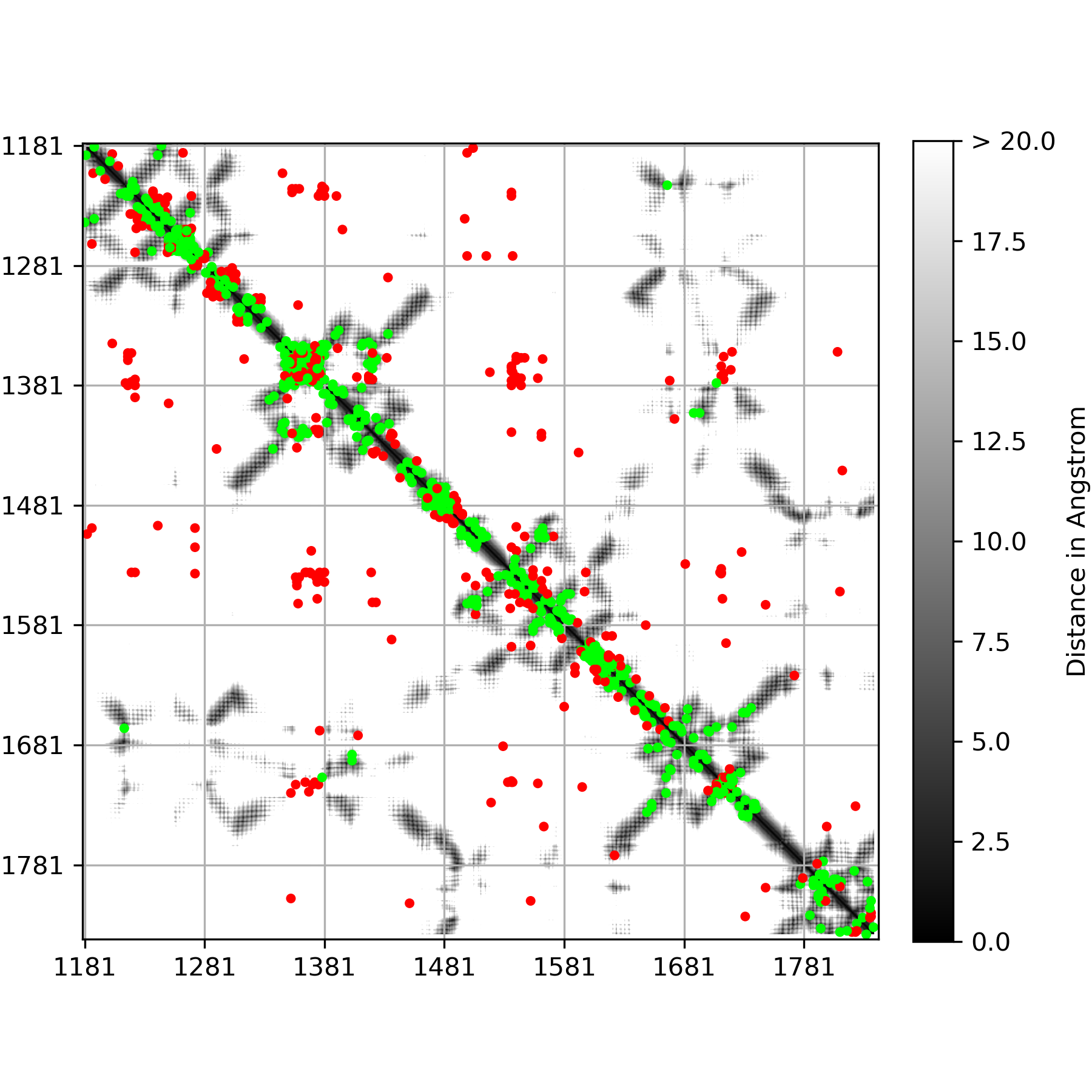}}
    \caption{6A90 - interval 1181-1851}
   \end{subfigure}
   \begin{subfigure}{0.40\textwidth}     \centering{\includegraphics[width=\linewidth]{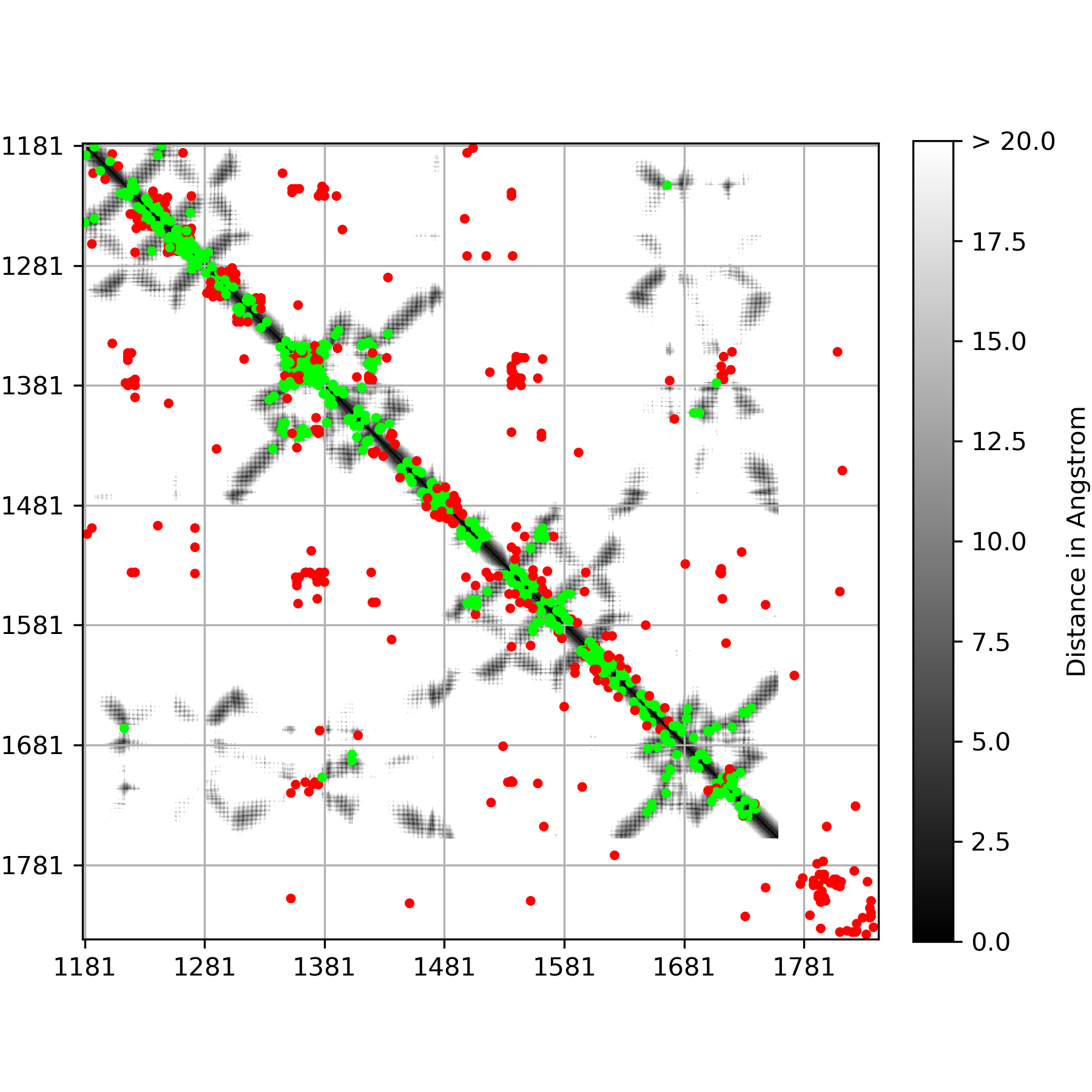}}
    \caption{6J8J - interval 1181-1851}
   \end{subfigure}
   \caption{DCA analysis for templates 6A90 and 6J8J. (a) and (b): full sequence. (c) and (d): interval 1181-1851}
    \label{fig:DCA_app}
\end{figure}

Figure \ref{fig:DCA_app} shows the results of the DCA analysis for templates 6A90 and 6J8J, commented in the main text.

DCA analysis was obtained by performing an initial alignment comprising sequences Q15858, Q62205, O08562 and Q28644. The alignment has been used as a seed for searching homologous sequences in Uniprot and Metaclust databases.
An hidden markov model profile of the seed has been built with hmmbuild with default parameters.
We searched the profile using hmmsearch with default parameters. We found 45447 matches on Uniprot and 104741 on Metaclust.
We concatenated the sequences obtained from the two databases into a single multiple sequence analysis (MSA) and filtered out all those sequences with more tha 50\% of gaps.
The filtered MSA only contained 6637 sequences.
In order to select a region of the protein for wich we could obtain a lower fraction of gaps we selected from the full MSA only those columns with less than 30\% of gaps, and the filtered out again those sequences with more than 70\% gaps. This restricted MSA contains 19045 sequences.

Uniprot database has been downloaded on 2018/11/27, while we used Metaclust release $2018-06$ for metagenomic data.

We performed pseudolikelihood DCA \cite{Malinverni2019,Ekeberg} on the full and restricted MSAs with both field and coupling regularizer equal to 0.01. The software used is available from https://gitlab.com/LBS-EPFL/code/lbsDCA/tree/v1.0

\section{Graph Kernels}
\label{S3_Appendix}

Formal introduction to graph kernels can ve found e.g. in \cite{NSV2019,shervashidze2011weisfeiler,SB2015}.
In this section we briefly illustrate the two kernels applied for RINs comparison:

 \begin{itemize}
     \item \textbf{Vertex Histogram kernel} (VH):  
     Given a graph $G$, its vertex histogram is defined as a vector counting, for each 
     possible node label, the number of nodes in $G$ having exactly that label.
     Hence, all input graphs on the same set of node labels can be compared through
     their vertex histrograms and the kernel function summarizes the result of
     such comparison;
     \item \textbf{Weisfeiler-Lehman subtree kernel} (WL): 
     the key idea of the Weisfeiler-Lehman algorithm is to 
     replace the label of
     each vertex with a multiset label consisting of the original label of the vertex
     and the sorted set of labels of its neighbors. The resultant multiset is then
     compressed into a new, short label. Such new label reflects the knowledge of the node and its
     neighborhood. This relabeling process is then repeated for $h$ iterations.  
     By performing this procedure simultaneously on all input graphs, it follows
     that two vertices from different graphs will get identical new labels if and only if they have identical multiset labels. 
     The kernel function in this case compare the node labels of the graphs resulting after 
     each iteration and summarizes the comparison with a real number. It can be shown that this is equivalent to comparing the number of shared subtrees between the two input graphs (the kernel considers all subtrees up to height $h$).
\end{itemize}

\section{Support Vector Machine}
\label{S4_Appendix}

 In the basic configuration, Support Vector Machine (SVM) works with two classes and considers linearly separable problems, that is data points which can be separated by an hyperplane. Considering data as n-dimensional labeled points, SVM searches the hyperplane that separates points of different classes, maximizing the distance between the boundary and data points. 
For unsupervised learning, model's prediction accuracy must be evaluated. To this aim we use the cross validation technique, as implemented in 
the \textit{Scikit learn} library \cite{sklearn_api}. 
Roughly speaking, given a dataset of known data (the \emph{training set}), cross validation divides the  dataset into $k$ partitions and, for $k$ times, one partition at a time is used as validation set, while the rest of the dataset remains as training set. At each iteration a value of prediction accuracy is measured and, at the end, the mean value of such measures will be the final prediction accuracy result. 
In our case the training set consists of (the representation of) the 85 considered mutations and we set $k = 10$, that is we repeat the partition into validation set and training set $10$ times.

Cross-validation is useful to avoid overfitting problems in accuracy evaluation. The boundary found by SVM to divide two classes is maintained by a small subset of points, called support vectors. These points are the closest to the boundary, so adding more data does not affect predictions unless the new data is considerably closer to the boundary than the support vectors. Since we don't have a large data set, the use of cross-validation helps to perturb the boundary. In fact, for each step, the support vectors are composed of different points. If cross validation produces a good result for accuracy it means that the problem is highly linearly separable and the model has a good level of generalization.

\section{Quality assessment: detailed analysis}
\label{SM:quality}

In this section we report further results concerning the quality assessment of the three-dimensional structures obtained with our computational pipeline.

Each structure obtained by FG-MD was subjected to an quality evaluation with the tool QMEANBrane, which has a special scoring function designed for membrane proteins. 
As reported in the paper, QMEANBrane shows that the produced models are of high quality within each domain area, while, in the inter-domains loops area, the reliability of the models are significantly lower.

\begin{figure}[ht]
   \begin{subfigure}{0.30\textwidth}
     \centering{\includegraphics[angle=90,width=\linewidth]{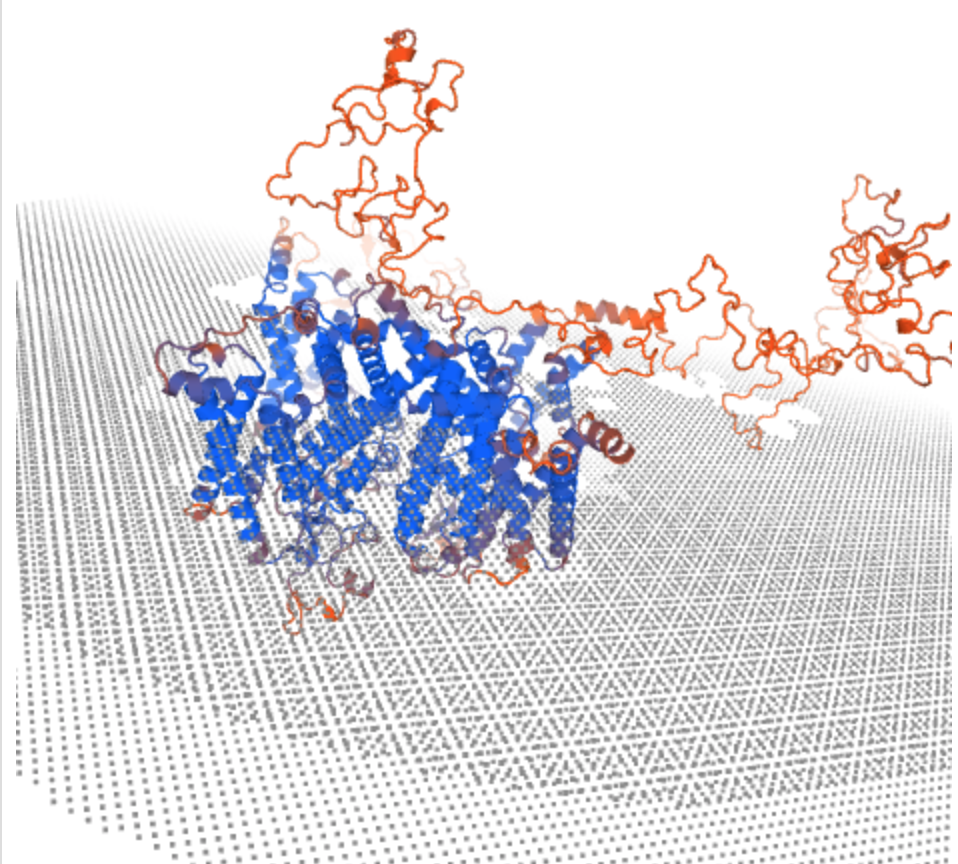}}
    \caption{Structure}
   \end{subfigure}
   \begin{subfigure}{0.68\textwidth}
     \centering{\includegraphics[width=\linewidth]{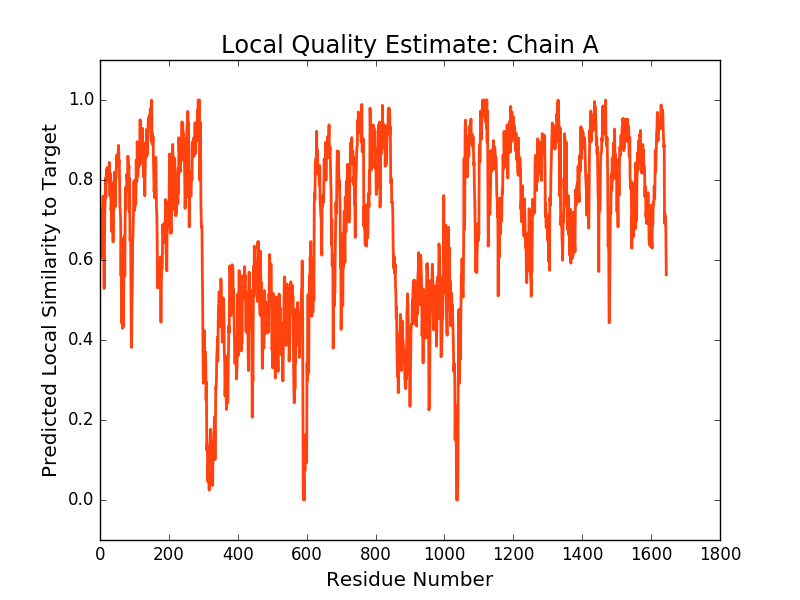}}
     \caption{Quality values along the peptide sequence}
   \end{subfigure}
   \begin{subfigure}{0.7\textwidth}
     \centering{\includegraphics[width=0.70\linewidth]{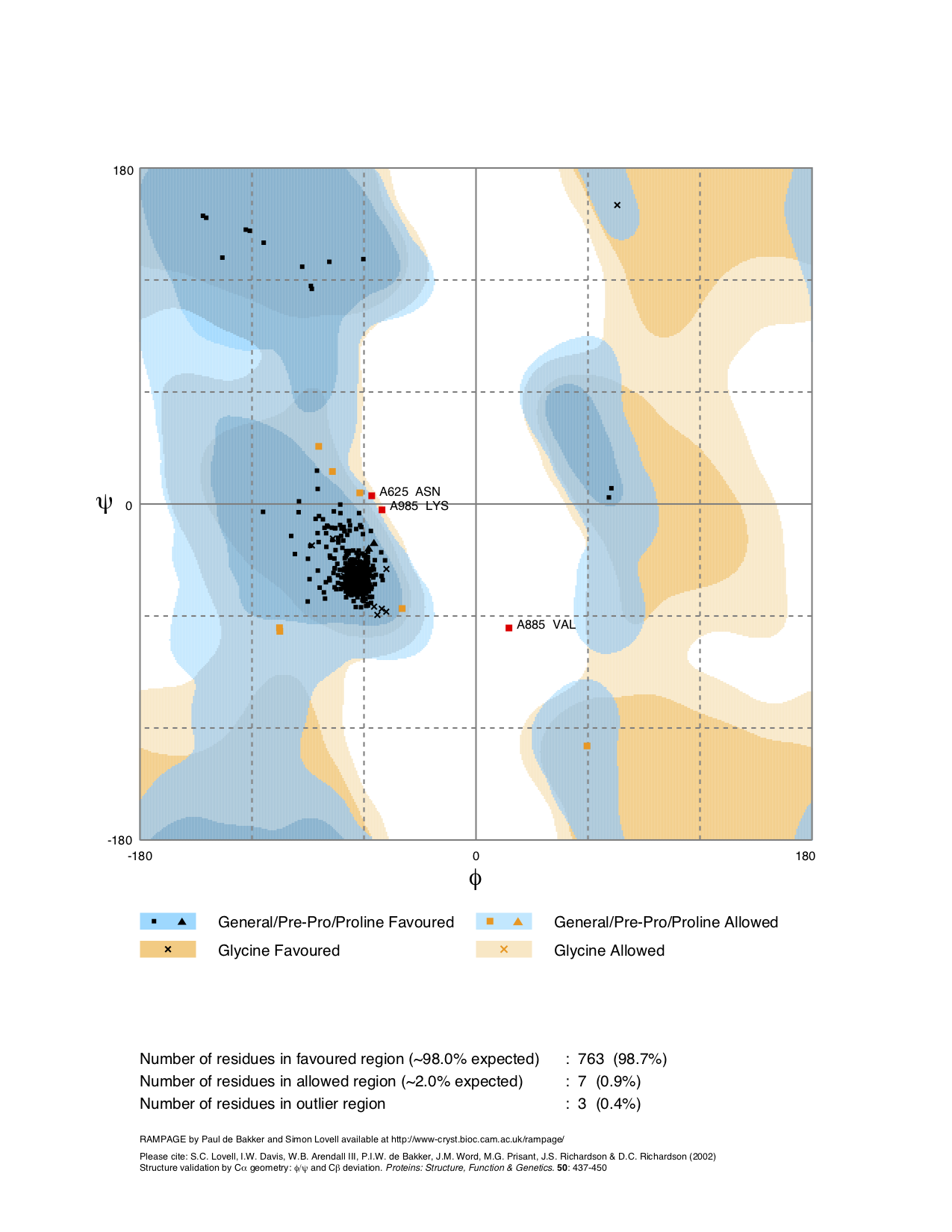}}
     \caption{Ramachandran plot analysis}
   \end{subfigure}
   \caption{Quality results for MOESM3 Wild Type}
   \label{fig:qualityMOESM3}
\end{figure}

Figure \ref{fig:qualityMOESM3} shows the quality results for the MOESM3 template. 
In particular, part (a) depicts the structure of the WT and part (b) shows the quality values along peptide sequence: the higher values correspond to positions falling into the transmembrane region. Moreover, part (c) of Figure
\ref{fig:qualityMOESM3} shows the RAMPAGE results for the 6A90 WT. The analysis considers only the protein segments corresponding to the transmembrane region, i.e. all the $\alpha$-helices.
It turns out that 95.8\%  amino acids fall within a favorable region, 4\% fall within a permitted region and only one amino acid falls in the forbidden region.

\begin{figure}[ht]
   \begin{subfigure}{0.35\textwidth}
     \centering{\includegraphics[angle=90,width=\linewidth]{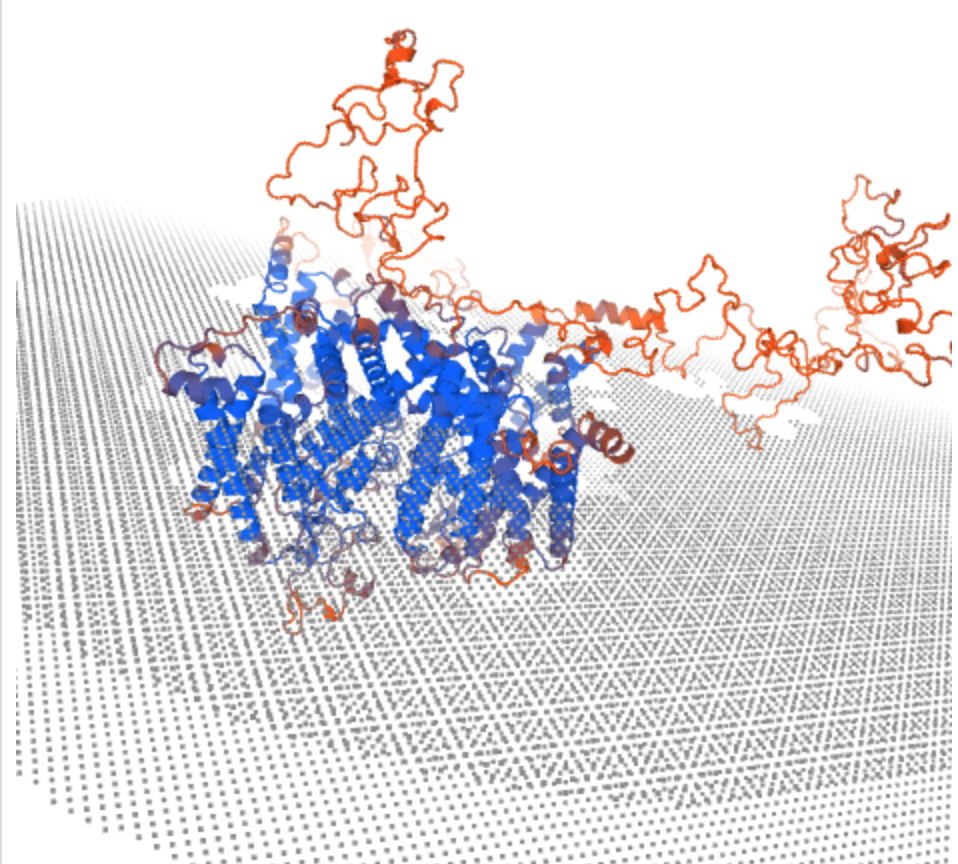}}
    \caption{Structure}
   \end{subfigure}
   \begin{subfigure}{0.63\textwidth}
     \centering{\includegraphics[width=\linewidth]{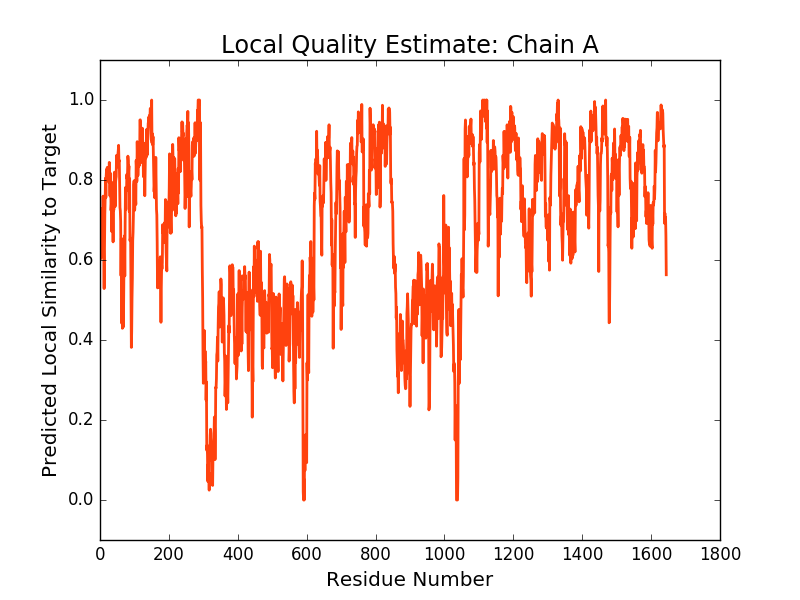}}
     \caption{Quality values along the peptide sequence}
   \end{subfigure}
   \begin{subfigure}{0.7\textwidth}
     \centering{\includegraphics[width=0.70\linewidth]{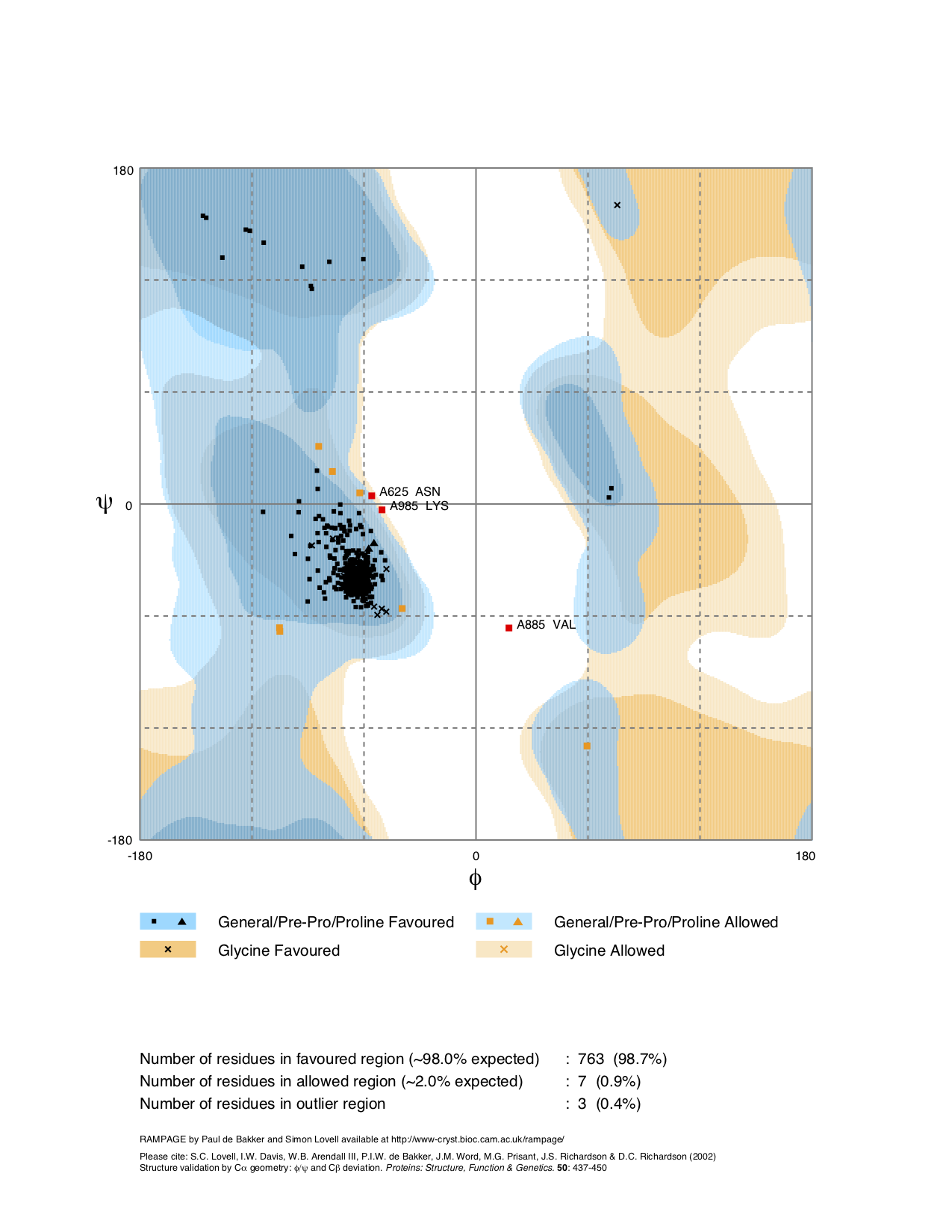}}
     \caption{Ramachandran plot analysis}
   \end{subfigure}
   \caption{Quality results for 6J8J Wild Type}
   \label{fig:quality6J8J}
\end{figure}

The same analysis is reported in Figure \ref{fig:quality6J8J} for template 6J8J.

To complete the quality assessment, the local quality values of all the considered point mutations are reported in Tables \ref{tab:MutQMEANBrane} and \ref{tab:hSNPM3Brane} for the MOESM3 template; in Tables \ref{tab:MutQMEANBrane6A} and \ref{tab:hSNPBrane6A} for the 6A90 template and 
in Tables \ref{tab:MutQMEANBrane6J8J} and \ref{tab:hSNPBrane6J8J} for the 6J8J template. 
\begin{table}[ht]
    \centering
    \scriptsize{
    \begin{tabular}{|p{1cm} c|c|c|}
      \hline
      \multicolumn{4}{|c|}{\textbf{QMEANBrane results: MOESM3} - pain related} \\
 \hline
& Mutation  & WT  & Mutated\\ 
 \hline
\multirow{18}{*}{\textbf{IEM}} 
  &  I136V & 0.86 & 0.9   \\
  & S211P  & 0.73 & 0.8   \\
  & F216S  & 0.8  & 0.82  \\
  & I234T  & 0.81 & 0.84  \\
  & S241T  & 0.88 & 0.92  \\
  & N395K  & 0.85 & 0.83  \\
  & V400M  & 0.92 & 0.94  \\
  & L823R  & 0.82 & 0.78  \\
  & I848T  & 0.79 & 0.79  \\
  & L858H  & 0.8  & 0.77  \\
  & L859F  & 0.8  & 0.61  \\
  & A863P  & 0.9  & 0.87  \\
  & V872G  & 0.87 & 0.88  \\
  & P1308L & 0.61 & 0.74  \\
  & V1316A & 0.85 & 0.88  \\
  & F1449V & 0.84 & 0.89  \\
  & W1538R & 0.6  & 0.57  \\
  & A1746G & 0.91 & 0.91  \\
  \hline
\multirow{6}{*}{\textbf{PEPD}}  
  & V1298D & 0.77 & 0.87  \\
  & V1298F & 0.77 & 0.72  \\
  & V1299F & 0.77 & 0.81  \\
  & G1607R & 0.83 & 0.76  \\
  & M1627K & 0.77 & 0.74  \\
  & A1632E & 0.82 & 0.77  \\
  \hline
\multirow{6}{*}{\textbf{SFN}}  
  & R185H  & 0.29 & 0.38  \\
  & I228M  & 0.77 & 0.77  \\
  & I739V  & 0.84 & 0.87  \\
  & G856D  & 0.72 & 0.71  \\
  & M932L  & 0.76 & 0.73  \\
  & M1532I & 0.62 & 0.49 \\
  \hline
  \end{tabular}
}    
    \caption{QMEANBrane results for the models generated by the MOESM3 template: quality value of the the point mutation before (WT) and after (mutated) the amino acid change. All the mutations related to pain disorders are considered.}
    \label{tab:MutQMEANBrane}
\end{table}

\begin{table}[htbp]
    \centering
   \scriptsize{
    \begin{tabular}{|c| c|c || c| c|c|}
      \hline
      \multicolumn{6}{|c|}{\textbf{QMEANBrane results: MOESM3} - non pain related} \\
 \hline
Mut & WT(FG-MD)  & Mut(FG-MD) & Mut   & WT(FG-MD)  & Mut(FG-MD)\\ 
 \hline
 S126A   & 0.81 & 0.77 & A1505V  & 0.73 & 0.72 \\
 L127A   & 0.81 & 0.84 & S1509A  & 0.9  & 0.93 \\
 M145L   & 0.54 & 0.77 & S1509T  & 0.9  & 0.87 \\
 N146S   & 0.53 & 0.73 & Q1530D  & 0.69 & 0.56 \\
 V194I   & 0.85 & 0.84 & Q1530K  & 0.69 & 0.42 \\
 L201V   & 0.8  & 0.82 & Q1530P  & 0.69 & 0.62 \\
 N206D   & 0.73 & 0.6  & H531Y   & 0.7  & 0.47 \\
 T370M   & 0.64 & 0.56 & M1532V  & 0.62 & 0.62 \\
 E759D   & 0.65 & 0.74 & E1534D  & 0.47 & 0.58 \\
 A766T   & 0.76 & 0.78 & Y1537N  & 0.55 & 0.64 \\
 A766V   & 0.76 & 0.81 & T1548S  & 0.88 & 0.89 \\
 I767V   & 0.76 & 0.85 & H1560C  & 0.67 & 0.77 \\
 T773S   & 0.82 & 0.83 & H1560Y  & 0.67 & 0.58 \\
 V795I   & 0.84 & 0.77 & V1565I  & 0.8  & 0.71 \\
 A815S   & 0.76 & 0.88 & I1577L  & 0.81 & 0.85 \\
 D890E   & 0.54 & 0.51 & D1586E  & 0.75 & 0.6  \\
 D890V   & 0.54 & 0.51 & T1590K  & 0.5  & 0.53 \\
 T920N   & 0.64 & 0.73 & T1590R  & 0.5  & 0.58 \\
 K1176R  & 0.32 & 0.54 & T1596I  & 0.84 & 0.75 \\
 R1207K  & 0.66 & 0.68 & V1613I  & 0.78 & 0.78 \\
 T1210N  & 0.75 & 0.77 & D1662A  & 0.52 & 0.58 \\
 I1235V  & 0.71 & 0.86 & G1674A  & 0.84 & 0.84 \\
 N1245S  & 0.72 & 0.74 & K1700A  & 0.63 & 0.67 \\
 L1267V  & 0.79 & 0.89 & &&\\
 T1398M  & 0.77 & 0.7  & &&\\
 I1399D  & 0.74 & 0.79 & &&\\
 D1411N  & 0.55 & 0.55 & &&\\
 K1412E  & 0.55 & 0.62 & &&\\
 K1412I  & 0.55 & 0.67 & &&\\
 K1415I  & 0.61 & 0.67 & &&\\
 S1419N  & 0.84 & 0.84 & &&\\
 V1428I  & 0.88 & 0.86 & &&\\
 \hline
    \end{tabular}
}
    \caption{QMEANBrane results for the models generated by the MOESM3 template: quality value of the the point mutation before (WT) and after (mutated) the amino acid change. All mutations not related to pain disorders are considered.}
    \label{tab:hSNPM3Brane}
\end{table}

\begin{table}[htbp]
    \centering
    \scriptsize{
    \begin{tabular}{|p{1cm} |c| c|c|}
      \hline
 \multicolumn{4}{|c|}{\textbf{QMEANBrane results: 6A90} - pain related} \\
 
 \hline
& Mut   & WT(FG-MD)  & Mut(FG-MD)\\ 
 \hline
\multirow{18}{*}{\textbf{IEM}}
  & I136V   & 0.85 & 0.89  \\
  & S211P   & 0.68 & 0.78  \\
  & F216S   & 0.78 & 0.83  \\
  & I234T   & 0.86 & 0.86  \\
  & S241T   & 0.93 & 0.94  \\
  & N395K   & 0.86 & 0.88  \\
  & V400M   & 0.99 & 0.94  \\
  & L823R   & 0.74 & 0.74  \\
  & I848T   & 0.78 & 0.75  \\
  & L858H   & 0.88 & 0.82  \\
  & L859F   & 0.88 & 0.89  \\
  & A863P   & 0.98 & 0.97  \\
  & V872G   & 0.86 & 0.88  \\
  & P1308L  & 0.8  & 0.76  \\
  & V1316A  & 0.9  & 0.93  \\
  & F1449V  & 0.95 & 0.89  \\
  & W1538R  & 0.96 & 1.0   \\
  & A1746G  & 0.96 & 0.99  \\
  \hline 
\multirow{6}{*}{\textbf{PEPD}}  
  & V1298D  & 0.77 & 0.87  \\
  & V1298F  & 0.92 & 0.89  \\
  & V1299F  & 0.92 & 0.93  \\
  & G1607R  & 0.85 & 0.76  \\
  & M1627K  & 0.88 & 0.9   \\
  & A1632E  & 0.9  & 0.84  \\
  \hline
\multirow{6}{*}{\textbf{SFN}}  
  & R185H   & 0.46 & 0.6   \\
  & I228M   & 0.72 & 0.72  \\
  & I739V   & 0.92 & 0.96  \\
  & G856D   & 0.86 & 0.8   \\
  & M932L   & 0.77 & 0.79  \\
  & M1532I  & 0.84 & 0.79  \\
  \hline
    \end{tabular}
}
    \caption{QMEANBrane results for the models generated by the 6A90 template: quality value of the the point mutation before (WT) and after (mutated) the amino acid change. All pain related mutations are considered.}    
    \label{tab:MutQMEANBrane6A}
\end{table}

\begin{table}[htbp]
    \centering
  \scriptsize{
    \begin{tabular}{|c|c|c || c|c|c|}
      \hline
 \multicolumn{6}{|c|}{\textbf{QMEANBrane results: 6A90} - non pain related} \\
 \hline
Mut    & WT(FG-MD) & Mut(FG-MD) & Mut   & WT(FG-MD)  & Mut(FG-MD)\\ 
 \hline
 S126A    & 0.72 & 0.77 & A1505V  & 0.87 & 0.82 \\
 L127A    & 0.78 & 0.77 & S1509A  & 0.93 & 0.89 \\
 M145L    & 0.73 & 0.77 & S1509T  & 0.93 & 0.97 \\
 N146S    & 0.79 & 0.69 & Q1530D  & 0.87 & 0.67 \\
 V194I    & 0.91 & 0.88 & Q1530K  & 0.87 & 0.73 \\
 L201V    & 0.69 & 0.75 & Q1530P  & 0.87 & 0.77 \\
 N206D    & 0.47 & 0.32 & H531Y   & 0.88 & 1.0  \\
 T370M    & 0.93 & 0.92 & M1532V  & 0.84 & 0.88 \\
 E759D    & 0.87 & 0.9  & E1534D  & 0.88 & 0.88 \\
 A766T    & 0.88 & 0.92 & Y1537N  & 0.91 & 0.91 \\
 A766V    & 0.88 & 0.92 & T1548S  & 0.93 & 0.94 \\
 I767V    & 0.84 & 0.93 & H1560C  & 0.64 & 0.65 \\
 T773S    & 0.9  & 0.93 & H1560Y  & 0.64 & 0.7  \\
 V795I    & 0.75 & 0.69 & V1565I  & 0.72 & 0.72 \\
 A815S    & 0.66 & 0.74 & I1577L  & 0.86 & 0.9  \\
 D890E    & 0.53 & 0.59 & D1586E  & 0.58 & 0.42 \\
 D890V    & 0.53 & 0.49 & T1590K  & 0.47 & 0.38 \\
 T920N    & 0.92 & 0.89 & T1590R  & 0.47 & 0.42 \\
 K1176R   & 0.74 & 0.64 & T1596I  & 0.8  & 0.81 \\
 R1207K   & 0.59 & 0.71 & V1613I  & 0.77 & 0.78 \\
 T1210N   & 0.82 & 0.69 & D1662A  & 0.7  & 0.63 \\
 I1235V   & 0.84 & 0.83 & G1674A  & 0.87 & 0.85 \\
 N1245S   & 0.74 & 0.76 & K1700A  & 0.64 & 0.68 \\
 L1267V   & 0.34 & 0.05 & &&\\
 T1398M   & 0.9  & 0.9  & &&\\
 I1399D   & 0.86 & 0.86 & &&\\
 D1411N   & 0.67 & 0.7  & &&\\
 K1412E   & 0.68 & 0.65 & &&\\
 K1412I   & 0.68 & 0.63 & &&\\
 K1415I   & 0.66 & 0.66 & &&\\
 S1419N   & 0.76 & 0.75 & &&\\
 V1428I   & 0.88 & 0.86 & &&\\
\hline 
    \end{tabular}}
    \caption{QMEANBrane results for the models generated by the 6A90 template: quality value of the the point mutation before (WT) and after (mutated) the amino acid change. All mutations not related to pain disorders are considered.}     
    \label{tab:hSNPBrane6A}
\end{table}

\begin{table}[htbp]
    \centering
    \scriptsize{
    \begin{tabular}{|p{1cm} |c| c|c|}
      \hline
 \multicolumn{4}{|c|}{\textbf{QMEANBrane results: 6J8J} - pain related} \\
 
 \hline
& Mut   & WT(FG-MD)  & Mut(FG-MD)\\ 
 \hline
\multirow{18}{*}{\textbf{IEM}}
  & I136V   & 0.85 & 0.90  \\
  & S211P   & 0.82 & 0.88  \\
  & F216S   & 0.80 & 0.81  \\
  & I234T   & 0.80 & 0.91  \\
  & S241T   & 0.89 & 0.93  \\
  & N395K   & 0.92 & 0.89  \\
  & V400M   & 0.95 & 0.93  \\
  & L823R   & 0.77 & 0.70  \\
  & I848T   & 0.85 & 0.82  \\
  & L858H   & 0.84 & 0.78  \\
  & L859F   & 0.84 & 0.85  \\
  & A863P   & 0.97 & 0.98  \\
  & V872G   & 0.90 & 0.91  \\
  & P1308L  & 0.78 & 0.81  \\
  & V1316A  & 0.92 & 0.92  \\
  & F1449V  & 0.85 & 0.92  \\
  & W1538R  & 1.00 & 1.0   \\
  & A1746G  & 0.95 & 0.97  \\
  \hline 
\multirow{6}{*}{\textbf{PEPD}}  
  & V1298D  & 0.82 & 0.81  \\
  & V1298F  & 0.82 & 0.86  \\
  & V1299F  & 0.85 & 0.87  \\
  & G1607R  & 0.92 & 0.82  \\
  & M1627K  & 0.90 & 0.87  \\
  & A1632E  & 0.90 & 0.88  \\
  \hline
\multirow{6}{*}{\textbf{SFN}}  
  & R185H   & 0.46 & 0.65  \\
  & I228M   & 0.75 & 0.73  \\
  & I739V   & 0.88 & 0.93  \\
  & G856D   & 0.79 & 0.73  \\
  & M932L   & 0.93 & 0.90  \\
  & M1532I  & 0.87 & 0.88  \\
  \hline
    \end{tabular}
}
    \caption{QMEANBrane results for the models generated by the 6J8J template: quality value of the the point mutation before (WT) and after (mutated) the amino acid change. All pain related mutations are considered.}    
    \label{tab:MutQMEANBrane6J8J}
\end{table}

\begin{table}[htbp]
    \centering
  \scriptsize{
    \begin{tabular}{|c|c|c || c|c|c|}
      \hline
 \multicolumn{6}{|c|}{\textbf{QMEANBrane results: 6J8J} - non pain related} \\
 \hline
Mut    & WT(FG-MD) & Mut(FG-MD) & Mut   & WT(FG-MD)  & Mut(FG-MD)\\ 
 \hline
 S126A    & 0.80 & 0.86 & A1505V  & 0.87 & 0.84 \\
 L127A    & 0.80 & 0.84 & S1509A  & 0.91 & 0.95 \\
 M145L    & 0.70 & 0.75 & S1509T  & 0.91 & 0.91 \\
 N146S    & 0.73 & 0.85 & Q1530D  & 0.84 & 0.96 \\
 V194I    & 0.87 & 0.86 & Q1530K  & 0.84 & 0.93 \\
 L201V    & 0.70 & 0.77 & Q1530P  & 0.84 & 0.78 \\
 N206D    & 0.52 & 0.66 & H531Y   & 0.89 & 1.0  \\
 T370M    & 0.94 & 0.94 & M1532V  & 0.87 & 0.97 \\
 E759D    & 0.72 & 1.00 & E1534D  & 0.90 & 0.76 \\
 A766T    & 0.90 & 0.86 & Y1537N  & 0.98 & 0.90 \\
 A766V    & 0.90 & 0.91 & T1548S  & 0.96 & 0.98 \\
 I767V    & 0.85 & 0.90 & H1560C  & 0.77 & 0.72 \\
 T773S    & 0.90 & 0.92 & H1560Y  & 0.77 & 0.54 \\
 V795I    & 0.76 & 0.70 & V1565I  & 0.77 & 0.70 \\
 A815S    & 0.75 & 0.77 & I1577L  & 0.91 & 0.94 \\
 D890E    & 0.72 & 0.69 & D1586E  & 0.74 & 0.78 \\
 D890V    & 0.72 & 0.64 & T1590K  & 0.73 & 0.76 \\
 T920N    & 0.93 & 0.91 & T1590R  & 0.73 & 0.82 \\
 K1176R   & 0.82 & 0.68 & T1596I  & 0.79 & 0.78 \\
 R1207K   & 0.59 & 0.70 & V1613I  & 0.80 & 0.80 \\
 T1210N   & 0.83 & 0.80 & D1662A  & 0.68 & 0.67 \\
 I1235V   & 0.87 & 0.92 & G1674A  & 0.88 & 0.88 \\
 N1245S   & 0.59 & 0.74 & K1700A  & 0.66 & 0.64 \\
 L1267V   & 0.74 & 0.74 & &&\\
 T1398M   & 0.92 & 0.88 & &&\\
 I1399D   & 0.88 & 0.87 & &&\\
 D1411N   & 0.66 & 0.67 & &&\\
 K1412E   & 0.68 & 0.71 & &&\\
 K1412I   & 0.68 & 0.64 & &&\\
 K1415I   & 0.66 & 0.62 & &&\\
 S1419N   & 0.75 & 0.64 & &&\\
 V1428I   & 0.93 & 0.92 & &&\\
\hline 
    \end{tabular}}
    \caption{QMEANBrane results for the models generated by the 6J8J template: quality value of the the point mutation before (WT) and after (mutated) the amino acid change. All mutations not related to pain disorders are considered.}  
    \label{tab:hSNPBrane6J8J}
\end{table}

\begin{table}[htbp]
    \centering
    \scalebox{0.65}{
    \begin{tabular}{|p{1cm} c|c | c| c|}
      \hline
 \multicolumn{5}{|c|}{\textbf{Ramachandran Plot Analysis of WT and mutations pain-related}} \\
 
 \hline
& Mutation & MOESM3(FG-MD)  & 6A90(FG-MD) & 6J8J(FG-MD)\\ 
 \hline
\multirow{18}{*}{\textbf{IEM}}
  & I136   & $\surd$  & $\surd$ & $\surd$  \\
  & S211   & $\surd$  & $\times\circ$ & $\surd$ \\
  & F216   & $\surd$  & $\surd$ & $\surd$ \\
  & I234   & $\surd$  & $\surd$ & $\surd$ \\
  & S241   & $\surd$  & $\surd$ & $\surd$ \\
  & N395   & $\surd$  & $\surd$ & $\surd$ \\
  & V400   & $\surd$  & $\surd$ & $\surd$ \\
  & L823   & $\surd$  & $\surd$ & $\surd$ \\
  & I848   & $\surd$  & $\surd$ & $\surd$ \\
  & L858   & $\surd$  & $\surd$ & $\surd$ \\
  & A863   & $\surd$  & $\surd$ & $\surd$ \\
  & V872   & $\surd$  & $\surd$ & $\surd$ \\
  & P1308  & $\surd$  & $\surd$ & $\surd$ \\
  & V1316  & $\surd$  & $\surd$ & $\surd$ \\
  & F1449  & $\surd$  & $\surd$ & $\surd$ \\
  & W1538  & $\surd$  & $\surd$ & $\surd$ \\
  & A1746  & $\surd$  & $\surd$ & $\surd$ \\
  \hline
\multirow{6}{*}{\textbf{PEPD}}  
  & V1298  & $\surd$ & $\surd$ & $\surd$ \\
  & V1299  & $\surd$ & $\surd$ & $\surd$ \\
  & G1607  & $\surd$ & $\surd$ & $\surd$ \\
  & M1627  & $\surd$ & $\surd$ & $\surd$ \\
  & A1632  & $\surd$ & $\surd$ & $\surd$ \\
  \hline
\multirow{6}{*}{\textbf{SFN}}  
  & R185  & $\times$  & $\surd$ & $\times\circ$ \\
  & I228  & $\times\circ$  & $\times\circ$ & $\times\circ$ \\
  & I739  & $\surd$  & $\surd$ & $\surd$ \\
  & G856  & $\surd$  & $\surd$ & $\surd$ \\
  & M932  & $\surd$  & $\surd$ & $\surd$ \\
  & M1532 & $\surd$  & $\surd$ & $\surd$ \\
  \hline
    \end{tabular}}
    \caption{$\surd$= residue present in the model containing only the sequence sections having a secondary helical structure, $\times$=residue not present and $\times\circ$= residual not present, but its first neighbor yes.}
    \label{tab:MutRamachandranPlot}
\end{table}

\begin{table}[htbp]
    \centering
    \scalebox{0.65}{
    \begin{tabular}{|c|c |c | c || c|c |c | c|}
      \hline
 \multicolumn{8}{|c|}{\textbf{Ramachandran Plot Analysis of WT and mutations not related with pain-syndrome}} \\
 \hline
Mutation  & M3(FG-MD)  & 6A(FG-MD) & 6J(FG-MD) & Mutation   & M3(FG-MD)  & 6A(FG-MD) & 6J(FG-MD) \\ 
 \hline
 S126    & $\surd$       & $\surd$  & $\surd$       & A1505 & $\surd$       & $\surd$ & $\surd$ \\
 L127    & $\surd$       & $\surd$   & $\surd$      & S1509 & $\surd$       & $\surd$ & $\surd$ \\
 M145    & $\times\circ$ & $\times\circ$  & $\surd$ & Q1530 & $\surd$       & $\surd$ & $\surd$ \\
 N146    & $\times$      & $\times$  & $\times\circ$     & H1531 & $\surd$       & $\surd$ & $\surd$ \\
 V194    & $\surd$       & $\surd$  & $\surd$       & M1532 & $\surd$       & $\surd$  & $\surd$\\
 L201    & $\surd$       & $\surd$  & $\surd$      & E1534 & $\times\circ$ & $\surd$  & $\surd$\\
 N206    & $\times$      & $\times$ & $\times$      & Y1537 & $\surd$       & $\surd$  & $\surd$\\
 T370    & $\times$      & $\surd$   & $\surd$      & T1548 & $\surd$       & $\surd$ & $\surd$\\
 E759    & $\surd$       & $\surd$    & $\surd$     & H1560 & $\surd$       & $\times$ & $\surd$\\
 A766    & $\surd$       & $\surd$    & $\surd$     & V1565 & $\times\circ$ & $\times\circ$ & $\surd$\\
 I767    & $\surd$       & $\surd$  & $\surd$       & I1577 & $\surd$       & $\surd$ & $\surd$\\
 T773    & $\surd$       & $\surd$   & $\surd$      & D1586 & $\surd$       & $\times$ & $\surd$\\
 V795    & $\times\circ$ & $\times\circ$ & $\times\circ$ & T1590 & $\times$      & $\times$ & $\surd$ \\
 A815    & $\times\circ$ & $\times$  & $\surd$       & T1596 & $\surd$       & $\surd$ & $\surd$\\
 D890    & $\times$      & $\times$  & $\times$      & V1613 & $\surd$       & $\times\circ$ & $\surd$ \\
 T920    & $\times$      & $\surd$   & $\surd$       & D1662 & $\surd$       & $\times$ & $\times$  \\
 K1176   & $\times$      & $\surd$   & $\surd$       & G1674 & $\surd$       & $\surd$ & $\surd$ \\
 R1207   & $\surd$       & $\times$   & $\times$    & K1700 & $\times$      & $\times$ & $\times$ \\
 T1210   & $\surd$       & $\surd$   & $\surd$  &  & & &\\
 I1235   & $\surd$       & $\surd$  & $\surd$  &  & & &\\
 N1245   & $\times\circ$ & $\times\circ$ & $\times\circ$ & & & &\\
 L1267   & $\surd$       & $\times$ & $\surd$ & & & &\\
 T1398   & $\surd$       & $\surd$ & $\surd$& & & &\\
 I1399   & $\surd$       & $\surd$ & $\surd$& & & &\\
 D1411   & $\times$      & $\times$& $\times$ & & & &\\
 K1412   & $\surd$       & $\times$ & $\times$ & & & & \\
 K1415   & $\surd$       & $\times$ & $\times$ & & & &\\
 S1419   & $\surd$       & $\times$ & $\times$ & & & &\\
 V1428   & $\surd$       & $\surd$   & $\surd$ & & & & \\
 \hline
    \end{tabular}}
    \caption{$\surd$= residue present in the model containing only the sequence sections having a secondary helical structure, $\times$=residue not present and $\times\circ$= residual not present, but its first neighbor yes.}
   \label{tab:hSNPRamachandranPlot}
\end{table}
\section{Energy Landscape}
\begin{figure}[ht]
    \centering
    \includegraphics[width=0.8\textwidth]{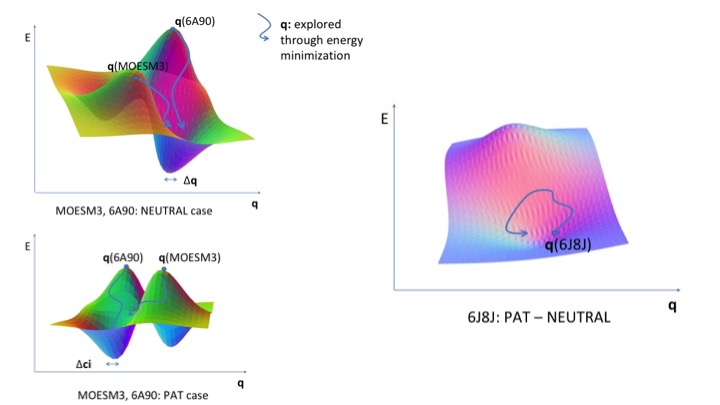}
    \caption{Hypothetical energy landscape explored in the energy minimization step for each template.}
    \label{fig:enrscape}
\end{figure}

A pivotal step, immediately following homology modeling, is the step involving the energy minimization. It is well known the connection between function and structure of a protein, therefore mutations that cause similar physiological changes will also have similar structures.
We used the same starting geometry (reference template) for all the protein sequences that we modeled. 
And the basic idea is that the energy minimization step, by taking into account the punctual amino acid differences, would have led the models to divide into two groups characterized by similar geometries. Indeed, these two groups of similar geometries correspond precisely to the two groups having different physiological behaviors (see Figure \ref{fig:enrscape}).

\section{Graph kernels and Dominant Set further results }

In this section we complete the presentation of the kernels results by showing the similarity matrices and the dominant set results not included in the main paper. 
We recall that each similarity matrix has rows and columns numbered in the range 0-84, which are the mutations ids already shown in the main paper: 
according to the ids list, note that ids 0--29 are relative to pathogenic mutations while ids 30--84 identify mutations not associated with pain disorders. 
Each cell (i,j) in a matrix shows the similarity value between the i-th and j-th RINs. The lighter is the cell color the more similar are the two graphs (the main diagonal shows always the lightest color, being the result of the comparison of a graph with itself).

\begin{figure}[ht]
   \begin{subfigure}[b]{0.329\textwidth}
     \includegraphics[width=\linewidth]{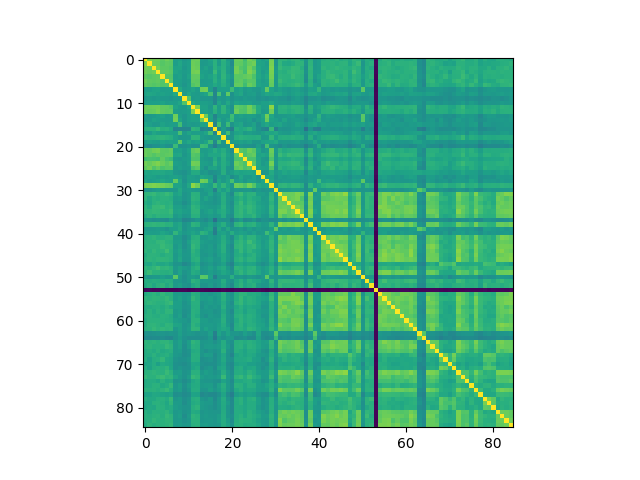}
     \caption{H-bonds}
   \end{subfigure}
   \begin{subfigure}[b]{0.329\textwidth}
     \includegraphics[width=\linewidth]{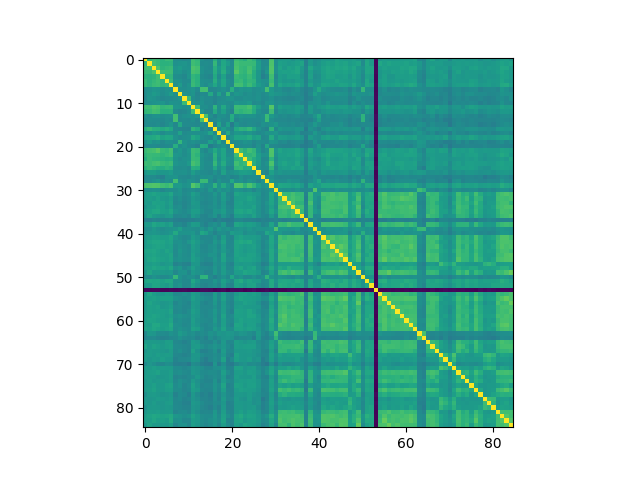}
     \caption{Van der Walls}
   \end{subfigure}
   \begin{subfigure}[b]{0.329\textwidth}
     \includegraphics[width=\linewidth]{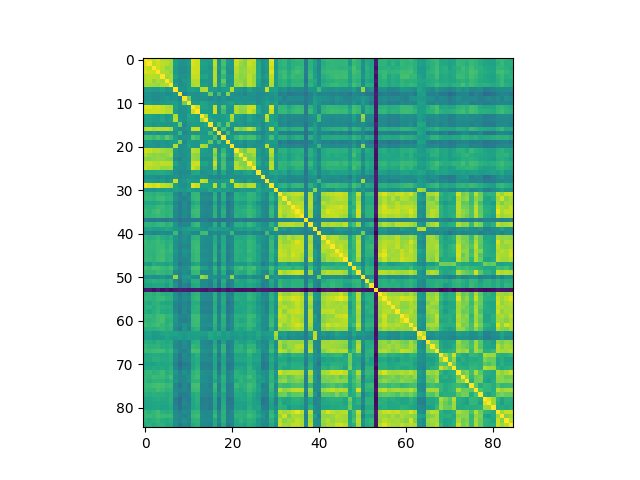}
     \caption{Ionic }
   \end{subfigure}
   \begin{subfigure}[b]{0.329\textwidth}
     \includegraphics[width=\linewidth]{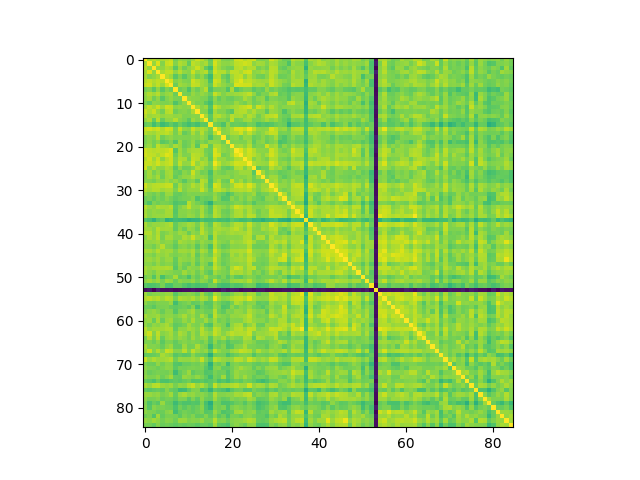}
     \caption{$\pi-\pi$ stack}
   \end{subfigure}
   \begin{subfigure}[b]{0.329\textwidth}
     \includegraphics[width=\linewidth]{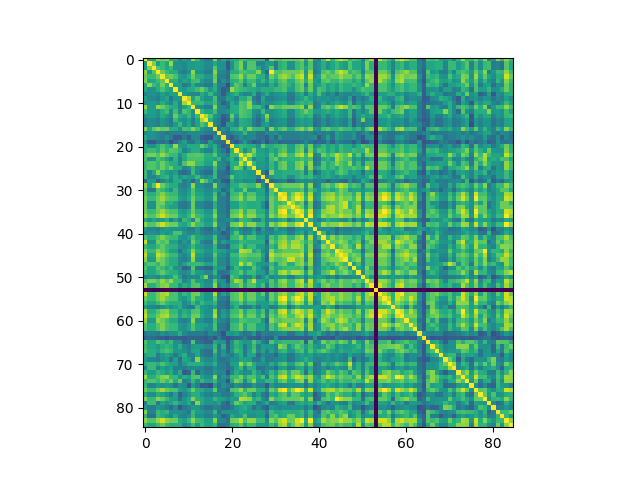}
     \caption{$\pi$-cations}
   \end{subfigure}
   \caption{Similarity matrices of the Weisfeiler-Lehman  (5 iterations) kernels applied to RINs resulting from MOESM3 template with 
   separated interactions }   
   \label{fig:kernelResultsMOESM3separated}   
\end{figure}

\begin{figure}[ht]
   \begin{subfigure}[b]{0.329\textwidth}
     \includegraphics[width=\linewidth]{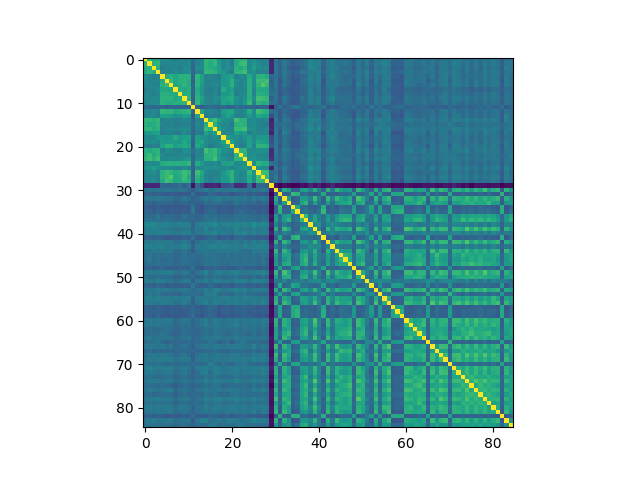}
     \caption{H-bonds}
   \end{subfigure}
   \begin {subfigure}[b]{0.329\textwidth}
     \includegraphics[width=\linewidth]{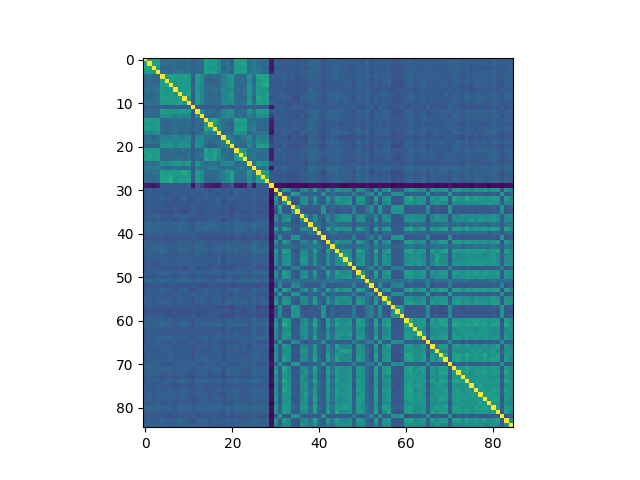}
     \caption{Van der Walls} 
   \end{subfigure} 
   \begin {subfigure}[b]{0.329\textwidth}
     \includegraphics[width=\linewidth]{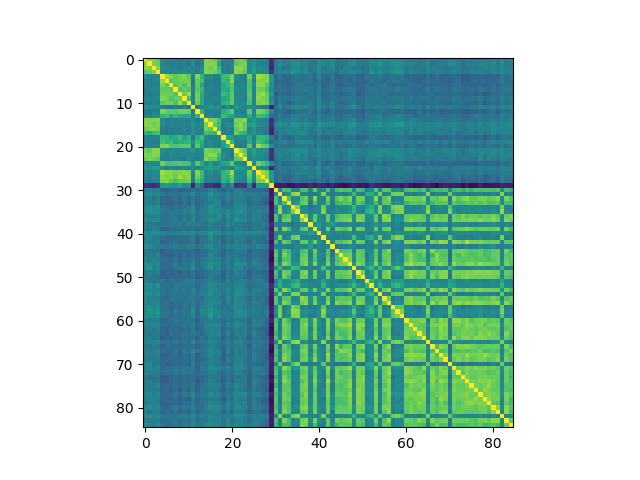}
     \caption{Ionic} 
   \end{subfigure} 
   \begin {subfigure}[b]{0.329\textwidth}
     \includegraphics[width=\linewidth]{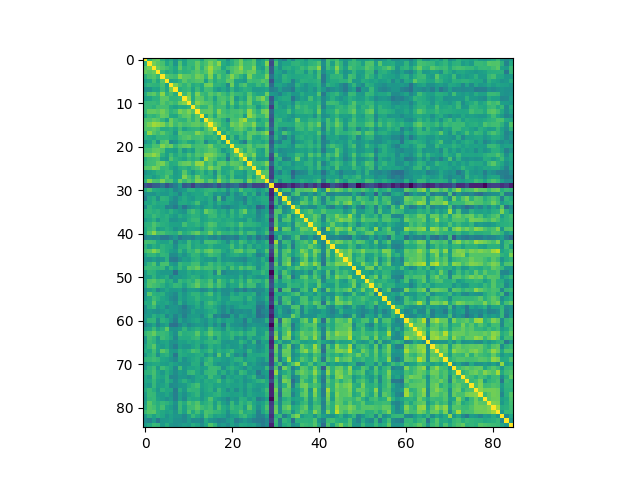}
     \caption{$\pi-\pi$ stack} 
   \end{subfigure} 
   \begin {subfigure}[b]{0.329\textwidth}
     \includegraphics[width=\linewidth]{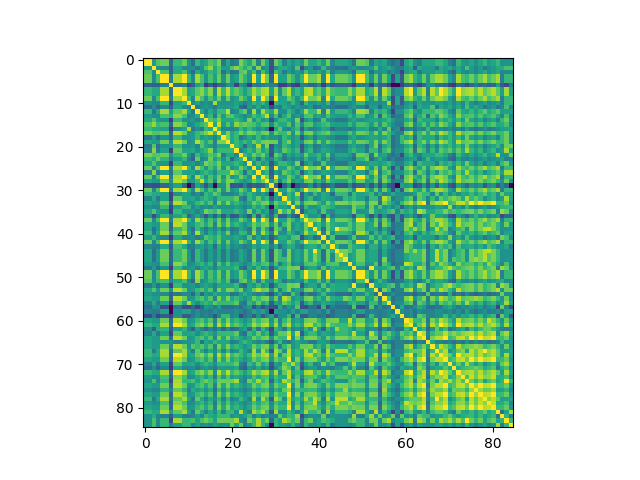}
     \caption{$\pi$ cations} 
   \end{subfigure} 
   \caption{Similarity matrices of the Weisfeiler-Lehman  (5 iterations) kernels applied to RINs resulting from 6A90 template with 
   separated interactions } 
   \label{fig:kernelResults6A90separated}
\end{figure}

The first analysis examines the role of each interaction separately, with the aim of checking their contribution to the pattern observed in the whole RINs comparison. Figures \ref{fig:kernelResultsMOESM3separated} and \ref{fig:kernelResults6A90separated} show the similarity matrices of the WL kernel 
for the MOESM3 template and 6A90 template, respectively. Note that interactions H-bond, Van Der Waals and Ionic, taken separately, are in agreement with the pattern observed in the comparison of the whole RINs as shown in the main paper. 

\begin{figure}[ht]
   \begin{subfigure}[b]{0.50\textwidth}
     \includegraphics[width=\linewidth]{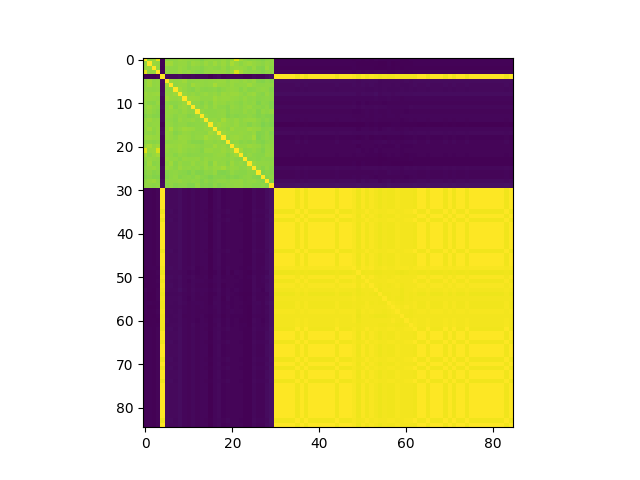}
     \caption{5HvX: VH}
   \end{subfigure}
   \begin{subfigure}[b]{0.50\textwidth}
     \includegraphics[width=\linewidth]{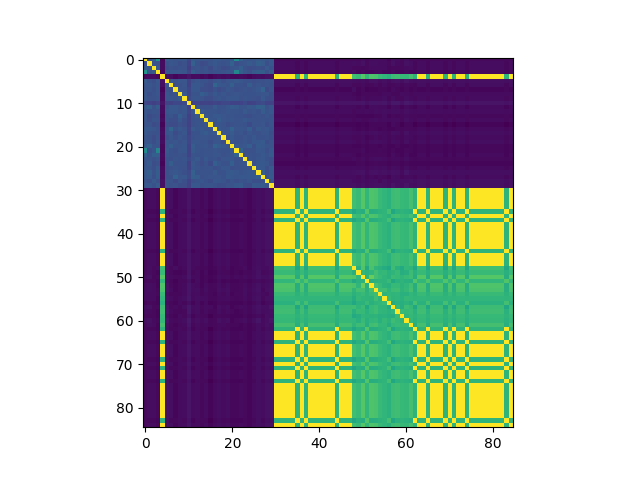}
     \caption{5HVX: WL}
   \end{subfigure}
   \caption{Similarity matrices of the VH and WL  (5 iterations) kernels applied to template 5HVX}
   \label{fig:kernelResults5HVX}
\end{figure}

Figure \ref{fig:kernelResults5HVX} shows the similarity matrices of the VH and WL kernels applied to the
5HVX template. 
This template depicts the sodium channel in the open state. It is interesting to observe that the kernel results are in line with the ones for MOESM3 and 6A90 also in this other configuration of the protein:  the two kernels are able to discriminate between pathogenic and not pathogenic mutations. In the open state, the separation of the two clusters is much more evident and it is also glaring that the non pathogenic mutations are much more similar one to each other than the pathogenic ones.

\begin{figure}[ht]
   \begin{subfigure}[b]{0.328\textwidth}
    \centering
     \includegraphics[angle=-90,origin=c, scale=0.17]{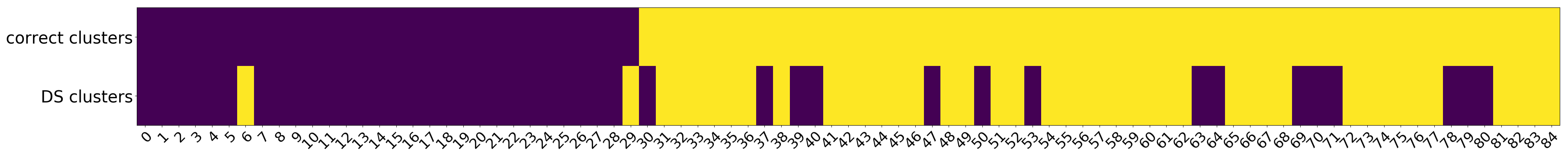}
     \caption{MOESM3}
   \end{subfigure}
   \begin{subfigure}[b]{0.328\textwidth}
    \centering     \includegraphics[angle=-90,origin=c, scale=0.17]{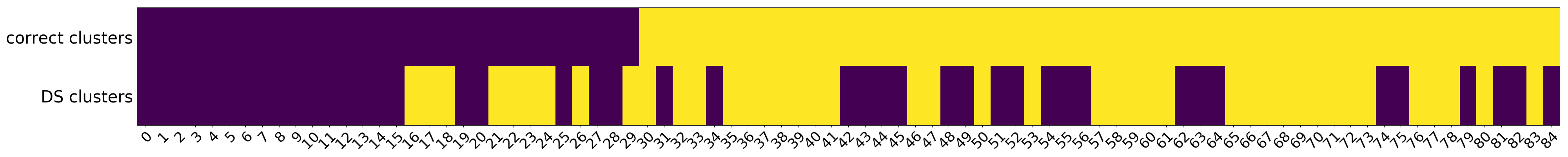}
     \caption{6J8J}
   \end{subfigure}   
   \begin{subfigure}[b]{0.328\textwidth}
    \centering     \includegraphics[angle=-90,origin=c, scale=0.17]{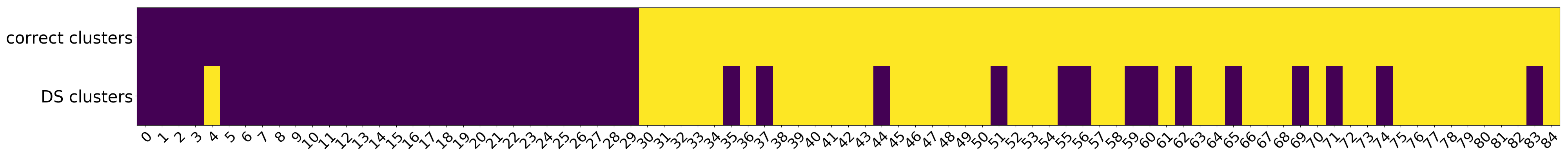}
     \caption{5HVX}
   \end{subfigure}
   \caption{DS classification for templates MOESM3, 6J8J and 5HVX with WL kernel.}
   \label{fig:dominantSetResults}
\end{figure}

Finally, Figure \ref{fig:dominantSetResults} shows the Dominant Set (DS) results for templates MOESM3, 6J8J and 5HVX. The diagrams  are composed of 85 rows labelled with the ids of considered mutations and two columns: one showing the classification resulting from the application of the Dominant Set method and the other one showing the correct classification, 
which is known, and distinguish between pain related mutations (dark color) and non-pain mutations (yellow). 
The DS classification of the three templates are in line with their WL kernel results. In particular, 
templates MOESM3 and 5HWX obtain a good classification: among the pathogenic mutations two are misclassified in MOESM3  and only one in 5HVX.  
Concerning template 6J8J, since the kernel is not able to discriminate the two clusters, it is not a surprise that also DS does not show a good result.
\section{Considering other human NaV1.7 templates}

\begin{figure}
    \centering
    \includegraphics[width=\textwidth]{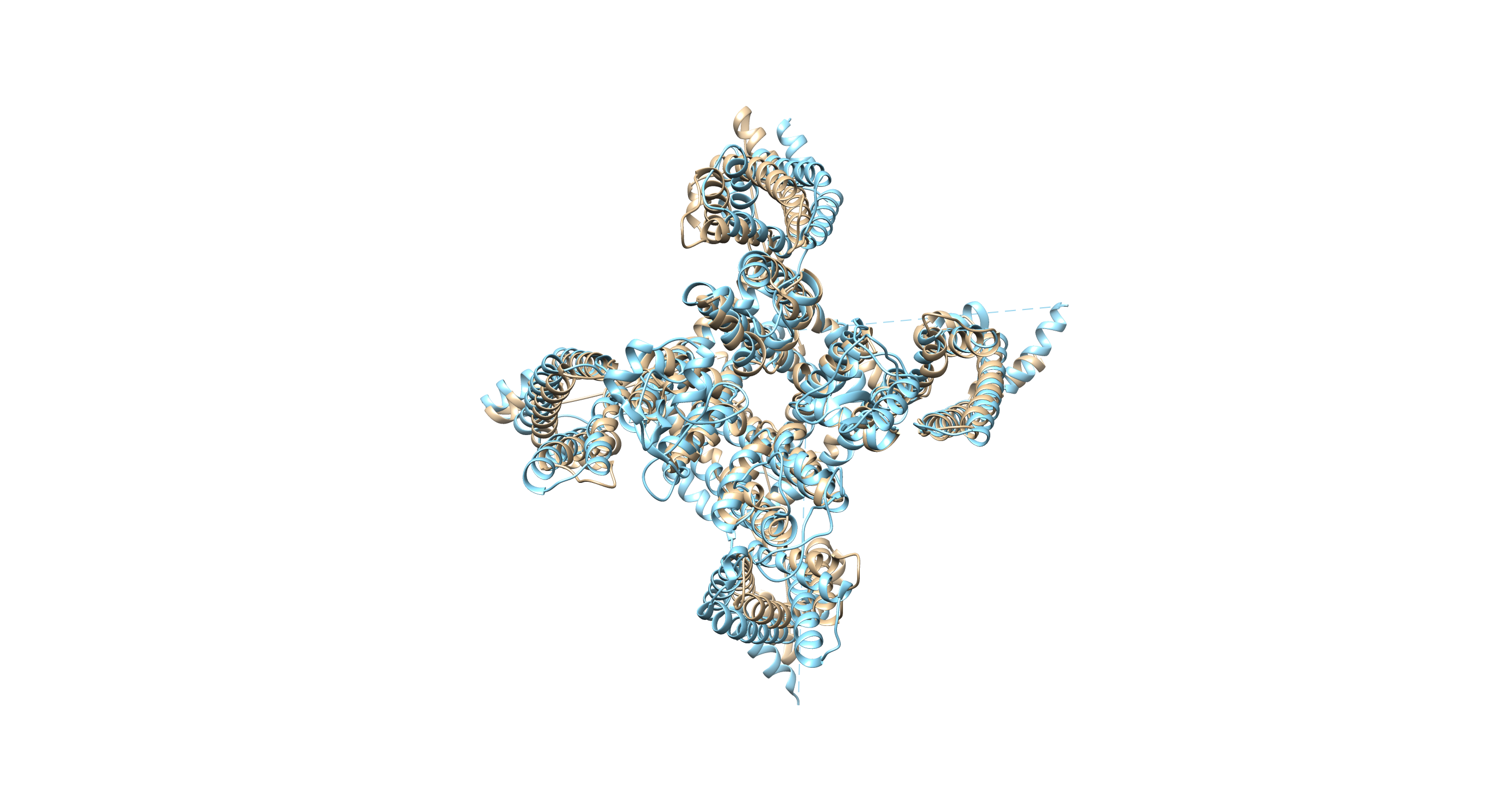}
    \caption{6J8J (in cyan) and 6N4I (in gold)}
    \label{fig:6J8J-6N4I}
\end{figure}

We considered three further templates that stem from the paper by Xu et al (Cell 2019) and refer to the following structures that model only the voltage-sensor domain II (VSD2) of Nav1.7. The PDB entry 6N4Q represents the protein in the activated state and is in complex with a spider toxin; the PDB entry 6N4R represents NaV1.7 in deactivated state, again in complex with a spider toxin. Neither of these two cases refer to the closed state and hence cannot be compared with those of the present study. 
Instead, the PDB entry 6N4I is is a chimeric structure that model the NaV1.7 protein in closed state.
As we further elaborated below, however, the quality of this template is significantly lower compared to the other human template 6J8J and it cannot be used efficiently in our computational pipeline.
We start by superimposing the two structures in Figure \ref{fig:6J8J-6N4I}   with 6J8J in cyan and 6N4I in gold.
The 6N4I template is obtained via X-ray crystallographic diffraction with 3.54 \AA ~ resolution and is formed by only 245 amino acids that belong to the second domain. Such structure has been then replicated by symmetry in order to model  the other three domains and complete the structure of the sodium channel.  Instead, the 6J8J template represents 1193  amino acids out of the nearly 2000 of the whole protein sequence. We compared the two templates 6N4I and 6J8J and calculated their Root Mean Squared Deviation (RMSD): while the RMSD between 128 selected and pruned amino acids  is 1.25 A, the total RMSD is larger that 35 A.

\begin{figure}
    \centering
    \includegraphics[width=\textwidth]{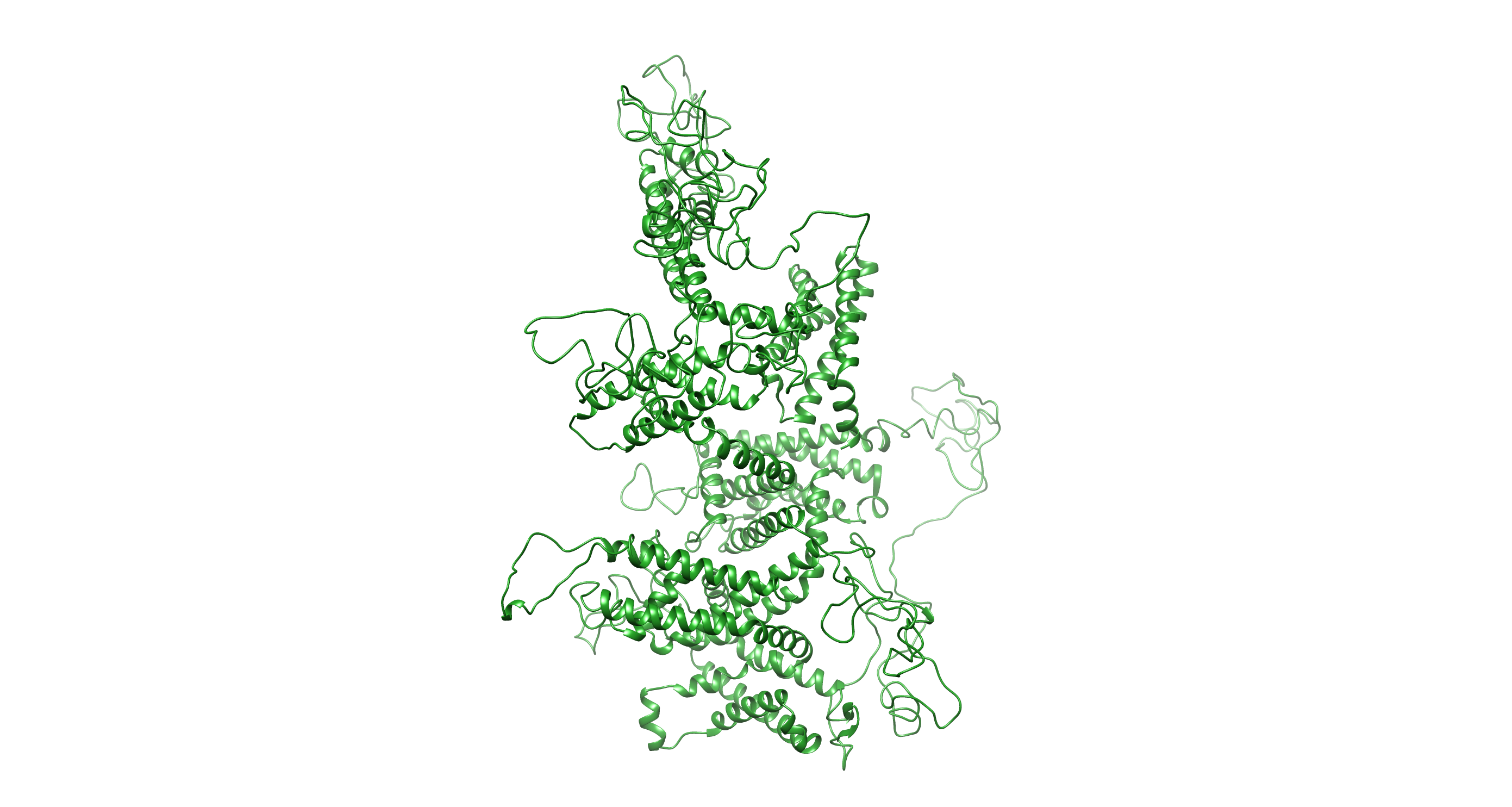}
    \caption{6N4I: homology modeling resulting structure for the WT sequence}
    \label{fig:6N4I-WT}
\end{figure}

This notwithstanding, we have used Swiss-model to perform homology modelling using the WT sequence and the 6N4I template, in a way akin to that we did for the 6J8J template in the original manuscript.  The resulting structure is depicted in Figure \ref{fig:6N4I-WT}. 

As noticeable, the geometry of the sodium channel is quite different with respect to the original template at variance with what was happening when using the 6J8J template. We ascribed this result both to the significantly lower quality of the original crystal of the 6N4I deposited structure and to the much lower number of represented amino acids. Hence, the structure of 6N4I is clearly insufficient to be used for homology modeling.
 
\clearpage

\end{document}